\documentclass[prl,aps,superscriptaddress,twocolumn,floatfix,showpacs]{revtex4-1}
\usepackage{graphicx}
\usepackage{amsfonts}
\usepackage{epsfig}
\usepackage[pdftex]{color}
\usepackage{epstopdf}
\usepackage{amsmath,graphicx,amssymb,braket,subfigure,upgreek}
\usepackage[colorlinks=true, linkcolor=blue, citecolor=blue, urlcolor=blue, breaklinks=true]{hyperref}
\usepackage{bigints}
\usepackage{dcolumn}
\usepackage{bm}
\usepackage{mathtools}
\usepackage{mathrsfs}
\usepackage{frcursive}
\usepackage{dsfont}
\usepackage{bbm}
\usepackage[bottom]{footmisc}
\newcommand{\R}{{\rm Re}}
\newcommand{\I}{{\rm Im}}

\newcommand{\mi}{{\rm i}}

\newcommand{\gamA}{{{\gamma}_{{A}}}}
\newcommand{\gamB}{{{\gamma}_{{B}}}}
\newcommand{\gameff}{{{\gamma}^{\text{eff}}_{{A}}}}
\newcommand{\gAeff}{g^{\text{eff}}_{{A}}}

\newcommand{\DB}{\Delta_{{B}}}
\newcommand{\Dc}{\Delta_{{c}}}
\newcommand{\gA}{g_{{A}}}
\newcommand{\gB}{g_{{B}}}
\newcommand{\keff}{\kappa^\text{eff}}

\begin{document}
\title{Ensemble-induced strong light-matter coupling of a single quantum emitter}

\author{S.~Sch\"utz}
\affiliation{ISIS (UMR 7006) and icFRC, University of Strasbourg and CNRS, 67000 Strasbourg, France}
\author{J.~Schachenmayer}
\affiliation{ISIS (UMR 7006) and icFRC, University of Strasbourg and CNRS, 67000 Strasbourg, France}
\author{D.~Hagenm\"uller}
\affiliation{ISIS (UMR 7006) and icFRC, University of Strasbourg and CNRS, 67000 Strasbourg, France}
\author{G.~K.~Brennen}
\affiliation{Department of Physics \& Astronomy and ARC Centre of Excellence for Engineered Quantum Systems, Macquarie University, NSW 2109, Australia}
\author{T.~Volz}
\affiliation{Department of Physics \& Astronomy and ARC Centre of Excellence for Engineered Quantum Systems, Macquarie University, NSW 2109, Australia}
\author{V.~Sandoghdar}
\affiliation{Max Planck Institute for the Science of Light, Staudtstra{\ss}e 2,
	D-91058 Erlangen, Germany}
\affiliation{Department of Physics, University of Erlangen-Nuremberg, Staudtstra{\ss}e 7,
	D-91058 Erlangen, Germany}
\author{T.~W.~Ebbesen}
\affiliation{ISIS (UMR 7006) and icFRC, University of Strasbourg and CNRS, 67000 Strasbourg, France}
\author{C.~Genes}
\affiliation{Max Planck Institute for the Science of Light, Staudtstra{\ss}e 2,
	D-91058 Erlangen, Germany}
\affiliation{Department of Physics, University of Erlangen-Nuremberg, Staudtstra{\ss}e 7,
	D-91058 Erlangen, Germany}
\author{G. Pupillo}
\affiliation{ISIS (UMR 7006) and icFRC, University of Strasbourg and CNRS, 67000 Strasbourg, France}

\date{\today}

\begin{abstract}
We discuss a technique to strongly couple a single target quantum emitter to a cavity mode, which is enabled by virtual excitations of a nearby mesoscopic ensemble of emitters. A collective coupling of the latter to both the cavity and the target emitter induces strong photon non-linearities in addition to polariton formation, in contrast to common schemes for ensemble strong coupling. We demonstrate that strong coupling at the level of a single emitter can be engineered via coherent and dissipative dipolar interactions with the ensemble, and provide realistic parameters for a possible implementation with SiV$^{-}$ defects in diamond. Our scheme can find applications, amongst others, in quantum information processing or in the field of cavity-assisted quantum chemistry.
\end{abstract}

\maketitle

In the strong coupling regime of light-matter interactions, the coherent exchange of energy between matter and light occurs at a rate faster than all dissipative processes~\cite{Kimble_1998,Haroche_review}. Strong coupling has been actively pursued in atomic~\cite{Line_shape_1983,Rydberg_1984,Kimble_1992,Haroche_PRL,Rempe,colombe} and molecular~\cite{doi:10.1063/1.443834_1982,Lidzey,PhysRevLett.93.036404,dintinger_2005,ebbesen4_2015,Baum,Wang2017_Coher,aizer_2017,PhysRevX.7.021014} physics, circuit quantum electrodynamics (QED)~\cite{Blais2004,Wall,chiorescu2004,scho2008,gu2017}, and condensed-matter physics~\cite{Yoshie,hennessy,RevModPhys.82.1489,RevModPhys.85.299}, with applications in, e.g., quantum information processing~\cite{Zoller_1995,PhysRevLett.75.4710,imam,PhysRevLett.83.5166,PhysRevLett.85.2392,Doherty_2002,brien}, transport~\cite{orgiu2015conductivity,PhysRevLett.114.196402,PhysRevLett.114.196403,doi:10.1002/anie.201703539,PhysRevLett.119.223601,doi:10.1021/acsphotonics.7b01332,magneto}, optomechanics~\cite{Brennecke235,Aspelmeyer,PhysRevLett.103.063005,PhysRevLett.109.223601,PhysRevLett.112.013601,Benz726}, and quantum chemistry~\cite{hutch_2012,thomas_2016,herrera_2016,galego_2016,Flick201615509,C8SC01043A,Thomas2018}.

For a single quantum emitter, strong coupling is accompanied by non-linear quantum effects leading to, e.g., the generation of nonclassical states of light used in photon quantum gates~\cite{boca,dirk,PhysRevLett.106.243601}. Reaching strong coupling, however, is often difficult as it relies on the integration of small volume resonators with large quality factors. For mesoscopic ensembles this requirement is relaxed as the coupling can be collectively enhanced by a factor $\sqrt{N}$ (for $N$-emitters)~\cite{PhysRev.93.99,PhysRev.170.379,Kimble_1992}. This allows for the realization of collective polariton dynamics~\cite{PhysRevLett.51.1175,PhysRevLett.63.240}, however, it does not enhance nonlinear quantum effects for the cavity mode~\cite{PhysRevLett.67.1727,CARMICHAEL199173,PhysRevA.98.013839,Trivedi2019}, limiting its use for, e.g., quantum optics and quantum information processing~\cite{10.1007/3-540-40894-0_18}. 

Here, we introduce an alternative approach where an ensemble of two-level quantum emitters with negligible populations modifies the cavity coupling of an individual beneficiary \textit{target} two-level system. This allows for bringing the single target emitter from the weak to the strong coupling regime, even when the mediating ensemble is only weakly coupled to the cavity. The induced strong coupling is visible in collectively-enhanced single emitter-cavity vacuum Rabi splittings. Remarkably, here the resulting photon non-linearities are also enhanced by collective effects. We identify two distinct mechanisms for reaching strong coupling in the target-cavity system, based on an increase of coherent coupling strength or a modification of dissipation, respectively. In the first case, we provide a possible implementation of our scheme with realistic parameters considering SiV$^{-}$ centers in diamond coupled to a microcavity. 
 
Our results can provide a viable path towards the realization of quantum non-linear optics experiments within cavity QED with moderate light-matter coupling strengths or cavity quality factors. In contrast to schemes using active dressing of, e.g., Rydberg states~\cite{PhysRevA.94.053830,Motzoi_2018}, our setup solely relies on virtual excitations of a nearby ensemble. 

\begin{figure*}[t]
\centering
\includegraphics[width=1\textwidth]{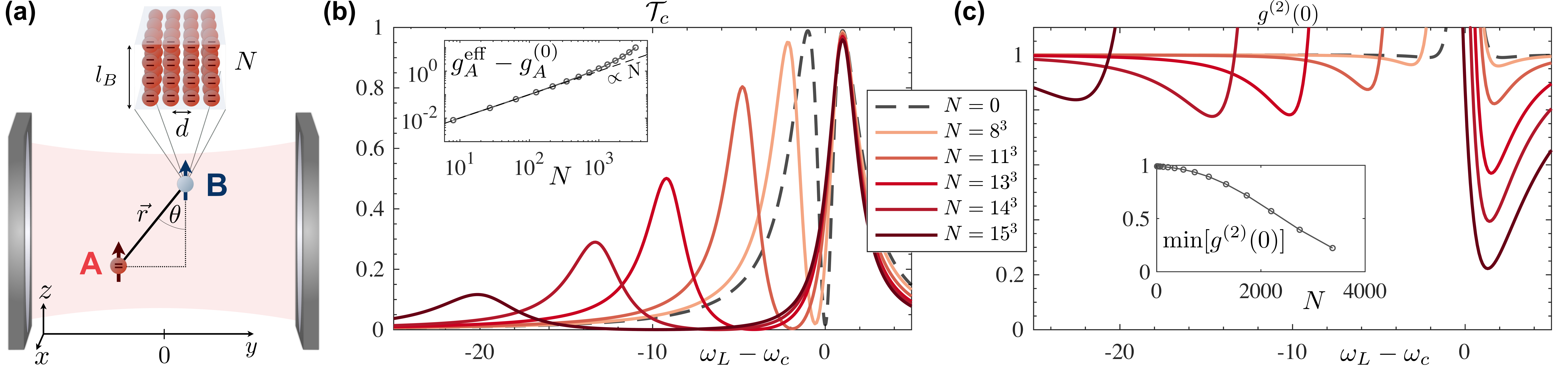}
\caption{(a) A single target emitter $A$ is weakly coupled to a resonant ($\omega_A = \omega_c$) cavity with strength $g_A^{(0)}$, and close to a small ensemble $B$. As $B$ we consider a cube of $N$ emitters identical to $A$ but with different transition frequencies $\omega_B>\omega_A$. When $l_B \ll r$, $B$ can be approximated by a single point-dipole at distance $|\vec{r}|\equiv r$. (b) The normalized cavity transmission spectrum $\mathcal{T}_c$ probed by a weak laser field with frequency $\omega_{L}$~\cite{SM} shows a splitting increasing with $N$, which results from an enhanced effective cavity-coupling strength $g_A^{\rm eff} > g_A^{(0)}$ of the {\em single} target emitter $A$. We find  $g_A^{\rm eff} - g_A^{(0)} \propto N$ (see inset and text). (c) Equal-time cavity photon-photon correlation function $g^{(2)}(0)$. The induced cavity-coupling strength occurs on a {\em single} emitter level, giving rise to photon blockade effects, $g^{(2)}(0) \ll 1$ and $\min[g^{(2)}(0)]$ decreasing with $N$ (inset). [Parameters for (b)/(c): $g_A^{(0)}\equiv 1$, $\gamma_A=0.01$,  $\kappa = 2$, $\omega_B- \omega_A = 1000$, cavity-wavelength $\lambda$, $\vec{r}_A = (0,0,0)$, $\vec{r}=(0,0,\lambda/20)$, $d = 10^{-3} \lambda$ (lattice spacing)].}\label{fig:intro}
\end{figure*}

The basic scheme is shown in Fig.~\ref{fig:intro}(a). The target emitter $A$ and $N$ ensemble emitters denoted as $B$ are coupled to a cavity mode via a Jaynes-Cummings interaction~\cite{JC_1963} $H_{\text{JC}} = a \left( g_A \sigma_A^+ + \sum_{\ell=1}^{N} g_\ell \sigma_\ell^+ \right) + {\rm h.c.}$ and to each other via coherent dipole-dipole interactions~\cite{Lehmberg1970_Radia} $H_{\rm DD} = \sum_{i\neq j} \sqrt{\gamma_i\gamma_j} g(\vec{r}_{ij}) \sigma_{i}^+ \sigma_{j}^-$ (where $i,j = A, 1,\dots,N)$. Here, $\sigma_{i}^\pm$ denote the raising/lowering operators for the emitters and $g_i = g_{i}^{(0)}\cos ( k y_{i}) $ their position-dependent coupling strength to the cavity mode (wavelength $\lambda$, $k = 2\pi/\lambda$, annihilation operator $a$, and cavity mode profile constant along $x$ and $z$). The coherent dipole-dipole coupling strengths depend on the relative positions of the emitters $\vec{r}_{ij} = \vec{r}_i - \vec{r}_j$ via the anisotropic function $g{(\vec{r}_{ij})}$ (cf.~\cite{AB_long_paper} for details) and on their decay rate $\gamma_A$ and $\gamma_{\ell} \equiv \gamma_B$ for all $\ell=1,\cdots,N$. For significant collective enhancement of the $A$-cavity coupling, we require $ |\vec {r}_{\ell A}| \lesssim \lambda$, so that the ensemble $B$ is in the near-field of $A$.  

The dynamics of the system with density operator  $\rho$ is governed by a standard master equation 
\begin{align}
\partial_t \rho = -\mi [H_0 + H_{\text{JC}} + H_{\text{DD}}, \rho] + \mathcal{L} \rho\, ,
\label{eq:QME}
\end{align}
where $\hbar \equiv 1$. In a frame rotating at the system $A$'s frequency $\omega_A$, $H_0 = \Dc a^{\dagger} a + \Delta_{B}\sum_{\ell} \sigma_{\ell}^+ \sigma_{\ell}^-$ represents the free Hamiltonian with detunings $\Delta_c = \omega_c - \omega_A$ and $\Delta_{B} = \omega_{B} - \omega_A$, where $\omega_c$ and $\omega_{B}$ are the bare cavity mode and ensemble emitter frequencies, respectively. 

Dissipation is described by $\mathcal{L} \rho$ $=$ $- \kappa \mathcal{D}(a^{\dagger},a) \rho - \sum_{i,j} \sqrt{\gamma_{i}\gamma_j} f{(\vec{r}_{ij})} \mathcal{D}(\sigma_i^+,\sigma_{j}^-) \rho\,$, with superoperator $\mathcal{D}(X,Y) \rho = [X,Y \rho] + [\rho X, Y]$. The first term on the right-hand side governs photon losses at rate $2\kappa$. Diagonal terms ($i = j$) in the second term correspond to standard spontaneous emission of the emitters. Off-diagonal ($i \neq j$) terms give rise to mutual decay into the environment depending on the separation of the emitters via the function $f(\vec{r}_{ij})$, ranging from $0$ at large separations to unity at zero separation~\cite{AB_long_paper}. In this near-field limit, they are responsible for the emergence of super- and subradiant modes~\cite{GROSS1982301}.

We are interested in obtaining an effective dynamics for the target A and the cavity, by adiabatically eliminating~\cite{AB_long_paper,doi:10.1063/1.1731409,PhysRevA.85.032111,Schuetz2013} the interacting ensemble of emitters $B$. When the latter is weakly populated, the equations of motion within the sub-system $B$ become linear, $\partial_t \langle \sigma_{\ell}^- \rangle = -\mi \sum_{m} (\mathbf{M})_{\ell m} \langle \sigma_{m}^- \rangle$, with $(\mathbf{M})_{\ell m} = (\Delta_B - \mi \gamma_B) \delta_{\ell m} + (1 - \delta_{\ell m}) \gamma_B \big[ g (\vec{r}_{\ell m}) - \mi f (\vec{r}_{\ell m}) \big]$ a $N \times N$ matrix~\cite{doi:10.1080/09500340.2011.594911,lesano}. Adiabatic elimination of $B$ generally requires that the eigenvalues $\lambda_\eta$ of $\mathbf{M}$ are the largest parameters (see~\cite{AB_long_paper} for details). This ensures that sub-system $B$ evolves fast compared to the reduced system including $A$ and the cavity. The coupling of the latter to $B$ also has to be perturbatively small such that $B$ remains weakly populated.
\begin{figure*}[t]
\centering
\includegraphics[width=1\textwidth]{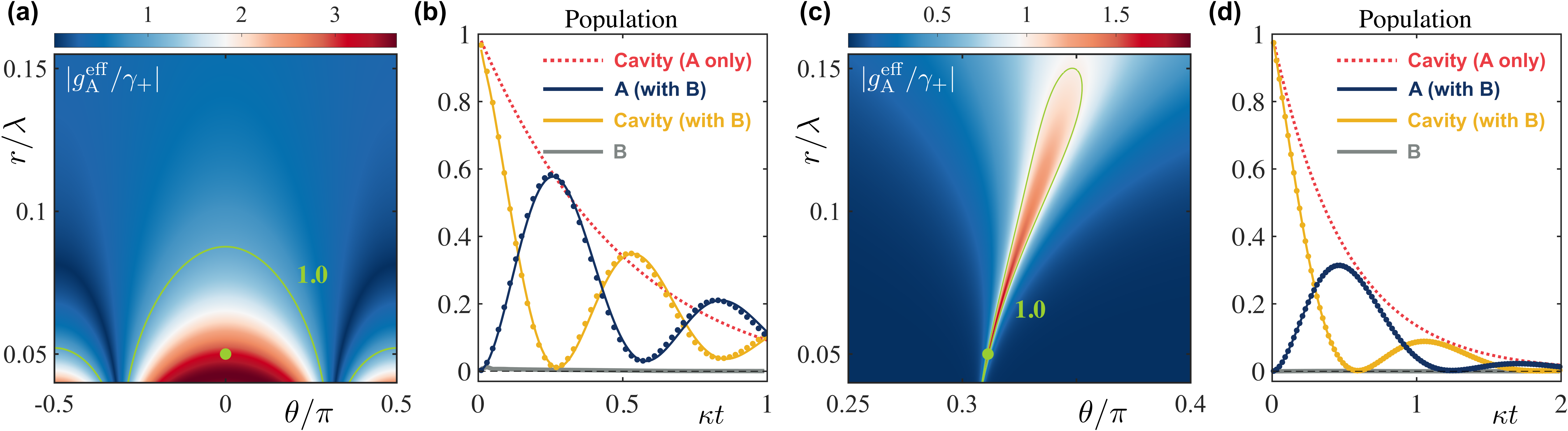}
\caption{(a)/(c) Contour plots of $|\gAeff/ \gamma_+|$ vs.~separation $\vec{r} (r,\theta)$ between $A$ and $B$ (point-dipole, see Fig.~\ref{fig:intro}). $|\gAeff/ \gamma_+|>1$ (inside the green contour line) indicates strong coupling. The parameters are chosen to illustrate either {\em coherent coupling inheritance} (a) or {\em dissipative linewidth narrowing} (c) for reaching strong coupling of $A$ (see text). (b)/(d) Time-evolution of population in $A$,  cavity (initial state with one photon) and $B$ (barely visible). Oscillations exemplify strong coupling [parameters for green dots in (a) and (c), we adapt $\omega_c$ such that $\Delta_A^{\rm eff}\simeq \Delta_c^{\rm eff}$]. Lines: full master equation simulation [Eq.~\eqref{eq:QME} with two emitters], points: effective model [Eq.~\eqref{eq:eqme}], dashed line: cavity decay without $B$. [Parameters (a)/(b): As in Fig.~\ref{fig:intro}, but choosing $g_B^{(0)} =100$ and thus $\gamma_B = \gamma_{A} (g^{(0)}_{B} /g^{(0)}_{A})^{2}=100$, $A$ and $B$ in the $x$-$z$ plane, $\vec{r} = r (\sin(\theta),0,\cos(\theta))$. (c)/(d): $\gamma_A =2.5$, $\kappa = 0.1$, $\omega_B = \omega_A$, $g_B^{(0)} = 20$ and thus $\gamma_B = 1000$, $\vec{r}_A = (0,-1/4,0)\lambda$, $A$ and $B$ in the $y$-$z$ plane, $\vec{r} = r (0,\sin(\theta),\cos(\theta))$.]}
\label{fig:geff_coh}
\end{figure*}

After elimination of $B$, the effective master equation for the reduced density operator $\rho^{\text{eff}}$ for target $A$ and cavity only reads
\begin{align}
{\partial_t} \rho^{\text{eff}} = - \mi [H_0^{\text{eff}} + H_{\rm JC}^{\rm eff}, \rho^{\text{eff}}] + \mathcal{L}^{\text{eff}} \rho^{\text{eff}}.
\label{eq:eqme}
\end{align}
The effective free Hamiltonian $H_0^{\text{eff}} = \Delta_A^{\text{eff}} \sigma_A^+ \sigma_A^- + \Delta_c^{\text{eff}} a^{\dagger} a$ contains renormalized detunings $\Delta_A^{\text{eff}}  = \omega_A^{\text{eff}} - \omega_A =   -  \R \big( \vec{v}^T \mathbf{M}^{-1} \vec{v} \big)$ and $\Delta_c^{\text{eff}} = \omega_c^{\text{eff}} - \omega_A = \Delta_c - \R \big( \vec{g}^T \mathbf{M}^{-1} \vec{g} \big)$, which can be generally compensated. 
Here, $(\vec{g})_{\ell} = g_{\ell}$ and  $(\vec{v})_{\ell} = \sqrt{\gamma_A \gamma_B} [ g (\vec{r}_{{\ell}A}) - \mi f (\vec{r}_{{\ell}A}) ]$.
The effective $A$--cavity interaction $H_{\text{JC}}^{\text{eff}} = g_A^{\text{eff}} \big( a^{\dagger} \sigma_A^- + \sigma_A^+ a \big)$ has the modified coherent coupling
\begin{align}
g_A^{\text{eff}} = g_A - \R \big( \vec{g}^T \mathbf{M}^{-1} \vec{v} \big) \,.
\label{eq:gAeff}
\end{align}
The effective dissipator for the reduced system in Lindblad form reads
$\mathcal{L}^{\text{eff}} \rho^{\text{eff}} =-\sum_{\nu = \pm}\gamma_{\nu} \mathcal{D}(L^\dag_\nu, L_\nu) \rho^{\rm eff}$ with rates
\begin{align}
\gamma_\pm = \frac{ \keff  + \gameff}{2} \pm \sqrt{\frac{ (\keff - \gameff)^2}{4} + \mu^2}.
\label{eq:gampm}
\end{align}
The main results of this work follow from Eqs.~\eqref{eq:gAeff} and \eqref{eq:gampm}.  We define the system to be strongly coupled when $|g_A^{\text{eff}}|$ exceeds the largest rate $\gamma_+$. The parameters $\kappa^{\rm eff} = \kappa + \I[\vec{g}^T  \mathbf{M}^{-1} \vec{g} ]$ and $\gameff = \gamA + \I[\vec{v}^T  \mathbf{M}^{-1} \vec{v} ]$ in Eq.~\eqref{eq:gampm} describe the modification of cavity decay and target emitter linewidth, respectively, due to the interaction with the $B$ ensemble. The rate $\mu = \I[\vec{g}^T  \mathbf{M}^{-1} \vec{v} ]$ gives rise to mutual decay mechanisms, dissipatively coupling $A$ and the cavity via Lindblad jump operators $L_+$ and $L_-$(see~\cite{AB_long_paper}). 

Using the mode-decomposition of {\bf M} in terms of its eigenvectors $\vec{x}_\eta$, the effective coupling modification in Eq.~\eqref{eq:gAeff} reads $\Delta g_{A}\equiv g^{\rm eff}_{A} - g_{A}=-\sum_\eta \R ( \vec{g}^T \vec{x}_\eta \vec{x}^T_\eta \vec{v} /\lambda_\eta)$, which depends both on the mode-overlaps $\vec{g}^T \vec{x}_\eta$, $\vec{v}^T \vec{x}_\eta$, and the complex eigenvalues $\lambda_\eta$, and can increase with $N$. For example, when the sum $\sum_\eta$ is dominated by a nearly homogeneous ($l_B\ll |\vec{r}_{\ell A}| \ll \lambda$) coupling to a symmetric mode $\vec{x}_S = (1,1,\dots,1)/\sqrt{N}$ of $B$ with eigenvalue $\lambda_{S}$, the enhancement scales as $\sim N$ since both mode-overlaps $\vec{g}^T \vec{x}_S$, $\vec{v}^T \vec{x}_S$ scale as $ \sim \sqrt{N}$. 

In Fig.~\ref{fig:intro}(b) we show the normalized cavity transmission spectrum $\mathcal{T}_c$ for the crystalline cubic structure in panel (a), where the symmetric mode $\vec{x}_S$ is found to dominate~\cite{SM}. Here, $\mathcal{T}_c(\omega_L)$ is computed numerically from Eq.~\eqref{eq:eqme} for a weak cavity drive (strength $\phi$) at frequency $\omega_L$, and is given by $\mathcal{T}_c(\omega_L)=\langle a^{\dagger}a\rangle (\kappa^2/\phi^2)$~\cite{SM}. We find a collective enhancement of the polariton-splitting that scales approximately $\propto N$, with super-linear corrections for large $N$ due to energy shifts in $\lambda_S$. We note that due to the small size of the cube ($l_B\ll |\vec{r}_{jA}| \ll \lambda$), the condition $\Delta_A^{\rm eff}\simeq \Delta_c^{\rm eff}$ is always satisfied. The polariton peaks display a marked asymmetry growing with $N$, which we attribute to the presence of the mutual decay rate $\mu$~\cite{AB_long_paper}.

Importantly, the collective enhancement leads to strong non-linear effects at the single photon level. Fig.~\ref{fig:intro}(c) shows the equal-time photon-photon correlation function $g^{(2)}(\tau=0)=\langle a^\dagger a^\dagger a a\rangle/\langle a^\dagger a\rangle^2$, computed for the same parameters as in panel (b) and using the steady state of Eq.~\eqref{eq:eqme} with a weak laser drive~\cite{SM}. The figure shows that $g^{(2)}(0)$ becomes $g^{(2)}(0)\lesssim 1$ at frequencies close to the polariton ones, and decreases with $N$. It can reach values $g^{(2)}(0)\ll 1$ for the upper polariton, reflecting the polariton asymmetry in $\mathcal{T}_c$ found above. This demonstrates that it is possible to achieve the ``photon blockade'' regime for a single target quantum emitter solely by increasing the number $N$ of dipoles making up the mediating mesoscopic ensemble.

When $l_B \ll |\vec{r}_{\ell A}|$, the collective dipole of $B$ can be approximated by a point dipole of strength $g_B^{(0)}=\sqrt{N}g_A^{(0)}$ and $\gamma_B=N\gamma_A$ positioned at ${\vec r}_B$ (each dipole moment in $B$ is chosen identical to that of $A$). Analytical expressions for the parameters of Eq.~\eqref{eq:eqme} can then be obtained in the single-mode limit. In this case the effective coupling strength reads~\cite{AB_long_paper}
\begin{align}
\gAeff= \gA^{(0)}  \Big\{\cos(k y_{\text{A}}) -    \frac{\gamB\DB g(\vec{r}) + \gamB^2 f(\vec{r})}{\DB^2+\gamB^2}\cos(k y_{\text{B}})\Big\},
\label{eq:gA_eff}
\end{align}
where ${\vec r}\equiv {\vec r}_B-{\vec r}_A$. We find an increased effective cavity decay, $\keff = \kappa + \gamB \gB^2({\vec{r}}_{\rm B})/(\DB^2 + \gamB^2)$, as well as a modified linewidth
\begin{align}
\gameff =
\gamA \Big\{ 1-f(\vec{r})^2+\frac{\left[g(\vec{r})\gamB- f(\vec{r}) \DB\right]^2}{\DB^2+\gamB^2} \Big\}.
\label{eq:gam_A_eff}
\end{align}
The mutual decay term entering Eq.~\eqref{eq:gampm} instead reads $\mu=\gB \sqrt{\gamA \gamB} ( g(\vec{r})\gamB - f(\vec{r}) \DB ) /  ( \DB^2 + \gamB^2 )$. The equations above show that the parameters in Eq.~\eqref{eq:eqme} are generally tuned by a combination of coherent/dispersive~\cite{PhysRevLett.112.213602} and incoherent/dissipative terms (e.g., $\sim \Delta_B g(\vec{r})$ and $\sim \gamB f(\vec{r})$ in Eq.~\eqref{eq:gA_eff}, respectively), and depend strongly on geometrical effects, due to the dipolar nature of the $A$-$B$ interaction. For a given geometry (i.e., positioning of $A$ and $B$, and dipole orientations), $\gAeff$ can be significantly increased by modifying the detuning $\Delta_B$, an effect that we term {\it coherent inheritance of coupling strength}. A competing effect that we dub {\it reduction of linewidth} in $A$ can be similarly obtained for $\gameff$ in Eq.~\eqref{eq:gam_A_eff} (even in the absence of a cavity, e.g. by putting the third term on the r.h.s. of the equation to zero, and for small separations where $f(\vec{r})^2\sim 1$). Such linewidth narrowing effect can be related to the formation of a subradiant mode.
\begin{figure*}[t]
\centering
\includegraphics[width=\textwidth]{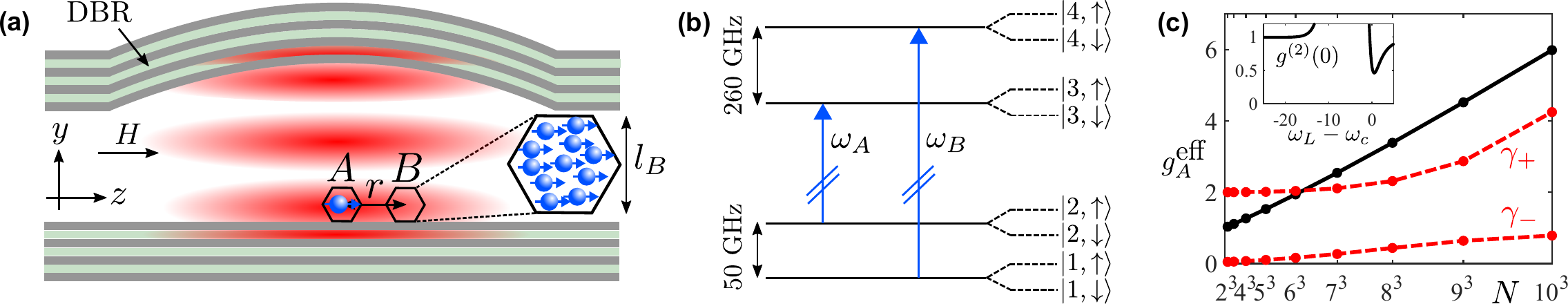}
\caption{Implementation with SiV$^{-}$ centers coupled to a microcavity mode (red). (a) A single SiV$^{-}$ center ($A$) is deposited above the bottom distributed Bragg reflector (DBR). A nanodiamond of size $l_{B}$ (modeled as a cube with lattice spacing $d=8$ nm $\simeq 0.01 \lambda$) separated from $A$ by a distance $\vec{r} = (0,0,60)$ nm contains an ensemble of $N$ SiV$^{-}$ centers ($B$). (b) Energy levels of an SiV$^{-}$ center under a magnetic field $H$ along $z$. The optical transitions $\ket{2}\to\ket{3}$ and $\ket{1}\to\ket{4}$ with frequencies $\omega_{A}$ (wavelength $\lambda=737$ nm) and $\omega_{B}$ correspond to $A$ and $B$, respectively. (c) Scaling of $g^{\rm eff}_{A}$ (black) showing that the system can be brought into strong coupling ($g^{\rm eff}_{A}>\gamma_{+}$) by increasing $N$. The $g^{(2)}(0)$ for $N=1000$ is shown in the inset as a function of the detuning $\omega_{L}-\omega_{c}$. Other parameters are $g^{(0)}_{A}\equiv 1$, $\omega_{A}=\omega_{c}$, $\gamma_{A}=\gamma_{B}=0.044$, $\kappa=2$, and $\Delta_{B}=265$, $\vec{r}_A = (0,0,0)$.} 
\label{fig:coll}
\end{figure*}

Figs.~\ref{fig:geff_coh}(a) and (c) show contour plots of $|\gAeff/\gamma_+|$ as a function of the relative distance $r$ between $A$ and $B$, and the dipole-orientation angle $\theta$ [see Fig.~\ref{fig:intro}(a)], for two choices of parameters where coherent inheritance of coupling strength [panel (a)] or reduction of linewidth [panel (c)] are the dominant effects. In the two panels, the bare parameters of Eq.~\eqref{eq:eqme} are chosen as $\gamma_A < g_A^{(0)} <  \kappa$ and $\kappa< g_A^{(0)} <  \gamma_A $, respectively. In (a) we further choose a large detuning $|\Delta_B| \gg \gamma_B$ (with $\gamma_B=10^{4}\gamma_{A}$). This implies an effective coherent coupling $g_A^{\text{eff}}/g_A^{(0)} \approx  \cos(ky_A) -  \frac{\gamB g(\vec{r}) }{\Delta_B} \cos(ky_B)$, when the positions of $A$ and $B$ are such that $|g(\vec{r})| > |f(\vec{r})|$. Note that while larger effective couplings can be obtained for smaller $\vert \Delta_{B}\vert$, the latter has to be large enough to satisfy the conditions for adiabatic elimination. In (c), instead, $A$ and $B$ are resonant ($\Delta_B=0$) and $\gamma_B$ is large, such that the effective coherent coupling and the linewidth simplify to $g_A^{\text{eff}}/g_A^{(0)} \approx  \cos(ky_A) -  f(\vec{r})\cos(ky_B)$ and $\gamma_A^{\rm eff}= \gamma_A [1 - f(\vec{r})^2 + g(\vec{r})^2]$, respectively. Furthermore, we position $A$ such that $g_A^{\text{eff}}/g_A^{(0)} \approx  -  f(\vec{r})\cos(ky_B)$ only stems from the dissipative dipole term $f(\vec{r})$. In both cases, the figure shows that a regime of strong coupling with $\vert g_A^{\text{eff}}/\gamma_+ \vert >1$ between the target $A$ and the cavity can be reached (area enclosed by the green line), e.g., in a ``head-to-tail'' configuration with $\theta\sim 0$ or close to the ``magic angle'' $\theta\sim 0.3\pi$ in panel (a) and (c), respectively. This is particularly surprising for panel (c), as coherent light-matter coupling is entirely due to dissipative effects.

Fig.~\ref{fig:geff_coh} confirms strong coupling with a simulation of the time-evolution for the mean populations of $A$, $B$, and the cavity, starting with a single photon. The presence of $B$ induces a clear coherent exchange of energy between the cavity and $A$ (negligible population in $B$), in contrast to the trivially damped behavior for the bare system. The validity of the effective master equation Eq.~\eqref{eq:eqme} is confirmed via a comparison to a full simulation of Eq.~\eqref{eq:QME}. We note that we have also demonstrated that our scheme is robust against inhomogeneous broadening and moderate positional disorder~\cite{SM}.

We now discuss a possible implementation of \emph{coherent coupling inheritance} with realistic parameters considering SiV$^{-}$ centers in diamond coupled to a microcavity of the kind described in Ref.~[\onlinecite{Najer2019}] [Fig.~\ref{fig:coll}(a)]. Two nanodiamonds separated by a distance $r=60$ nm containing respectively a single ($A$) and an ensemble ($B$) of SiV$^{-}$ defects are located above the cavity mirror. The nanodiamond with size $l_{B}$ containing $N$ dipoles $B$ could be a fragment of a bulk diamond that was grown with a high percentage of the centers oriented along the $y$ axis~\cite{Michl2014}, which coincides with the $\langle 111\rangle$ crystallographic direction~\footnote{Note that we are defining this direction as $y$ while defining it as $z$ is more common in the literature.}.

The energy levels of SiV$^{-}$ consist of electronic ground $E_g$ and excited $E_u$ state doublets with splittings of $\sim 50$ GHz and $\sim 260$ GHz, respectively~\cite{Hepp_Electronic_2014}. At temperatures below $4$ K and in the presence of a magnetic field $H$ aligned along $z$, the states are Zeeman-split and both $A$ and $B$ species can be prepared in the ground state $\ket{1,\downarrow}$ by spin-flipping optical pumping~\cite{Evans662}. The splitting of the $E_g$ doublet~\cite{Meesala2018} then allows for microwave resolvable addressing to map the dipoles $A$ to the state $\ket{2,\downarrow}$. $A$ and $B$ only couple to the optical transitions $\ket{2} \leftrightarrow \ket{3}$ and $\ket{1} \leftrightarrow \ket{4}$, respectively [Fig.~\ref{fig:coll}(b)] via photons polarized along $z$. A cavity mode with vacuum wavelength $\lambda =737$ nm is resonant with the target $A$ and detuned from $B$ by $\Delta_B = 2\pi \times 310$ GHz. 

The bare couplings $g^{(0)}_{A/B}$ can be estimated by comparing our setup to that of Ref.~[\onlinecite{Evans662}], where pairs of SiV$^{-}$ were coupled to a diamond photonic crystal cavity with strength $g_{\rm pcc}=2\pi \times 7.3$ GHz and mode volume $V_{\rm pcc}=0.5 (\lambda/2.4)^{3}$ ($2.4$ is the refractive index of diamond). In our setup [Fig.~\ref{fig:coll}(a)], we choose a microcavity with mode volume $V_{\rm mc}=1.4 \lambda^{3}$~\cite{Najer2019}, where the coupling strengths then read $g^{(0)}_{A/B} = g_{\rm pcc} \sqrt{V_{\rm pcc}/V_{\rm mc}} \approx 2\pi \times 1.17$ GHz. In Fig.~\ref{fig:coll} we model the nanodiamond $B$ as a cube with lattice spacing $d=8$ nm, which provides a density of emitters $\sim 2\times 10^{6} {\rm \mu m}^{-3}$~\cite{Bradac2017}. The decay rates are expected to be $\gamma_{A}=\gamma_{B}\sim 2\pi \times 51.5$ MHz~\cite{Jahnke_2015}. 
We choose $\kappa=2 g^{(0)}_{A}$ corresponding to a moderate cavity quality factor $Q \approx 8.7 \times 10^{4}$ (see, e.g., Refs.~\cite{doi:10.1021/acs.nanolett.7b05075,Najer2019,Riedel_Deterministic_2017}). This provides a finesse $F \approx 3 \times 10^4$ for a cavity of length $L = 3 \lambda/2 \approx 1.1 \mu $m.
Note that the non-radiative decay of the state $\ket{2}$ with rate $\sim 2\pi \times 13$ MHz~\cite{Jahnke_2015} can be overcome using a re-pumping scheme.

The effective coupling strength $g^{\rm eff}_{A}$ is displayed as a function of $N$ in Fig.~\ref{fig:coll}(c), showing that the single SiV$^{-}$ center $A$ can be brought to strong coupling with the microcavity mode ($g^{\rm eff}_{A} > \gamma_{+}$) due to the presence of the nanodiamond containing the $B$ emitters. Remarkably, this leads to strong non-linear effects $g^{(2)}(0)\approx 0.46$ for $N=1000$ and at a frequency close to that of the upper polariton.

{\em In conclusion}, we have investigated different mechanisms of how a {\em single} quantum emitter $A$, initially weakly coupled to a cavity mode, can be brought into strong coupling. The effect relies on both coherent and dissipative interactions with a nearby ensemble of identical emitters, and provides a viable path towards the realization of strong coupling at the level of a single SiV$^{-}$ center, which has, to the best of our knowledge, never been achieved. In contrast to usual collective strong coupling, our scheme gives rise to collectively-induced strong photon non-linearities and can find important applications in quantum information technologies with Si-V centers. Our work also emphasizes the important role played by optically active environments in cavity-QED. In particular, it is an interesting prospect to explore whether our model could be extended to vibrational strong coupling in molecular ensembles~\cite{Lather2018}, and to scalable microwave qubits magnetically coupled to a transmission line resonator on the verge of strong coupling~\cite{Viennot408}. 

{\textbf{Acknowledgements} --- }
We are grateful to C.~Genet and S.~Whitlock for stimulating discussions. Work in Strasbourg was supported by the ANR - ``ERA-NET QuantERA'' - Projet ``RouTe'' (ANR-18-QUAN-0005-01), IdEx Unistra - Projet ``STEMQuS'', LabEx NIE, and ERC Adv. Grant 788482 ``MOLUSC''. G.~B. and T.~V. acknowledge funding through the Australian Research Council Centre of Excellence for Engineered Quantum Systems EQUS (Project number CE170100009). C.~G. acknowledges support from the Max Planck Society and from the German Federal Ministry of Education and Research, co-funded by the European Commission (project RouTe), project number 13N14839 within the research program ``Photonik Forschung Deutschland''. This work is supported by IdEx Unistra (project STEMQuS) with funding managed by the French National Research Agency as part of the ``Investments for the future program''.

\bibliographystyle{apsrev4-1}
\bibliography{paperAB_arXiv}

\begin{thebibliography}{90}%
\makeatletter
\providecommand \@ifxundefined [1]{%
 \@ifx{#1\undefined}
}%
\providecommand \@ifnum [1]{%
 \ifnum #1\expandafter \@firstoftwo
 \else \expandafter \@secondoftwo
 \fi
}%
\providecommand \@ifx [1]{%
 \ifx #1\expandafter \@firstoftwo
 \else \expandafter \@secondoftwo
 \fi
}%
\providecommand \natexlab [1]{#1}%
\providecommand \enquote  [1]{``#1''}%
\providecommand \bibnamefont  [1]{#1}%
\providecommand \bibfnamefont [1]{#1}%
\providecommand \citenamefont [1]{#1}%
\providecommand \href@noop [0]{\@secondoftwo}%
\providecommand \href [0]{\begingroup \@sanitize@url \@href}%
\providecommand \@href[1]{\@@startlink{#1}\@@href}%
\providecommand \@@href[1]{\endgroup#1\@@endlink}%
\providecommand \@sanitize@url [0]{\catcode `\\12\catcode `\$12\catcode
  `\&12\catcode `\#12\catcode `\^12\catcode `\_12\catcode `\%12\relax}%
\providecommand \@@startlink[1]{}%
\providecommand \@@endlink[0]{}%
\providecommand \url  [0]{\begingroup\@sanitize@url \@url }%
\providecommand \@url [1]{\endgroup\@href {#1}{\urlprefix }}%
\providecommand \urlprefix  [0]{URL }%
\providecommand \Eprint [0]{\href }%
\providecommand \doibase [0]{http://dx.doi.org/}%
\providecommand \selectlanguage [0]{\@gobble}%
\providecommand \bibinfo  [0]{\@secondoftwo}%
\providecommand \bibfield  [0]{\@secondoftwo}%
\providecommand \translation [1]{[#1]}%
\providecommand \BibitemOpen [0]{}%
\providecommand \bibitemStop [0]{}%
\providecommand \bibitemNoStop [0]{.\EOS\space}%
\providecommand \EOS [0]{\spacefactor3000\relax}%
\providecommand \BibitemShut  [1]{\csname bibitem#1\endcsname}%
\let\auto@bib@innerbib\@empty
\bibitem [{\citenamefont {Kimble}(1998)}]{Kimble_1998}%
  \BibitemOpen
  \bibfield  {author} {\bibinfo {author} {\bibfnamefont {H.~J.}\ \bibnamefont
  {Kimble}},\ }\href {http://stacks.iop.org/1402-4896/1998/i=T76/a=019}
  {\bibfield  {journal} {\bibinfo  {journal} {Physica Scripta}\ }\textbf
  {\bibinfo {volume} {T76}},\ \bibinfo {pages} {127} (\bibinfo {year}
  {1998})}\BibitemShut {NoStop}%
\bibitem [{\citenamefont {Raimond}\ \emph {et~al.}(2001)\citenamefont
  {Raimond}, \citenamefont {Brune},\ and\ \citenamefont
  {Haroche}}]{Haroche_review}%
  \BibitemOpen
  \bibfield  {author} {\bibinfo {author} {\bibfnamefont {J.~M.}\ \bibnamefont
  {Raimond}}, \bibinfo {author} {\bibfnamefont {M.}~\bibnamefont {Brune}}, \
  and\ \bibinfo {author} {\bibfnamefont {S.}~\bibnamefont {Haroche}},\ }\href
  {\doibase 10.1103/RevModPhys.73.565} {\bibfield  {journal} {\bibinfo
  {journal} {Rev. Mod. Phys.}\ }\textbf {\bibinfo {volume} {73}},\ \bibinfo
  {pages} {565} (\bibinfo {year} {2001})}\BibitemShut {NoStop}%
\bibitem [{\citenamefont {Sanchez-Mondragon}\ \emph {et~al.}(1983)\citenamefont
  {Sanchez-Mondragon}, \citenamefont {Narozhny},\ and\ \citenamefont
  {Eberly}}]{Line_shape_1983}%
  \BibitemOpen
  \bibfield  {author} {\bibinfo {author} {\bibfnamefont {J.~J.}\ \bibnamefont
  {Sanchez-Mondragon}}, \bibinfo {author} {\bibfnamefont {N.~B.}\ \bibnamefont
  {Narozhny}}, \ and\ \bibinfo {author} {\bibfnamefont {J.~H.}\ \bibnamefont
  {Eberly}},\ }\href {\doibase 10.1103/PhysRevLett.51.550} {\bibfield
  {journal} {\bibinfo  {journal} {Phys. Rev. Lett.}\ }\textbf {\bibinfo
  {volume} {51}},\ \bibinfo {pages} {550} (\bibinfo {year} {1983})}\BibitemShut
  {NoStop}%
\bibitem [{\citenamefont {Agarwal}(1984)}]{Rydberg_1984}%
  \BibitemOpen
  \bibfield  {author} {\bibinfo {author} {\bibfnamefont {G.~S.}\ \bibnamefont
  {Agarwal}},\ }\href {\doibase 10.1103/PhysRevLett.53.1732} {\bibfield
  {journal} {\bibinfo  {journal} {Phys. Rev. Lett.}\ }\textbf {\bibinfo
  {volume} {53}},\ \bibinfo {pages} {1732} (\bibinfo {year}
  {1984})}\BibitemShut {NoStop}%
\bibitem [{\citenamefont {Thompson}\ \emph {et~al.}(1992)\citenamefont
  {Thompson}, \citenamefont {Rempe},\ and\ \citenamefont
  {Kimble}}]{Kimble_1992}%
  \BibitemOpen
  \bibfield  {author} {\bibinfo {author} {\bibfnamefont {R.~J.}\ \bibnamefont
  {Thompson}}, \bibinfo {author} {\bibfnamefont {G.}~\bibnamefont {Rempe}}, \
  and\ \bibinfo {author} {\bibfnamefont {H.~J.}\ \bibnamefont {Kimble}},\
  }\href {\doibase 10.1103/PhysRevLett.68.1132} {\bibfield  {journal} {\bibinfo
   {journal} {Phys. Rev. Lett.}\ }\textbf {\bibinfo {volume} {68}},\ \bibinfo
  {pages} {1132} (\bibinfo {year} {1992})}\BibitemShut {NoStop}%
\bibitem [{\citenamefont {Brune}\ \emph {et~al.}(1996)\citenamefont {Brune},
  \citenamefont {Schmidt-Kaler}, \citenamefont {Maali}, \citenamefont {Dreyer},
  \citenamefont {Hagley}, \citenamefont {Raimond},\ and\ \citenamefont
  {Haroche}}]{Haroche_PRL}%
  \BibitemOpen
  \bibfield  {author} {\bibinfo {author} {\bibfnamefont {M.}~\bibnamefont
  {Brune}}, \bibinfo {author} {\bibfnamefont {F.}~\bibnamefont
  {Schmidt-Kaler}}, \bibinfo {author} {\bibfnamefont {A.}~\bibnamefont
  {Maali}}, \bibinfo {author} {\bibfnamefont {J.}~\bibnamefont {Dreyer}},
  \bibinfo {author} {\bibfnamefont {E.}~\bibnamefont {Hagley}}, \bibinfo
  {author} {\bibfnamefont {J.~M.}\ \bibnamefont {Raimond}}, \ and\ \bibinfo
  {author} {\bibfnamefont {S.}~\bibnamefont {Haroche}},\ }\href {\doibase
  10.1103/PhysRevLett.76.1800} {\bibfield  {journal} {\bibinfo  {journal}
  {Phys. Rev. Lett.}\ }\textbf {\bibinfo {volume} {76}},\ \bibinfo {pages}
  {1800} (\bibinfo {year} {1996})}\BibitemShut {NoStop}%
\bibitem [{\citenamefont {Pinkse}\ \emph {et~al.}(2000)\citenamefont {Pinkse},
  \citenamefont {Fischer}, \citenamefont {Maunz},\ and\ \citenamefont
  {Rempe}}]{Rempe}%
  \BibitemOpen
  \bibfield  {author} {\bibinfo {author} {\bibfnamefont {P.~W.~H.}\
  \bibnamefont {Pinkse}}, \bibinfo {author} {\bibfnamefont {T.}~\bibnamefont
  {Fischer}}, \bibinfo {author} {\bibfnamefont {P.}~\bibnamefont {Maunz}}, \
  and\ \bibinfo {author} {\bibfnamefont {G.}~\bibnamefont {Rempe}},\ }\href
  {\doibase 10.1038/35006006} {\bibfield  {journal} {\bibinfo  {journal}
  {Nature}\ }\textbf {\bibinfo {volume} {404}},\ \bibinfo {pages} {365}
  (\bibinfo {year} {2000})}\BibitemShut {NoStop}%
\bibitem [{\citenamefont {Colombe}\ \emph {et~al.}(2007)\citenamefont
  {Colombe}, \citenamefont {Steinmetz}, \citenamefont {Dubois}, \citenamefont
  {Linke}, \citenamefont {Hunger},\ and\ \citenamefont {Reichel}}]{colombe}%
  \BibitemOpen
  \bibfield  {author} {\bibinfo {author} {\bibfnamefont {Y.}~\bibnamefont
  {Colombe}}, \bibinfo {author} {\bibfnamefont {T.}~\bibnamefont {Steinmetz}},
  \bibinfo {author} {\bibfnamefont {G.}~\bibnamefont {Dubois}}, \bibinfo
  {author} {\bibfnamefont {F.}~\bibnamefont {Linke}}, \bibinfo {author}
  {\bibfnamefont {D.}~\bibnamefont {Hunger}}, \ and\ \bibinfo {author}
  {\bibfnamefont {J.}~\bibnamefont {Reichel}},\ }\href {\doibase
  10.1038/nature06331} {\bibfield  {journal} {\bibinfo  {journal} {Nature}\
  }\textbf {\bibinfo {volume} {450}},\ \bibinfo {pages} {272} (\bibinfo {year}
  {2007})}\BibitemShut {NoStop}%
\bibitem [{\citenamefont {Pockrand}\ \emph {et~al.}(1982)\citenamefont
  {Pockrand}, \citenamefont {Brillante},\ and\ \citenamefont
  {M\"{o}bius}}]{doi:10.1063/1.443834_1982}%
  \BibitemOpen
  \bibfield  {author} {\bibinfo {author} {\bibfnamefont {I.}~\bibnamefont
  {Pockrand}}, \bibinfo {author} {\bibfnamefont {A.}~\bibnamefont {Brillante}},
  \ and\ \bibinfo {author} {\bibfnamefont {D.}~\bibnamefont {M\"{o}bius}},\
  }\href {\doibase 10.1063/1.443834} {\bibfield  {journal} {\bibinfo  {journal}
  {The Journal of Chemical Physics}\ }\textbf {\bibinfo {volume} {77}},\
  \bibinfo {pages} {6289} (\bibinfo {year} {1982})}\BibitemShut {NoStop}%
\bibitem [{\citenamefont {Lidzey}\ \emph {et~al.}(1998)\citenamefont {Lidzey},
  \citenamefont {Bradley}, \citenamefont {Skolnick}, \citenamefont {Virgili},
  \citenamefont {Walker},\ and\ \citenamefont {Whittaker}}]{Lidzey}%
  \BibitemOpen
  \bibfield  {author} {\bibinfo {author} {\bibfnamefont {D.~G.}\ \bibnamefont
  {Lidzey}}, \bibinfo {author} {\bibfnamefont {D.~D.~C.}\ \bibnamefont
  {Bradley}}, \bibinfo {author} {\bibfnamefont {M.~S.}\ \bibnamefont
  {Skolnick}}, \bibinfo {author} {\bibfnamefont {T.}~\bibnamefont {Virgili}},
  \bibinfo {author} {\bibfnamefont {S.}~\bibnamefont {Walker}}, \ and\ \bibinfo
  {author} {\bibfnamefont {D.~M.}\ \bibnamefont {Whittaker}},\ }\href {\doibase
  10.1038/25692} {\bibfield  {journal} {\bibinfo  {journal} {Nature}\ }\textbf
  {\bibinfo {volume} {395}},\ \bibinfo {pages} {53} (\bibinfo {year}
  {1998})}\BibitemShut {NoStop}%
\bibitem [{\citenamefont {Bellessa}\ \emph {et~al.}(2004)\citenamefont
  {Bellessa}, \citenamefont {Bonnand}, \citenamefont {Plenet},\ and\
  \citenamefont {Mugnier}}]{PhysRevLett.93.036404}%
  \BibitemOpen
  \bibfield  {author} {\bibinfo {author} {\bibfnamefont {J.}~\bibnamefont
  {Bellessa}}, \bibinfo {author} {\bibfnamefont {C.}~\bibnamefont {Bonnand}},
  \bibinfo {author} {\bibfnamefont {J.~C.}\ \bibnamefont {Plenet}}, \ and\
  \bibinfo {author} {\bibfnamefont {J.}~\bibnamefont {Mugnier}},\ }\href
  {\doibase 10.1103/PhysRevLett.93.036404} {\bibfield  {journal} {\bibinfo
  {journal} {Phys. Rev. Lett.}\ }\textbf {\bibinfo {volume} {93}},\ \bibinfo
  {pages} {036404} (\bibinfo {year} {2004})}\BibitemShut {NoStop}%
\bibitem [{\citenamefont {Dintinger}\ \emph {et~al.}(2005)\citenamefont
  {Dintinger}, \citenamefont {Klein}, \citenamefont {Bustos}, \citenamefont
  {Barnes},\ and\ \citenamefont {Ebbesen}}]{dintinger_2005}%
  \BibitemOpen
  \bibfield  {author} {\bibinfo {author} {\bibfnamefont {J.}~\bibnamefont
  {Dintinger}}, \bibinfo {author} {\bibfnamefont {S.}~\bibnamefont {Klein}},
  \bibinfo {author} {\bibfnamefont {F.}~\bibnamefont {Bustos}}, \bibinfo
  {author} {\bibfnamefont {W.~L.}\ \bibnamefont {Barnes}}, \ and\ \bibinfo
  {author} {\bibfnamefont {T.~W.}\ \bibnamefont {Ebbesen}},\ }\href {\doibase
  10.1103/PhysRevB.71.035424} {\bibfield  {journal} {\bibinfo  {journal} {Phys.
  Rev. B}\ }\textbf {\bibinfo {volume} {71}},\ \bibinfo {pages} {035424}
  (\bibinfo {year} {2005})}\BibitemShut {NoStop}%
\bibitem [{\citenamefont {Shalabney}\ \emph {et~al.}(2015)\citenamefont
  {Shalabney}, \citenamefont {George}, \citenamefont {Hutchison}, \citenamefont
  {Pupillo}, \citenamefont {Genet},\ and\ \citenamefont
  {Ebbesen}}]{ebbesen4_2015}%
  \BibitemOpen
  \bibfield  {author} {\bibinfo {author} {\bibfnamefont {A.}~\bibnamefont
  {Shalabney}}, \bibinfo {author} {\bibfnamefont {J.}~\bibnamefont {George}},
  \bibinfo {author} {\bibfnamefont {J.~A.}\ \bibnamefont {Hutchison}}, \bibinfo
  {author} {\bibfnamefont {G.}~\bibnamefont {Pupillo}}, \bibinfo {author}
  {\bibfnamefont {C.}~\bibnamefont {Genet}}, \ and\ \bibinfo {author}
  {\bibfnamefont {T.~W.}\ \bibnamefont {Ebbesen}},\ }\href {\doibase
  10.1038/ncomms6981} {\bibfield  {journal} {\bibinfo  {journal} {Nature
  Communications}\ }\textbf {\bibinfo {volume} {6}},\ \bibinfo {pages} {5981}
  (\bibinfo {year} {2015})}\BibitemShut {NoStop}%
\bibitem [{\citenamefont {Chikkaraddy}\ \emph {et~al.}(2016)\citenamefont
  {Chikkaraddy}, \citenamefont {de~Nijs}, \citenamefont {Benz}, \citenamefont
  {Barrow}, \citenamefont {Scherman}, \citenamefont {Rosta}, \citenamefont
  {Demetriadou}, \citenamefont {Fox}, \citenamefont {Hess},\ and\ \citenamefont
  {Baumberg}}]{Baum}%
  \BibitemOpen
  \bibfield  {author} {\bibinfo {author} {\bibfnamefont {R.}~\bibnamefont
  {Chikkaraddy}}, \bibinfo {author} {\bibfnamefont {B.}~\bibnamefont
  {de~Nijs}}, \bibinfo {author} {\bibfnamefont {F.}~\bibnamefont {Benz}},
  \bibinfo {author} {\bibfnamefont {S.~J.}\ \bibnamefont {Barrow}}, \bibinfo
  {author} {\bibfnamefont {O.~A.}\ \bibnamefont {Scherman}}, \bibinfo {author}
  {\bibfnamefont {E.}~\bibnamefont {Rosta}}, \bibinfo {author} {\bibfnamefont
  {A.}~\bibnamefont {Demetriadou}}, \bibinfo {author} {\bibfnamefont
  {P.}~\bibnamefont {Fox}}, \bibinfo {author} {\bibfnamefont {O.}~\bibnamefont
  {Hess}}, \ and\ \bibinfo {author} {\bibfnamefont {J.~J.}\ \bibnamefont
  {Baumberg}},\ }\href {\doibase 10.1038/nature17974} {\bibfield  {journal}
  {\bibinfo  {journal} {Nature}\ }\textbf {\bibinfo {volume} {535}},\ \bibinfo
  {pages} {127} (\bibinfo {year} {2016})}\BibitemShut {NoStop}%
\bibitem [{\citenamefont {Wang}\ \emph {et~al.}(2019)\citenamefont {Wang},
  \citenamefont {Kelkar}, \citenamefont {Martin-Cano}, \citenamefont
  {Rattenbacher}, \citenamefont {Shkarin}, \citenamefont {Utikal},
  \citenamefont {G\"{o}tzinger},\ and\ \citenamefont
  {Sandoghdar}}]{Wang2017_Coher}%
  \BibitemOpen
  \bibfield  {author} {\bibinfo {author} {\bibfnamefont {D.}~\bibnamefont
  {Wang}}, \bibinfo {author} {\bibfnamefont {H.}~\bibnamefont {Kelkar}},
  \bibinfo {author} {\bibfnamefont {D.}~\bibnamefont {Martin-Cano}}, \bibinfo
  {author} {\bibfnamefont {D.}~\bibnamefont {Rattenbacher}}, \bibinfo {author}
  {\bibfnamefont {A.}~\bibnamefont {Shkarin}}, \bibinfo {author} {\bibfnamefont
  {T.}~\bibnamefont {Utikal}}, \bibinfo {author} {\bibfnamefont
  {S.}~\bibnamefont {G\"{o}tzinger}}, \ and\ \bibinfo {author} {\bibfnamefont
  {V.}~\bibnamefont {Sandoghdar}},\ }\href
  {https://doi.org/10.1038/s41567-019-0436-5} {\bibfield  {journal} {\bibinfo
  {journal} {Nature Physics}\ } (\bibinfo {year} {2019})}\BibitemShut {NoStop}%
\bibitem [{\citenamefont {Zhang}\ \emph {et~al.}(2017)\citenamefont {Zhang},
  \citenamefont {Meng}, \citenamefont {Zhang}, \citenamefont {Luo},
  \citenamefont {Yu}, \citenamefont {Yang}, \citenamefont {Zhang},
  \citenamefont {Esteban}, \citenamefont {Aizpurua}, \citenamefont {Luo},
  \citenamefont {Yang}, \citenamefont {Dong},\ and\ \citenamefont
  {Hou}}]{aizer_2017}%
  \BibitemOpen
  \bibfield  {author} {\bibinfo {author} {\bibfnamefont {Y.}~\bibnamefont
  {Zhang}}, \bibinfo {author} {\bibfnamefont {Q.-S.}\ \bibnamefont {Meng}},
  \bibinfo {author} {\bibfnamefont {L.}~\bibnamefont {Zhang}}, \bibinfo
  {author} {\bibfnamefont {Y.}~\bibnamefont {Luo}}, \bibinfo {author}
  {\bibfnamefont {Y.-J.}\ \bibnamefont {Yu}}, \bibinfo {author} {\bibfnamefont
  {B.}~\bibnamefont {Yang}}, \bibinfo {author} {\bibfnamefont {Y.}~\bibnamefont
  {Zhang}}, \bibinfo {author} {\bibfnamefont {R.}~\bibnamefont {Esteban}},
  \bibinfo {author} {\bibfnamefont {J.}~\bibnamefont {Aizpurua}}, \bibinfo
  {author} {\bibfnamefont {Y.}~\bibnamefont {Luo}}, \bibinfo {author}
  {\bibfnamefont {J.-L.}\ \bibnamefont {Yang}}, \bibinfo {author}
  {\bibfnamefont {Z.-C.}\ \bibnamefont {Dong}}, \ and\ \bibinfo {author}
  {\bibfnamefont {J.~G.}\ \bibnamefont {Hou}},\ }\href
  {http://dx.doi.org/10.1038/ncomms15225} {\bibfield  {journal} {\bibinfo
  {journal} {Nature Communications}\ }\textbf {\bibinfo {volume} {8}},\
  \bibinfo {pages} {15225} (\bibinfo {year} {2017})}\BibitemShut {NoStop}%
\bibitem [{\citenamefont {Wang}\ \emph {et~al.}(2017)\citenamefont {Wang},
  \citenamefont {Kelkar}, \citenamefont {Martin-Cano}, \citenamefont {Utikal},
  \citenamefont {G\"otzinger},\ and\ \citenamefont
  {Sandoghdar}}]{PhysRevX.7.021014}%
  \BibitemOpen
  \bibfield  {author} {\bibinfo {author} {\bibfnamefont {D.}~\bibnamefont
  {Wang}}, \bibinfo {author} {\bibfnamefont {H.}~\bibnamefont {Kelkar}},
  \bibinfo {author} {\bibfnamefont {D.}~\bibnamefont {Martin-Cano}}, \bibinfo
  {author} {\bibfnamefont {T.}~\bibnamefont {Utikal}}, \bibinfo {author}
  {\bibfnamefont {S.}~\bibnamefont {G\"otzinger}}, \ and\ \bibinfo {author}
  {\bibfnamefont {V.}~\bibnamefont {Sandoghdar}},\ }\href {\doibase
  10.1103/PhysRevX.7.021014} {\bibfield  {journal} {\bibinfo  {journal} {Phys.
  Rev. X}\ }\textbf {\bibinfo {volume} {7}},\ \bibinfo {pages} {021014}
  (\bibinfo {year} {2017})}\BibitemShut {NoStop}%
\bibitem [{\citenamefont {Blais}\ \emph {et~al.}(2004)\citenamefont {Blais},
  \citenamefont {Huang}, \citenamefont {Wallraff}, \citenamefont {Girvin},\
  and\ \citenamefont {Schoelkopf}}]{Blais2004}%
  \BibitemOpen
  \bibfield  {author} {\bibinfo {author} {\bibfnamefont {A.}~\bibnamefont
  {Blais}}, \bibinfo {author} {\bibfnamefont {R.-S.}\ \bibnamefont {Huang}},
  \bibinfo {author} {\bibfnamefont {A.}~\bibnamefont {Wallraff}}, \bibinfo
  {author} {\bibfnamefont {S.~M.}\ \bibnamefont {Girvin}}, \ and\ \bibinfo
  {author} {\bibfnamefont {R.~J.}\ \bibnamefont {Schoelkopf}},\ }\href
  {\doibase 10.1103/PhysRevA.69.062320} {\bibfield  {journal} {\bibinfo
  {journal} {Phys. Rev. A}\ }\textbf {\bibinfo {volume} {69}},\ \bibinfo
  {pages} {062320} (\bibinfo {year} {2004})}\BibitemShut {NoStop}%
\bibitem [{\citenamefont {Wallraff}\ \emph {et~al.}(2004)\citenamefont
  {Wallraff}, \citenamefont {Schuster}, \citenamefont {Blais}, \citenamefont
  {Frunzio}, \citenamefont {Huang}, \citenamefont {Majer}, \citenamefont
  {Kumar}, \citenamefont {Girvin},\ and\ \citenamefont {Schoelkopf}}]{Wall}%
  \BibitemOpen
  \bibfield  {author} {\bibinfo {author} {\bibfnamefont {A.}~\bibnamefont
  {Wallraff}}, \bibinfo {author} {\bibfnamefont {D.~I.}\ \bibnamefont
  {Schuster}}, \bibinfo {author} {\bibfnamefont {A.}~\bibnamefont {Blais}},
  \bibinfo {author} {\bibfnamefont {L.}~\bibnamefont {Frunzio}}, \bibinfo
  {author} {\bibfnamefont {R.-S.}\ \bibnamefont {Huang}}, \bibinfo {author}
  {\bibfnamefont {J.}~\bibnamefont {Majer}}, \bibinfo {author} {\bibfnamefont
  {S.}~\bibnamefont {Kumar}}, \bibinfo {author} {\bibfnamefont {S.~M.}\
  \bibnamefont {Girvin}}, \ and\ \bibinfo {author} {\bibfnamefont {R.~J.}\
  \bibnamefont {Schoelkopf}},\ }\href {\doibase 10.1038/nature02851} {\bibfield
   {journal} {\bibinfo  {journal} {Nature}\ }\textbf {\bibinfo {volume}
  {431}},\ \bibinfo {pages} {162} (\bibinfo {year} {2004})}\BibitemShut
  {NoStop}%
\bibitem [{\citenamefont {Chiorescu}\ \emph {et~al.}(2004)\citenamefont
  {Chiorescu}, \citenamefont {Bertet}, \citenamefont {Semba}, \citenamefont
  {Nakamura}, \citenamefont {Harmans},\ and\ \citenamefont
  {Mooij}}]{chiorescu2004}%
  \BibitemOpen
  \bibfield  {author} {\bibinfo {author} {\bibfnamefont {I.}~\bibnamefont
  {Chiorescu}}, \bibinfo {author} {\bibfnamefont {P.}~\bibnamefont {Bertet}},
  \bibinfo {author} {\bibfnamefont {K.}~\bibnamefont {Semba}}, \bibinfo
  {author} {\bibfnamefont {Y.}~\bibnamefont {Nakamura}}, \bibinfo {author}
  {\bibfnamefont {C.~J. P.~M.}\ \bibnamefont {Harmans}}, \ and\ \bibinfo
  {author} {\bibfnamefont {J.~E.}\ \bibnamefont {Mooij}},\ }\href {\doibase
  10.1038/nature02831} {\bibfield  {journal} {\bibinfo  {journal} {Nature}\
  }\textbf {\bibinfo {volume} {431}},\ \bibinfo {pages} {159} (\bibinfo {year}
  {2004})}\BibitemShut {NoStop}%
\bibitem [{\citenamefont {Schoelkopf}\ and\ \citenamefont
  {Girvin}(2008)}]{scho2008}%
  \BibitemOpen
  \bibfield  {author} {\bibinfo {author} {\bibfnamefont {R.~J.}\ \bibnamefont
  {Schoelkopf}}\ and\ \bibinfo {author} {\bibfnamefont {S.~M.}\ \bibnamefont
  {Girvin}},\ }\href {\doibase 10.1038/451664a} {\bibfield  {journal} {\bibinfo
   {journal} {Nature}\ }\textbf {\bibinfo {volume} {451}},\ \bibinfo {pages}
  {664} (\bibinfo {year} {2008})}\BibitemShut {NoStop}%
\bibitem [{\citenamefont {Gu}\ \emph {et~al.}(2017)\citenamefont {Gu},
  \citenamefont {Kockum}, \citenamefont {Miranowicz}, \citenamefont {Liu},\
  and\ \citenamefont {Nori}}]{gu2017}%
  \BibitemOpen
  \bibfield  {author} {\bibinfo {author} {\bibfnamefont {X.}~\bibnamefont
  {Gu}}, \bibinfo {author} {\bibfnamefont {A.~F.}\ \bibnamefont {Kockum}},
  \bibinfo {author} {\bibfnamefont {A.}~\bibnamefont {Miranowicz}}, \bibinfo
  {author} {\bibfnamefont {Y.-x.}\ \bibnamefont {Liu}}, \ and\ \bibinfo
  {author} {\bibfnamefont {F.}~\bibnamefont {Nori}},\ }\href {\doibase
  https://doi.org/10.1016/j.physrep.2017.10.002} {\bibfield  {journal}
  {\bibinfo  {journal} {Physics Reports}\ }\textbf {\bibinfo {volume}
  {718-719}},\ \bibinfo {pages} {1 } (\bibinfo {year} {2017})}\BibitemShut
  {NoStop}%
\bibitem [{\citenamefont {Yoshie}\ \emph {et~al.}(2004)\citenamefont {Yoshie},
  \citenamefont {Scherer}, \citenamefont {Hendrickson}, \citenamefont
  {Khitrova}, \citenamefont {Gibbs}, \citenamefont {Rupper}, \citenamefont
  {Ell}, \citenamefont {Shchekin},\ and\ \citenamefont {Deppe}}]{Yoshie}%
  \BibitemOpen
  \bibfield  {author} {\bibinfo {author} {\bibfnamefont {T.}~\bibnamefont
  {Yoshie}}, \bibinfo {author} {\bibfnamefont {A.}~\bibnamefont {Scherer}},
  \bibinfo {author} {\bibfnamefont {J.}~\bibnamefont {Hendrickson}}, \bibinfo
  {author} {\bibfnamefont {G.}~\bibnamefont {Khitrova}}, \bibinfo {author}
  {\bibfnamefont {H.~M.}\ \bibnamefont {Gibbs}}, \bibinfo {author}
  {\bibfnamefont {G.}~\bibnamefont {Rupper}}, \bibinfo {author} {\bibfnamefont
  {C.}~\bibnamefont {Ell}}, \bibinfo {author} {\bibfnamefont {O.~B.}\
  \bibnamefont {Shchekin}}, \ and\ \bibinfo {author} {\bibfnamefont {D.~G.}\
  \bibnamefont {Deppe}},\ }\href {\doibase 10.1038/nature03119} {\bibfield
  {journal} {\bibinfo  {journal} {Nature}\ }\textbf {\bibinfo {volume} {432}},\
  \bibinfo {pages} {200} (\bibinfo {year} {2004})}\BibitemShut {NoStop}%
\bibitem [{\citenamefont {Hennessy}\ \emph {et~al.}(2007)\citenamefont
  {Hennessy}, \citenamefont {Badolato}, \citenamefont {Winger}, \citenamefont
  {Gerace}, \citenamefont {Atat\"{u}re}, \citenamefont {Gulde}, \citenamefont
  {F\"{a}lt}, \citenamefont {Hu},\ and\ \citenamefont
  {Imamo\u{g}lu}}]{hennessy}%
  \BibitemOpen
  \bibfield  {author} {\bibinfo {author} {\bibfnamefont {K.}~\bibnamefont
  {Hennessy}}, \bibinfo {author} {\bibfnamefont {A.}~\bibnamefont {Badolato}},
  \bibinfo {author} {\bibfnamefont {M.}~\bibnamefont {Winger}}, \bibinfo
  {author} {\bibfnamefont {D.}~\bibnamefont {Gerace}}, \bibinfo {author}
  {\bibfnamefont {M.}~\bibnamefont {Atat\"{u}re}}, \bibinfo {author}
  {\bibfnamefont {S.}~\bibnamefont {Gulde}}, \bibinfo {author} {\bibfnamefont
  {S.}~\bibnamefont {F\"{a}lt}}, \bibinfo {author} {\bibfnamefont {E.~L.}\
  \bibnamefont {Hu}}, \ and\ \bibinfo {author} {\bibfnamefont {A.}~\bibnamefont
  {Imamo\u{g}lu}},\ }\href {\doibase 10.1038/nature05586} {\bibfield  {journal}
  {\bibinfo  {journal} {Nature}\ }\textbf {\bibinfo {volume} {445}},\ \bibinfo
  {pages} {896} (\bibinfo {year} {2007})}\BibitemShut {NoStop}%
\bibitem [{\citenamefont {Deng}\ \emph {et~al.}(2010)\citenamefont {Deng},
  \citenamefont {Haug},\ and\ \citenamefont {Yamamoto}}]{RevModPhys.82.1489}%
  \BibitemOpen
  \bibfield  {author} {\bibinfo {author} {\bibfnamefont {H.}~\bibnamefont
  {Deng}}, \bibinfo {author} {\bibfnamefont {H.}~\bibnamefont {Haug}}, \ and\
  \bibinfo {author} {\bibfnamefont {Y.}~\bibnamefont {Yamamoto}},\ }\href
  {\doibase 10.1103/RevModPhys.82.1489} {\bibfield  {journal} {\bibinfo
  {journal} {Rev. Mod. Phys.}\ }\textbf {\bibinfo {volume} {82}},\ \bibinfo
  {pages} {1489} (\bibinfo {year} {2010})}\BibitemShut {NoStop}%
\bibitem [{\citenamefont {Carusotto}\ and\ \citenamefont
  {Ciuti}(2013)}]{RevModPhys.85.299}%
  \BibitemOpen
  \bibfield  {author} {\bibinfo {author} {\bibfnamefont {I.}~\bibnamefont
  {Carusotto}}\ and\ \bibinfo {author} {\bibfnamefont {C.}~\bibnamefont
  {Ciuti}},\ }\href {\doibase 10.1103/RevModPhys.85.299} {\bibfield  {journal}
  {\bibinfo  {journal} {Rev. Mod. Phys.}\ }\textbf {\bibinfo {volume} {85}},\
  \bibinfo {pages} {299} (\bibinfo {year} {2013})}\BibitemShut {NoStop}%
\bibitem [{\citenamefont {Pellizzari}\ \emph {et~al.}(1995)\citenamefont
  {Pellizzari}, \citenamefont {Gardiner}, \citenamefont {Cirac},\ and\
  \citenamefont {Zoller}}]{Zoller_1995}%
  \BibitemOpen
  \bibfield  {author} {\bibinfo {author} {\bibfnamefont {T.}~\bibnamefont
  {Pellizzari}}, \bibinfo {author} {\bibfnamefont {S.~A.}\ \bibnamefont
  {Gardiner}}, \bibinfo {author} {\bibfnamefont {J.~I.}\ \bibnamefont {Cirac}},
  \ and\ \bibinfo {author} {\bibfnamefont {P.}~\bibnamefont {Zoller}},\ }\href
  {\doibase 10.1103/PhysRevLett.75.3788} {\bibfield  {journal} {\bibinfo
  {journal} {Phys. Rev. Lett.}\ }\textbf {\bibinfo {volume} {75}},\ \bibinfo
  {pages} {3788} (\bibinfo {year} {1995})}\BibitemShut {NoStop}%
\bibitem [{\citenamefont {Turchette}\ \emph {et~al.}(1995)\citenamefont
  {Turchette}, \citenamefont {Hood}, \citenamefont {Lange}, \citenamefont
  {Mabuchi},\ and\ \citenamefont {Kimble}}]{PhysRevLett.75.4710}%
  \BibitemOpen
  \bibfield  {author} {\bibinfo {author} {\bibfnamefont {Q.~A.}\ \bibnamefont
  {Turchette}}, \bibinfo {author} {\bibfnamefont {C.~J.}\ \bibnamefont {Hood}},
  \bibinfo {author} {\bibfnamefont {W.}~\bibnamefont {Lange}}, \bibinfo
  {author} {\bibfnamefont {H.}~\bibnamefont {Mabuchi}}, \ and\ \bibinfo
  {author} {\bibfnamefont {H.~J.}\ \bibnamefont {Kimble}},\ }\href {\doibase
  10.1103/PhysRevLett.75.4710} {\bibfield  {journal} {\bibinfo  {journal}
  {Phys. Rev. Lett.}\ }\textbf {\bibinfo {volume} {75}},\ \bibinfo {pages}
  {4710} (\bibinfo {year} {1995})}\BibitemShut {NoStop}%
\bibitem [{\citenamefont {Imamo\u{g}lu}\ \emph {et~al.}(1999)\citenamefont
  {Imamo\u{g}lu}, \citenamefont {Awschalom}, \citenamefont {Burkard},
  \citenamefont {DiVincenzo}, \citenamefont {Loss}, \citenamefont {Sherwin},\
  and\ \citenamefont {Small}}]{imam}%
  \BibitemOpen
  \bibfield  {author} {\bibinfo {author} {\bibfnamefont {A.}~\bibnamefont
  {Imamo\u{g}lu}}, \bibinfo {author} {\bibfnamefont {D.~D.}\ \bibnamefont
  {Awschalom}}, \bibinfo {author} {\bibfnamefont {G.}~\bibnamefont {Burkard}},
  \bibinfo {author} {\bibfnamefont {D.~P.}\ \bibnamefont {DiVincenzo}},
  \bibinfo {author} {\bibfnamefont {D.}~\bibnamefont {Loss}}, \bibinfo {author}
  {\bibfnamefont {M.}~\bibnamefont {Sherwin}}, \ and\ \bibinfo {author}
  {\bibfnamefont {A.}~\bibnamefont {Small}},\ }\href {\doibase
  10.1103/PhysRevLett.83.4204} {\bibfield  {journal} {\bibinfo  {journal}
  {Phys. Rev. Lett.}\ }\textbf {\bibinfo {volume} {83}},\ \bibinfo {pages}
  {4204} (\bibinfo {year} {1999})}\BibitemShut {NoStop}%
\bibitem [{\citenamefont {Rauschenbeutel}\ \emph {et~al.}(1999)\citenamefont
  {Rauschenbeutel}, \citenamefont {Nogues}, \citenamefont {Osnaghi},
  \citenamefont {Bertet}, \citenamefont {Brune}, \citenamefont {Raimond},\ and\
  \citenamefont {Haroche}}]{PhysRevLett.83.5166}%
  \BibitemOpen
  \bibfield  {author} {\bibinfo {author} {\bibfnamefont {A.}~\bibnamefont
  {Rauschenbeutel}}, \bibinfo {author} {\bibfnamefont {G.}~\bibnamefont
  {Nogues}}, \bibinfo {author} {\bibfnamefont {S.}~\bibnamefont {Osnaghi}},
  \bibinfo {author} {\bibfnamefont {P.}~\bibnamefont {Bertet}}, \bibinfo
  {author} {\bibfnamefont {M.}~\bibnamefont {Brune}}, \bibinfo {author}
  {\bibfnamefont {J.~M.}\ \bibnamefont {Raimond}}, \ and\ \bibinfo {author}
  {\bibfnamefont {S.}~\bibnamefont {Haroche}},\ }\href {\doibase
  10.1103/PhysRevLett.83.5166} {\bibfield  {journal} {\bibinfo  {journal}
  {Phys. Rev. Lett.}\ }\textbf {\bibinfo {volume} {83}},\ \bibinfo {pages}
  {5166} (\bibinfo {year} {1999})}\BibitemShut {NoStop}%
\bibitem [{\citenamefont {Zheng}\ and\ \citenamefont
  {Guo}(2000)}]{PhysRevLett.85.2392}%
  \BibitemOpen
  \bibfield  {author} {\bibinfo {author} {\bibfnamefont {S.-B.}\ \bibnamefont
  {Zheng}}\ and\ \bibinfo {author} {\bibfnamefont {G.-C.}\ \bibnamefont
  {Guo}},\ }\href {\doibase 10.1103/PhysRevLett.85.2392} {\bibfield  {journal}
  {\bibinfo  {journal} {Phys. Rev. Lett.}\ }\textbf {\bibinfo {volume} {85}},\
  \bibinfo {pages} {2392} (\bibinfo {year} {2000})}\BibitemShut {NoStop}%
\bibitem [{\citenamefont {Mabuchi}\ and\ \citenamefont
  {Doherty}(2002)}]{Doherty_2002}%
  \BibitemOpen
  \bibfield  {author} {\bibinfo {author} {\bibfnamefont {H.}~\bibnamefont
  {Mabuchi}}\ and\ \bibinfo {author} {\bibfnamefont {A.~C.}\ \bibnamefont
  {Doherty}},\ }\href {\doibase 10.1126/science.1078446} {\bibfield  {journal}
  {\bibinfo  {journal} {Science}\ }\textbf {\bibinfo {volume} {298}},\ \bibinfo
  {pages} {1372} (\bibinfo {year} {2002})}\BibitemShut {NoStop}%
\bibitem [{\citenamefont {O'Brien}\ \emph {et~al.}(2009)\citenamefont
  {O'Brien}, \citenamefont {Furusawa},\ and\ \citenamefont
  {Vu\v{c}kovi\'{c}}}]{brien}%
  \BibitemOpen
  \bibfield  {author} {\bibinfo {author} {\bibfnamefont {J.~L.}\ \bibnamefont
  {O'Brien}}, \bibinfo {author} {\bibfnamefont {A.}~\bibnamefont {Furusawa}}, \
  and\ \bibinfo {author} {\bibfnamefont {J.}~\bibnamefont {Vu\v{c}kovi\'{c}}},\
  }\href {\doibase 10.1038/nphoton.2009.229} {\bibfield  {journal} {\bibinfo
  {journal} {Nature Photonics}\ }\textbf {\bibinfo {volume} {3}},\ \bibinfo
  {pages} {687} (\bibinfo {year} {2009})}\BibitemShut {NoStop}%
\bibitem [{\citenamefont {Orgiu}\ \emph {et~al.}(2015)\citenamefont {Orgiu},
  \citenamefont {George}, \citenamefont {Hutchison}, \citenamefont {Devaux},
  \citenamefont {Dayen}, \citenamefont {Doudin}, \citenamefont {Stellacci},
  \citenamefont {Genet}, \citenamefont {Schachenmayer}, \citenamefont {Genes},
  \citenamefont {Pupillo}, \citenamefont {Samor\`{i}},\ and\ \citenamefont
  {Ebbesen}}]{orgiu2015conductivity}%
  \BibitemOpen
  \bibfield  {author} {\bibinfo {author} {\bibfnamefont {E.}~\bibnamefont
  {Orgiu}}, \bibinfo {author} {\bibfnamefont {J.}~\bibnamefont {George}},
  \bibinfo {author} {\bibfnamefont {J.~A.}\ \bibnamefont {Hutchison}}, \bibinfo
  {author} {\bibfnamefont {E.}~\bibnamefont {Devaux}}, \bibinfo {author}
  {\bibfnamefont {J.~F.}\ \bibnamefont {Dayen}}, \bibinfo {author}
  {\bibfnamefont {B.}~\bibnamefont {Doudin}}, \bibinfo {author} {\bibfnamefont
  {F.}~\bibnamefont {Stellacci}}, \bibinfo {author} {\bibfnamefont
  {C.}~\bibnamefont {Genet}}, \bibinfo {author} {\bibfnamefont
  {J.}~\bibnamefont {Schachenmayer}}, \bibinfo {author} {\bibfnamefont
  {C.}~\bibnamefont {Genes}}, \bibinfo {author} {\bibfnamefont
  {G.}~\bibnamefont {Pupillo}}, \bibinfo {author} {\bibfnamefont
  {P.}~\bibnamefont {Samor\`{i}}}, \ and\ \bibinfo {author} {\bibfnamefont
  {T.~W.}\ \bibnamefont {Ebbesen}},\ }\href
  {http://www.nature.com/nmat/journal/vaop/ncurrent/full/nmat4392.html}
  {\bibfield  {journal} {\bibinfo  {journal} {Nature Materials}\ }\textbf
  {\bibinfo {volume} {14}},\ \bibinfo {pages} {1123 } (\bibinfo {year}
  {2015})}\BibitemShut {NoStop}%
\bibitem [{\citenamefont {Feist}\ and\ \citenamefont
  {Garcia-Vidal}(2015)}]{PhysRevLett.114.196402}%
  \BibitemOpen
  \bibfield  {author} {\bibinfo {author} {\bibfnamefont {J.}~\bibnamefont
  {Feist}}\ and\ \bibinfo {author} {\bibfnamefont {F.~J.}\ \bibnamefont
  {Garcia-Vidal}},\ }\href {\doibase 10.1103/PhysRevLett.114.196402} {\bibfield
   {journal} {\bibinfo  {journal} {Phys. Rev. Lett.}\ }\textbf {\bibinfo
  {volume} {114}},\ \bibinfo {pages} {196402} (\bibinfo {year}
  {2015})}\BibitemShut {NoStop}%
\bibitem [{\citenamefont {Schachenmayer}\ \emph {et~al.}(2015)\citenamefont
  {Schachenmayer}, \citenamefont {Genes}, \citenamefont {Tignone},\ and\
  \citenamefont {Pupillo}}]{PhysRevLett.114.196403}%
  \BibitemOpen
  \bibfield  {author} {\bibinfo {author} {\bibfnamefont {J.}~\bibnamefont
  {Schachenmayer}}, \bibinfo {author} {\bibfnamefont {C.}~\bibnamefont
  {Genes}}, \bibinfo {author} {\bibfnamefont {E.}~\bibnamefont {Tignone}}, \
  and\ \bibinfo {author} {\bibfnamefont {G.}~\bibnamefont {Pupillo}},\ }\href
  {\doibase 10.1103/PhysRevLett.114.196403} {\bibfield  {journal} {\bibinfo
  {journal} {Phys. Rev. Lett.}\ }\textbf {\bibinfo {volume} {114}},\ \bibinfo
  {pages} {196403} (\bibinfo {year} {2015})}\BibitemShut {NoStop}%
\bibitem [{\citenamefont {Zhong}\ \emph {et~al.}(2017)\citenamefont {Zhong},
  \citenamefont {Chervy}, \citenamefont {Zhang}, \citenamefont {Thomas},
  \citenamefont {George}, \citenamefont {Genet}, \citenamefont {Hutchison},\
  and\ \citenamefont {Ebbesen}}]{doi:10.1002/anie.201703539}%
  \BibitemOpen
  \bibfield  {author} {\bibinfo {author} {\bibfnamefont {X.}~\bibnamefont
  {Zhong}}, \bibinfo {author} {\bibfnamefont {T.}~\bibnamefont {Chervy}},
  \bibinfo {author} {\bibfnamefont {L.}~\bibnamefont {Zhang}}, \bibinfo
  {author} {\bibfnamefont {A.}~\bibnamefont {Thomas}}, \bibinfo {author}
  {\bibfnamefont {J.}~\bibnamefont {George}}, \bibinfo {author} {\bibfnamefont
  {C.}~\bibnamefont {Genet}}, \bibinfo {author} {\bibfnamefont {J.~A.}\
  \bibnamefont {Hutchison}}, \ and\ \bibinfo {author} {\bibfnamefont {T.~W.}\
  \bibnamefont {Ebbesen}},\ }\href {\doibase 10.1002/anie.201703539} {\bibfield
   {journal} {\bibinfo  {journal} {Angewandte Chemie International Edition}\
  }\textbf {\bibinfo {volume} {56}},\ \bibinfo {pages} {9034} (\bibinfo {year}
  {2017})}\BibitemShut {NoStop}%
\bibitem [{\citenamefont {Hagenm\"uller}\ \emph {et~al.}(2017)\citenamefont
  {Hagenm\"uller}, \citenamefont {Schachenmayer}, \citenamefont {Sch\"utz},
  \citenamefont {Genes},\ and\ \citenamefont
  {Pupillo}}]{PhysRevLett.119.223601}%
  \BibitemOpen
  \bibfield  {author} {\bibinfo {author} {\bibfnamefont {D.}~\bibnamefont
  {Hagenm\"uller}}, \bibinfo {author} {\bibfnamefont {J.}~\bibnamefont
  {Schachenmayer}}, \bibinfo {author} {\bibfnamefont {S.}~\bibnamefont
  {Sch\"utz}}, \bibinfo {author} {\bibfnamefont {C.}~\bibnamefont {Genes}}, \
  and\ \bibinfo {author} {\bibfnamefont {G.}~\bibnamefont {Pupillo}},\ }\href
  {\doibase 10.1103/PhysRevLett.119.223601} {\bibfield  {journal} {\bibinfo
  {journal} {Phys. Rev. Lett.}\ }\textbf {\bibinfo {volume} {119}},\ \bibinfo
  {pages} {223601} (\bibinfo {year} {2017})}\BibitemShut {NoStop}%
\bibitem [{\citenamefont {Rozenman}\ \emph {et~al.}(2018)\citenamefont
  {Rozenman}, \citenamefont {Akulov}, \citenamefont {Golombek},\ and\
  \citenamefont {Schwartz}}]{doi:10.1021/acsphotonics.7b01332}%
  \BibitemOpen
  \bibfield  {author} {\bibinfo {author} {\bibfnamefont {G.~G.}\ \bibnamefont
  {Rozenman}}, \bibinfo {author} {\bibfnamefont {K.}~\bibnamefont {Akulov}},
  \bibinfo {author} {\bibfnamefont {A.}~\bibnamefont {Golombek}}, \ and\
  \bibinfo {author} {\bibfnamefont {T.}~\bibnamefont {Schwartz}},\ }\href
  {\doibase 10.1021/acsphotonics.7b01332} {\bibfield  {journal} {\bibinfo
  {journal} {ACS Photonics}\ }\textbf {\bibinfo {volume} {5}},\ \bibinfo
  {pages} {105} (\bibinfo {year} {2018})}\BibitemShut {NoStop}%
\bibitem [{\citenamefont {Paravicini-Bagliani}\ \emph
  {et~al.}(2019)\citenamefont {Paravicini-Bagliani}, \citenamefont
  {Appugliese}, \citenamefont {Richter}, \citenamefont {Valmorra},
  \citenamefont {Keller}, \citenamefont {Beck}, \citenamefont {Bartolo},
  \citenamefont {R\"{o}ssler}, \citenamefont {Ihn}, \citenamefont {Ensslin},
  \citenamefont {Ciuti}, \citenamefont {Scalari},\ and\ \citenamefont
  {Faist}}]{magneto}%
  \BibitemOpen
  \bibfield  {author} {\bibinfo {author} {\bibfnamefont {G.~L.}\ \bibnamefont
  {Paravicini-Bagliani}}, \bibinfo {author} {\bibfnamefont {F.}~\bibnamefont
  {Appugliese}}, \bibinfo {author} {\bibfnamefont {E.}~\bibnamefont {Richter}},
  \bibinfo {author} {\bibfnamefont {F.}~\bibnamefont {Valmorra}}, \bibinfo
  {author} {\bibfnamefont {J.}~\bibnamefont {Keller}}, \bibinfo {author}
  {\bibfnamefont {M.}~\bibnamefont {Beck}}, \bibinfo {author} {\bibfnamefont
  {N.}~\bibnamefont {Bartolo}}, \bibinfo {author} {\bibfnamefont
  {C.}~\bibnamefont {R\"{o}ssler}}, \bibinfo {author} {\bibfnamefont
  {T.}~\bibnamefont {Ihn}}, \bibinfo {author} {\bibfnamefont {K.}~\bibnamefont
  {Ensslin}}, \bibinfo {author} {\bibfnamefont {C.}~\bibnamefont {Ciuti}},
  \bibinfo {author} {\bibfnamefont {G.}~\bibnamefont {Scalari}}, \ and\
  \bibinfo {author} {\bibfnamefont {J.}~\bibnamefont {Faist}},\ }\href
  {https://doi.org/10.1038/s41567-018-0346-y} {\bibfield  {journal} {\bibinfo
  {journal} {Nature Physics}\ }\textbf {\bibinfo {volume} {15}},\ \bibinfo
  {pages} {186} (\bibinfo {year} {2019})}\BibitemShut {NoStop}%
\bibitem [{\citenamefont {Brennecke}\ \emph {et~al.}(2008)\citenamefont
  {Brennecke}, \citenamefont {Ritter}, \citenamefont {Donner},\ and\
  \citenamefont {Esslinger}}]{Brennecke235}%
  \BibitemOpen
  \bibfield  {author} {\bibinfo {author} {\bibfnamefont {F.}~\bibnamefont
  {Brennecke}}, \bibinfo {author} {\bibfnamefont {S.}~\bibnamefont {Ritter}},
  \bibinfo {author} {\bibfnamefont {T.}~\bibnamefont {Donner}}, \ and\ \bibinfo
  {author} {\bibfnamefont {T.}~\bibnamefont {Esslinger}},\ }\href {\doibase
  10.1126/science.1163218} {\bibfield  {journal} {\bibinfo  {journal}
  {Science}\ }\textbf {\bibinfo {volume} {322}},\ \bibinfo {pages} {235}
  (\bibinfo {year} {2008})}\BibitemShut {NoStop}%
\bibitem [{\citenamefont {Gr\"{o}blacher}\ \emph {et~al.}(2009)\citenamefont
  {Gr\"{o}blacher}, \citenamefont {Hammerer}, \citenamefont {Vanner},\ and\
  \citenamefont {Aspelmeyer}}]{Aspelmeyer}%
  \BibitemOpen
  \bibfield  {author} {\bibinfo {author} {\bibfnamefont {S.}~\bibnamefont
  {Gr\"{o}blacher}}, \bibinfo {author} {\bibfnamefont {K.}~\bibnamefont
  {Hammerer}}, \bibinfo {author} {\bibfnamefont {M.~R.}\ \bibnamefont
  {Vanner}}, \ and\ \bibinfo {author} {\bibfnamefont {M.}~\bibnamefont
  {Aspelmeyer}},\ }\href {\doibase 10.1038/nature08171} {\bibfield  {journal}
  {\bibinfo  {journal} {Nature}\ }\textbf {\bibinfo {volume} {460}},\ \bibinfo
  {pages} {724} (\bibinfo {year} {2009})}\BibitemShut {NoStop}%
\bibitem [{\citenamefont {Hammerer}\ \emph {et~al.}(2009)\citenamefont
  {Hammerer}, \citenamefont {Wallquist}, \citenamefont {Genes}, \citenamefont
  {Ludwig}, \citenamefont {Marquardt}, \citenamefont {Treutlein}, \citenamefont
  {Zoller}, \citenamefont {Ye},\ and\ \citenamefont
  {Kimble}}]{PhysRevLett.103.063005}%
  \BibitemOpen
  \bibfield  {author} {\bibinfo {author} {\bibfnamefont {K.}~\bibnamefont
  {Hammerer}}, \bibinfo {author} {\bibfnamefont {M.}~\bibnamefont {Wallquist}},
  \bibinfo {author} {\bibfnamefont {C.}~\bibnamefont {Genes}}, \bibinfo
  {author} {\bibfnamefont {M.}~\bibnamefont {Ludwig}}, \bibinfo {author}
  {\bibfnamefont {F.}~\bibnamefont {Marquardt}}, \bibinfo {author}
  {\bibfnamefont {P.}~\bibnamefont {Treutlein}}, \bibinfo {author}
  {\bibfnamefont {P.}~\bibnamefont {Zoller}}, \bibinfo {author} {\bibfnamefont
  {J.}~\bibnamefont {Ye}}, \ and\ \bibinfo {author} {\bibfnamefont {H.~J.}\
  \bibnamefont {Kimble}},\ }\href {\doibase 10.1103/PhysRevLett.103.063005}
  {\bibfield  {journal} {\bibinfo  {journal} {Phys. Rev. Lett.}\ }\textbf
  {\bibinfo {volume} {103}},\ \bibinfo {pages} {063005} (\bibinfo {year}
  {2009})}\BibitemShut {NoStop}%
\bibitem [{\citenamefont {Xuereb}\ \emph {et~al.}(2012)\citenamefont {Xuereb},
  \citenamefont {Genes},\ and\ \citenamefont
  {Dantan}}]{PhysRevLett.109.223601}%
  \BibitemOpen
  \bibfield  {author} {\bibinfo {author} {\bibfnamefont {A.}~\bibnamefont
  {Xuereb}}, \bibinfo {author} {\bibfnamefont {C.}~\bibnamefont {Genes}}, \
  and\ \bibinfo {author} {\bibfnamefont {A.}~\bibnamefont {Dantan}},\ }\href
  {\doibase 10.1103/PhysRevLett.109.223601} {\bibfield  {journal} {\bibinfo
  {journal} {Phys. Rev. Lett.}\ }\textbf {\bibinfo {volume} {109}},\ \bibinfo
  {pages} {223601} (\bibinfo {year} {2012})}\BibitemShut {NoStop}%
\bibitem [{\citenamefont {Restrepo}\ \emph {et~al.}(2014)\citenamefont
  {Restrepo}, \citenamefont {Ciuti},\ and\ \citenamefont
  {Favero}}]{PhysRevLett.112.013601}%
  \BibitemOpen
  \bibfield  {author} {\bibinfo {author} {\bibfnamefont {J.}~\bibnamefont
  {Restrepo}}, \bibinfo {author} {\bibfnamefont {C.}~\bibnamefont {Ciuti}}, \
  and\ \bibinfo {author} {\bibfnamefont {I.}~\bibnamefont {Favero}},\ }\href
  {\doibase 10.1103/PhysRevLett.112.013601} {\bibfield  {journal} {\bibinfo
  {journal} {Phys. Rev. Lett.}\ }\textbf {\bibinfo {volume} {112}},\ \bibinfo
  {pages} {013601} (\bibinfo {year} {2014})}\BibitemShut {NoStop}%
\bibitem [{\citenamefont {Benz}\ \emph {et~al.}(2016)\citenamefont {Benz},
  \citenamefont {Schmidt}, \citenamefont {Dreismann}, \citenamefont
  {Chikkaraddy}, \citenamefont {Zhang}, \citenamefont {Demetriadou},
  \citenamefont {Carnegie}, \citenamefont {Ohadi}, \citenamefont {de~Nijs},
  \citenamefont {Esteban}, \citenamefont {Aizpurua},\ and\ \citenamefont
  {Baumberg}}]{Benz726}%
  \BibitemOpen
  \bibfield  {author} {\bibinfo {author} {\bibfnamefont {F.}~\bibnamefont
  {Benz}}, \bibinfo {author} {\bibfnamefont {M.~K.}\ \bibnamefont {Schmidt}},
  \bibinfo {author} {\bibfnamefont {A.}~\bibnamefont {Dreismann}}, \bibinfo
  {author} {\bibfnamefont {R.}~\bibnamefont {Chikkaraddy}}, \bibinfo {author}
  {\bibfnamefont {Y.}~\bibnamefont {Zhang}}, \bibinfo {author} {\bibfnamefont
  {A.}~\bibnamefont {Demetriadou}}, \bibinfo {author} {\bibfnamefont
  {C.}~\bibnamefont {Carnegie}}, \bibinfo {author} {\bibfnamefont
  {H.}~\bibnamefont {Ohadi}}, \bibinfo {author} {\bibfnamefont
  {B.}~\bibnamefont {de~Nijs}}, \bibinfo {author} {\bibfnamefont
  {R.}~\bibnamefont {Esteban}}, \bibinfo {author} {\bibfnamefont
  {J.}~\bibnamefont {Aizpurua}}, \ and\ \bibinfo {author} {\bibfnamefont
  {J.~J.}\ \bibnamefont {Baumberg}},\ }\href {\doibase 10.1126/science.aah5243}
  {\bibfield  {journal} {\bibinfo  {journal} {Science}\ }\textbf {\bibinfo
  {volume} {354}},\ \bibinfo {pages} {726} (\bibinfo {year}
  {2016})}\BibitemShut {NoStop}%
\bibitem [{\citenamefont {Hutchison}\ \emph {et~al.}(2012)\citenamefont
  {Hutchison}, \citenamefont {Schwartz}, \citenamefont {Genet}, \citenamefont
  {Devaux},\ and\ \citenamefont {Ebbesen}}]{hutch_2012}%
  \BibitemOpen
  \bibfield  {author} {\bibinfo {author} {\bibfnamefont {J.~A.}\ \bibnamefont
  {Hutchison}}, \bibinfo {author} {\bibfnamefont {T.}~\bibnamefont {Schwartz}},
  \bibinfo {author} {\bibfnamefont {C.}~\bibnamefont {Genet}}, \bibinfo
  {author} {\bibfnamefont {E.}~\bibnamefont {Devaux}}, \ and\ \bibinfo {author}
  {\bibfnamefont {T.~W.}\ \bibnamefont {Ebbesen}},\ }\href {\doibase
  10.1002/anie.201107033} {\bibfield  {journal} {\bibinfo  {journal} {Angew.
  Chem.}\ }\textbf {\bibinfo {volume} {51}},\ \bibinfo {pages} {1592} (\bibinfo
  {year} {2012})}\BibitemShut {NoStop}%
\bibitem [{\citenamefont {Thomas}\ \emph {et~al.}(2016)\citenamefont {Thomas},
  \citenamefont {George}, \citenamefont {Shalabney}, \citenamefont {Dryzhakov},
  \citenamefont {Varma}, \citenamefont {Moran}, \citenamefont {Chervy},
  \citenamefont {Zhong}, \citenamefont {Devaux}, \citenamefont {Genet},
  \citenamefont {Hutchison},\ and\ \citenamefont {Ebbesen}}]{thomas_2016}%
  \BibitemOpen
  \bibfield  {author} {\bibinfo {author} {\bibfnamefont {A.}~\bibnamefont
  {Thomas}}, \bibinfo {author} {\bibfnamefont {J.}~\bibnamefont {George}},
  \bibinfo {author} {\bibfnamefont {A.}~\bibnamefont {Shalabney}}, \bibinfo
  {author} {\bibfnamefont {M.}~\bibnamefont {Dryzhakov}}, \bibinfo {author}
  {\bibfnamefont {S.~J.}\ \bibnamefont {Varma}}, \bibinfo {author}
  {\bibfnamefont {J.}~\bibnamefont {Moran}}, \bibinfo {author} {\bibfnamefont
  {T.}~\bibnamefont {Chervy}}, \bibinfo {author} {\bibfnamefont
  {X.}~\bibnamefont {Zhong}}, \bibinfo {author} {\bibfnamefont
  {E.}~\bibnamefont {Devaux}}, \bibinfo {author} {\bibfnamefont
  {C.}~\bibnamefont {Genet}}, \bibinfo {author} {\bibfnamefont {J.~A.}\
  \bibnamefont {Hutchison}}, \ and\ \bibinfo {author} {\bibfnamefont {T.~W.}\
  \bibnamefont {Ebbesen}},\ }\href {\doibase 10.1002/anie.201605504} {\bibfield
   {journal} {\bibinfo  {journal} {Angew. Chem.}\ }\textbf {\bibinfo {volume}
  {55}},\ \bibinfo {pages} {11462} (\bibinfo {year} {2016})}\BibitemShut
  {NoStop}%
\bibitem [{\citenamefont {Herrera}\ and\ \citenamefont
  {Spano}(2016)}]{herrera_2016}%
  \BibitemOpen
  \bibfield  {author} {\bibinfo {author} {\bibfnamefont {F.}~\bibnamefont
  {Herrera}}\ and\ \bibinfo {author} {\bibfnamefont {F.~C.}\ \bibnamefont
  {Spano}},\ }\href {\doibase 10.1103/PhysRevLett.116.238301} {\bibfield
  {journal} {\bibinfo  {journal} {Phys. Rev. Lett.}\ }\textbf {\bibinfo
  {volume} {116}},\ \bibinfo {pages} {238301} (\bibinfo {year}
  {2016})}\BibitemShut {NoStop}%
\bibitem [{\citenamefont {Galego}\ \emph {et~al.}(2016)\citenamefont {Galego},
  \citenamefont {Garcia-Vidal},\ and\ \citenamefont {Feist}}]{galego_2016}%
  \BibitemOpen
  \bibfield  {author} {\bibinfo {author} {\bibfnamefont {J.}~\bibnamefont
  {Galego}}, \bibinfo {author} {\bibfnamefont {F.~J.}\ \bibnamefont
  {Garcia-Vidal}}, \ and\ \bibinfo {author} {\bibfnamefont {J.}~\bibnamefont
  {Feist}},\ }\href {\doibase 10.1038/ncomms13841} {\bibfield  {journal}
  {\bibinfo  {journal} {Nature Communications}\ }\textbf {\bibinfo {volume}
  {7}},\ \bibinfo {pages} {13841} (\bibinfo {year} {2016})}\BibitemShut
  {NoStop}%
\bibitem [{\citenamefont {Flick}\ \emph {et~al.}(2017)\citenamefont {Flick},
  \citenamefont {Ruggenthaler}, \citenamefont {Appel},\ and\ \citenamefont
  {Rubio}}]{Flick201615509}%
  \BibitemOpen
  \bibfield  {author} {\bibinfo {author} {\bibfnamefont {J.}~\bibnamefont
  {Flick}}, \bibinfo {author} {\bibfnamefont {M.}~\bibnamefont {Ruggenthaler}},
  \bibinfo {author} {\bibfnamefont {H.}~\bibnamefont {Appel}}, \ and\ \bibinfo
  {author} {\bibfnamefont {A.}~\bibnamefont {Rubio}},\ }\href
  {http://www.pnas.org/content/early/2017/03/07/1615509114} {\bibfield
  {journal} {\bibinfo  {journal} {Proceedings of the National Academy of
  Sciences}\ }\textbf {\bibinfo {volume} {114}},\ \bibinfo {pages} {3026}
  (\bibinfo {year} {2017})}\BibitemShut {NoStop}%
\bibitem [{\citenamefont {Ribeiro}\ \emph {et~al.}(2018)\citenamefont
  {Ribeiro}, \citenamefont {Martínez-Martínez}, \citenamefont {Du},
  \citenamefont {Campos-Gonzalez-Angulo},\ and\ \citenamefont
  {Yuen-Zhou}}]{C8SC01043A}%
  \BibitemOpen
  \bibfield  {author} {\bibinfo {author} {\bibfnamefont {R.~F.}\ \bibnamefont
  {Ribeiro}}, \bibinfo {author} {\bibfnamefont {L.~A.}\ \bibnamefont
  {Martínez-Martínez}}, \bibinfo {author} {\bibfnamefont {M.}~\bibnamefont
  {Du}}, \bibinfo {author} {\bibfnamefont {J.}~\bibnamefont
  {Campos-Gonzalez-Angulo}}, \ and\ \bibinfo {author} {\bibfnamefont
  {J.}~\bibnamefont {Yuen-Zhou}},\ }\href {\doibase 10.1039/C8SC01043A}
  {\bibfield  {journal} {\bibinfo  {journal} {Chem. Sci.}\ }\textbf {\bibinfo
  {volume} {9}},\ \bibinfo {pages} {6325} (\bibinfo {year} {2018})}\BibitemShut
  {NoStop}%
\bibitem [{\citenamefont {Thomas}\ \emph {et~al.}(2019)\citenamefont {Thomas},
  \citenamefont {Lethuillier-Karl}, \citenamefont {Nagarajan}, \citenamefont
  {Vergauwe}, \citenamefont {George}, \citenamefont {Chervy}, \citenamefont
  {Shalabney}, \citenamefont {Devaux}, \citenamefont {Genet}, \citenamefont
  {Moran},\ and\ \citenamefont {Ebbesen}}]{Thomas2018}%
  \BibitemOpen
  \bibfield  {author} {\bibinfo {author} {\bibfnamefont {A.}~\bibnamefont
  {Thomas}}, \bibinfo {author} {\bibfnamefont {L.}~\bibnamefont
  {Lethuillier-Karl}}, \bibinfo {author} {\bibfnamefont {K.}~\bibnamefont
  {Nagarajan}}, \bibinfo {author} {\bibfnamefont {R.~M.~A.}\ \bibnamefont
  {Vergauwe}}, \bibinfo {author} {\bibfnamefont {J.}~\bibnamefont {George}},
  \bibinfo {author} {\bibfnamefont {T.}~\bibnamefont {Chervy}}, \bibinfo
  {author} {\bibfnamefont {A.}~\bibnamefont {Shalabney}}, \bibinfo {author}
  {\bibfnamefont {E.}~\bibnamefont {Devaux}}, \bibinfo {author} {\bibfnamefont
  {C.}~\bibnamefont {Genet}}, \bibinfo {author} {\bibfnamefont
  {J.}~\bibnamefont {Moran}}, \ and\ \bibinfo {author} {\bibfnamefont {T.~W.}\
  \bibnamefont {Ebbesen}},\ }\href {\doibase 10.1126/science.aau7742}
  {\bibfield  {journal} {\bibinfo  {journal} {Science}\ }\textbf {\bibinfo
  {volume} {363}},\ \bibinfo {pages} {615} (\bibinfo {year}
  {2019})}\BibitemShut {NoStop}%
\bibitem [{\citenamefont {Birnbaum}\ \emph {et~al.}(2005)\citenamefont
  {Birnbaum}, \citenamefont {Boca}, \citenamefont {Miller}, \citenamefont
  {Boozer}, \citenamefont {Northup},\ and\ \citenamefont {Kimble}}]{boca}%
  \BibitemOpen
  \bibfield  {author} {\bibinfo {author} {\bibfnamefont {K.~M.}\ \bibnamefont
  {Birnbaum}}, \bibinfo {author} {\bibfnamefont {A.}~\bibnamefont {Boca}},
  \bibinfo {author} {\bibfnamefont {R.}~\bibnamefont {Miller}}, \bibinfo
  {author} {\bibfnamefont {A.~D.}\ \bibnamefont {Boozer}}, \bibinfo {author}
  {\bibfnamefont {T.~E.}\ \bibnamefont {Northup}}, \ and\ \bibinfo {author}
  {\bibfnamefont {H.~J.}\ \bibnamefont {Kimble}},\ }\href {\doibase
  10.1038/nature03804} {\bibfield  {journal} {\bibinfo  {journal} {Nature}\
  }\textbf {\bibinfo {volume} {436}},\ \bibinfo {pages} {87} (\bibinfo {year}
  {2005})}\BibitemShut {NoStop}%
\bibitem [{\citenamefont {Englund}\ \emph {et~al.}(2007)\citenamefont
  {Englund}, \citenamefont {Faraon}, \citenamefont {Fushman}, \citenamefont
  {Stoltz}, \citenamefont {Petroff},\ and\ \citenamefont
  {Vu\v{c}kovi\'{c}}}]{dirk}%
  \BibitemOpen
  \bibfield  {author} {\bibinfo {author} {\bibfnamefont {D.}~\bibnamefont
  {Englund}}, \bibinfo {author} {\bibfnamefont {A.}~\bibnamefont {Faraon}},
  \bibinfo {author} {\bibfnamefont {I.}~\bibnamefont {Fushman}}, \bibinfo
  {author} {\bibfnamefont {N.}~\bibnamefont {Stoltz}}, \bibinfo {author}
  {\bibfnamefont {P.}~\bibnamefont {Petroff}}, \ and\ \bibinfo {author}
  {\bibfnamefont {J.}~\bibnamefont {Vu\v{c}kovi\'{c}}},\ }\href {\doibase
  10.1038/nature06234} {\bibfield  {journal} {\bibinfo  {journal} {Nature}\
  }\textbf {\bibinfo {volume} {450}},\ \bibinfo {pages} {857} (\bibinfo {year}
  {2007})}\BibitemShut {NoStop}%
\bibitem [{\citenamefont {Lang}\ \emph {et~al.}(2011)\citenamefont {Lang},
  \citenamefont {Bozyigit}, \citenamefont {Eichler}, \citenamefont {Steffen},
  \citenamefont {Fink}, \citenamefont {Abdumalikov}, \citenamefont {Baur},
  \citenamefont {Filipp}, \citenamefont {da~Silva}, \citenamefont {Blais},\
  and\ \citenamefont {Wallraff}}]{PhysRevLett.106.243601}%
  \BibitemOpen
  \bibfield  {author} {\bibinfo {author} {\bibfnamefont {C.}~\bibnamefont
  {Lang}}, \bibinfo {author} {\bibfnamefont {D.}~\bibnamefont {Bozyigit}},
  \bibinfo {author} {\bibfnamefont {C.}~\bibnamefont {Eichler}}, \bibinfo
  {author} {\bibfnamefont {L.}~\bibnamefont {Steffen}}, \bibinfo {author}
  {\bibfnamefont {J.~M.}\ \bibnamefont {Fink}}, \bibinfo {author}
  {\bibfnamefont {A.~A.}\ \bibnamefont {Abdumalikov}}, \bibinfo {author}
  {\bibfnamefont {M.}~\bibnamefont {Baur}}, \bibinfo {author} {\bibfnamefont
  {S.}~\bibnamefont {Filipp}}, \bibinfo {author} {\bibfnamefont {M.~P.}\
  \bibnamefont {da~Silva}}, \bibinfo {author} {\bibfnamefont {A.}~\bibnamefont
  {Blais}}, \ and\ \bibinfo {author} {\bibfnamefont {A.}~\bibnamefont
  {Wallraff}},\ }\href {\doibase 10.1103/PhysRevLett.106.243601} {\bibfield
  {journal} {\bibinfo  {journal} {Phys. Rev. Lett.}\ }\textbf {\bibinfo
  {volume} {106}},\ \bibinfo {pages} {243601} (\bibinfo {year}
  {2011})}\BibitemShut {NoStop}%
\bibitem [{\citenamefont {Dicke}(1954)}]{PhysRev.93.99}%
  \BibitemOpen
  \bibfield  {author} {\bibinfo {author} {\bibfnamefont {R.~H.}\ \bibnamefont
  {Dicke}},\ }\href {\doibase 10.1103/PhysRev.93.99} {\bibfield  {journal}
  {\bibinfo  {journal} {Phys. Rev.}\ }\textbf {\bibinfo {volume} {93}},\
  \bibinfo {pages} {99} (\bibinfo {year} {1954})}\BibitemShut {NoStop}%
\bibitem [{\citenamefont {Tavis}\ and\ \citenamefont
  {Cummings}(1968)}]{PhysRev.170.379}%
  \BibitemOpen
  \bibfield  {author} {\bibinfo {author} {\bibfnamefont {M.}~\bibnamefont
  {Tavis}}\ and\ \bibinfo {author} {\bibfnamefont {F.~W.}\ \bibnamefont
  {Cummings}},\ }\href {\doibase 10.1103/PhysRev.170.379} {\bibfield  {journal}
  {\bibinfo  {journal} {Phys. Rev.}\ }\textbf {\bibinfo {volume} {170}},\
  \bibinfo {pages} {379} (\bibinfo {year} {1968})}\BibitemShut {NoStop}%
\bibitem [{\citenamefont {Kaluzny}\ \emph {et~al.}(1983)\citenamefont
  {Kaluzny}, \citenamefont {Goy}, \citenamefont {Gross}, \citenamefont
  {Raimond},\ and\ \citenamefont {Haroche}}]{PhysRevLett.51.1175}%
  \BibitemOpen
  \bibfield  {author} {\bibinfo {author} {\bibfnamefont {Y.}~\bibnamefont
  {Kaluzny}}, \bibinfo {author} {\bibfnamefont {P.}~\bibnamefont {Goy}},
  \bibinfo {author} {\bibfnamefont {M.}~\bibnamefont {Gross}}, \bibinfo
  {author} {\bibfnamefont {J.~M.}\ \bibnamefont {Raimond}}, \ and\ \bibinfo
  {author} {\bibfnamefont {S.}~\bibnamefont {Haroche}},\ }\href {\doibase
  10.1103/PhysRevLett.51.1175} {\bibfield  {journal} {\bibinfo  {journal}
  {Phys. Rev. Lett.}\ }\textbf {\bibinfo {volume} {51}},\ \bibinfo {pages}
  {1175} (\bibinfo {year} {1983})}\BibitemShut {NoStop}%
\bibitem [{\citenamefont {Raizen}\ \emph {et~al.}(1989)\citenamefont {Raizen},
  \citenamefont {Thompson}, \citenamefont {Brecha}, \citenamefont {Kimble},\
  and\ \citenamefont {Carmichael}}]{PhysRevLett.63.240}%
  \BibitemOpen
  \bibfield  {author} {\bibinfo {author} {\bibfnamefont {M.~G.}\ \bibnamefont
  {Raizen}}, \bibinfo {author} {\bibfnamefont {R.~J.}\ \bibnamefont
  {Thompson}}, \bibinfo {author} {\bibfnamefont {R.~J.}\ \bibnamefont
  {Brecha}}, \bibinfo {author} {\bibfnamefont {H.~J.}\ \bibnamefont {Kimble}},
  \ and\ \bibinfo {author} {\bibfnamefont {H.~J.}\ \bibnamefont {Carmichael}},\
  }\href {\doibase 10.1103/PhysRevLett.63.240} {\bibfield  {journal} {\bibinfo
  {journal} {Phys. Rev. Lett.}\ }\textbf {\bibinfo {volume} {63}},\ \bibinfo
  {pages} {240} (\bibinfo {year} {1989})}\BibitemShut {NoStop}%
\bibitem [{\citenamefont {Rempe}\ \emph {et~al.}(1991)\citenamefont {Rempe},
  \citenamefont {Thompson}, \citenamefont {Brecha}, \citenamefont {Lee},\ and\
  \citenamefont {Kimble}}]{PhysRevLett.67.1727}%
  \BibitemOpen
  \bibfield  {author} {\bibinfo {author} {\bibfnamefont {G.}~\bibnamefont
  {Rempe}}, \bibinfo {author} {\bibfnamefont {R.~J.}\ \bibnamefont {Thompson}},
  \bibinfo {author} {\bibfnamefont {R.~J.}\ \bibnamefont {Brecha}}, \bibinfo
  {author} {\bibfnamefont {W.~D.}\ \bibnamefont {Lee}}, \ and\ \bibinfo
  {author} {\bibfnamefont {H.~J.}\ \bibnamefont {Kimble}},\ }\href {\doibase
  10.1103/PhysRevLett.67.1727} {\bibfield  {journal} {\bibinfo  {journal}
  {Phys. Rev. Lett.}\ }\textbf {\bibinfo {volume} {67}},\ \bibinfo {pages}
  {1727} (\bibinfo {year} {1991})}\BibitemShut {NoStop}%
\bibitem [{\citenamefont {Carmichael}\ \emph {et~al.}(1991)\citenamefont
  {Carmichael}, \citenamefont {Brecha},\ and\ \citenamefont
  {Rice}}]{CARMICHAEL199173}%
  \BibitemOpen
  \bibfield  {author} {\bibinfo {author} {\bibfnamefont {H.~J.}\ \bibnamefont
  {Carmichael}}, \bibinfo {author} {\bibfnamefont {R.~J.}\ \bibnamefont
  {Brecha}}, \ and\ \bibinfo {author} {\bibfnamefont {P.~R.}\ \bibnamefont
  {Rice}},\ }\href {\doibase 10.1016/0030-4018(91)90194-I} {\bibfield
  {journal} {\bibinfo  {journal} {Optics Communications}\ }\textbf {\bibinfo
  {volume} {82}},\ \bibinfo {pages} {73 } (\bibinfo {year} {1991})}\BibitemShut
  {NoStop}%
\bibitem [{\citenamefont {S\'aez-Bl\'azquez}\ \emph {et~al.}(2018)\citenamefont
  {S\'aez-Bl\'azquez}, \citenamefont {Feist}, \citenamefont
  {Garc\'{\i}a-Vidal},\ and\ \citenamefont
  {Fern\'andez-Dom\'{\i}nguez}}]{PhysRevA.98.013839}%
  \BibitemOpen
  \bibfield  {author} {\bibinfo {author} {\bibfnamefont {R.}~\bibnamefont
  {S\'aez-Bl\'azquez}}, \bibinfo {author} {\bibfnamefont {J.}~\bibnamefont
  {Feist}}, \bibinfo {author} {\bibfnamefont {F.~J.}\ \bibnamefont
  {Garc\'{\i}a-Vidal}}, \ and\ \bibinfo {author} {\bibfnamefont {A.~I.}\
  \bibnamefont {Fern\'andez-Dom\'{\i}nguez}},\ }\href {\doibase
  10.1103/PhysRevA.98.013839} {\bibfield  {journal} {\bibinfo  {journal} {Phys.
  Rev. A}\ }\textbf {\bibinfo {volume} {98}},\ \bibinfo {pages} {013839}
  (\bibinfo {year} {2018})}\BibitemShut {NoStop}%
\bibitem [{\citenamefont {Trivedi}\ \emph {et~al.}(2019)\citenamefont
  {Trivedi}, \citenamefont {Radulaski}, \citenamefont {Fischer}, \citenamefont
  {Fan},\ and\ \citenamefont {Vu\ifmmode \check{c}\else
  \v{c}\fi{}kovi\ifmmode~\acute{c}\else \'{c}\fi{}}}]{Trivedi2019}%
  \BibitemOpen
  \bibfield  {author} {\bibinfo {author} {\bibfnamefont {R.}~\bibnamefont
  {Trivedi}}, \bibinfo {author} {\bibfnamefont {M.}~\bibnamefont {Radulaski}},
  \bibinfo {author} {\bibfnamefont {K.~A.}\ \bibnamefont {Fischer}}, \bibinfo
  {author} {\bibfnamefont {S.}~\bibnamefont {Fan}}, \ and\ \bibinfo {author}
  {\bibfnamefont {J.}~\bibnamefont {Vu\ifmmode \check{c}\else
  \v{c}\fi{}kovi\ifmmode~\acute{c}\else \'{c}\fi{}}},\ }\href {\doibase
  10.1103/PhysRevLett.122.243602} {\bibfield  {journal} {\bibinfo  {journal}
  {Phys. Rev. Lett.}\ }\textbf {\bibinfo {volume} {122}},\ \bibinfo {pages}
  {243602} (\bibinfo {year} {2019})}\BibitemShut {NoStop}%
\bibitem [{\citenamefont {Lukin}\ \emph {et~al.}(2001)\citenamefont {Lukin},
  \citenamefont {Fleischhauer},\ and\ \citenamefont
  {Imamo{\u{g}}lu}}]{10.1007/3-540-40894-0_18}%
  \BibitemOpen
  \bibfield  {author} {\bibinfo {author} {\bibfnamefont {M.}~\bibnamefont
  {Lukin}}, \bibinfo {author} {\bibfnamefont {M.}~\bibnamefont {Fleischhauer}},
  \ and\ \bibinfo {author} {\bibfnamefont {A.}~\bibnamefont {Imamo{\u{g}}lu}},\
  }in\ \href@noop {} {\emph {\bibinfo {booktitle} {Directions in Quantum
  Optics}}},\ \bibinfo {editor} {edited by\ \bibinfo {editor} {\bibfnamefont
  {H.~J.}\ \bibnamefont {Carmichael}}, \bibinfo {editor} {\bibfnamefont
  {R.~J.}\ \bibnamefont {Glauber}}, \ and\ \bibinfo {editor} {\bibfnamefont
  {M.~O.}\ \bibnamefont {Scully}}}\ (\bibinfo  {publisher} {Springer Berlin
  Heidelberg},\ \bibinfo {address} {Berlin, Heidelberg},\ \bibinfo {year}
  {2001})\ pp.\ \bibinfo {pages} {193--203}\BibitemShut {NoStop}%
\bibitem [{\citenamefont {Wade}\ \emph {et~al.}(2016)\citenamefont {Wade},
  \citenamefont {Mattioli},\ and\ \citenamefont
  {M\o{}lmer}}]{PhysRevA.94.053830}%
  \BibitemOpen
  \bibfield  {author} {\bibinfo {author} {\bibfnamefont {A.~C.~J.}\
  \bibnamefont {Wade}}, \bibinfo {author} {\bibfnamefont {M.}~\bibnamefont
  {Mattioli}}, \ and\ \bibinfo {author} {\bibfnamefont {K.}~\bibnamefont
  {M\o{}lmer}},\ }\href {\doibase 10.1103/PhysRevA.94.053830} {\bibfield
  {journal} {\bibinfo  {journal} {Phys. Rev. A}\ }\textbf {\bibinfo {volume}
  {94}},\ \bibinfo {pages} {053830} (\bibinfo {year} {2016})}\BibitemShut
  {NoStop}%
\bibitem [{\citenamefont {Motzoi}\ and\ \citenamefont
  {M{\o}lmer}(2018)}]{Motzoi_2018}%
  \BibitemOpen
  \bibfield  {author} {\bibinfo {author} {\bibfnamefont {F.}~\bibnamefont
  {Motzoi}}\ and\ \bibinfo {author} {\bibfnamefont {K.}~\bibnamefont
  {M{\o}lmer}},\ }\href {\doibase 10.1088/1367-2630/aac0be} {\bibfield
  {journal} {\bibinfo  {journal} {New Journal of Physics}\ }\textbf {\bibinfo
  {volume} {20}},\ \bibinfo {pages} {053029} (\bibinfo {year}
  {2018})}\BibitemShut {NoStop}%
\bibitem [{\citenamefont {Sch\"utz}\ \emph {et~al.}(2019)\citenamefont
  {Sch\"utz}, \citenamefont {Schachenmayer}, \citenamefont {Hagenm\"uller},
  \citenamefont {Brennen}, \citenamefont {Volz}, \citenamefont {Sandoghdar},
  \citenamefont {Ebbesen}, \citenamefont {Genes},\ and\ \citenamefont
  {Pupillo.}}]{SM}%
  \BibitemOpen
  \bibfield  {author} {\bibinfo {author} {\bibfnamefont {S.}~\bibnamefont
  {Sch\"utz}}, \bibinfo {author} {\bibfnamefont {J.}~\bibnamefont
  {Schachenmayer}}, \bibinfo {author} {\bibfnamefont {D.}~\bibnamefont
  {Hagenm\"uller}}, \bibinfo {author} {\bibfnamefont {G.~K.}\ \bibnamefont
  {Brennen}}, \bibinfo {author} {\bibfnamefont {T.}~\bibnamefont {Volz}},
  \bibinfo {author} {\bibfnamefont {V.}~\bibnamefont {Sandoghdar}}, \bibinfo
  {author} {\bibfnamefont {T.~W.}\ \bibnamefont {Ebbesen}}, \bibinfo {author}
  {\bibfnamefont {C.}~\bibnamefont {Genes}}, \ and\ \bibinfo {author}
  {\bibfnamefont {G.}~\bibnamefont {Pupillo.}},\ }\href@noop {} {\bibfield
  {journal} {\bibinfo  {journal} {\textit{Supplemental material for
  ``Ensemble-induced strong light-matter coupling of a single quantum
  emitter''}}\ } (\bibinfo {year} {2019})}\BibitemShut {NoStop}%
\bibitem [{\citenamefont {Jaynes}\ and\ \citenamefont
  {Cummings}(1963)}]{JC_1963}%
  \BibitemOpen
  \bibfield  {author} {\bibinfo {author} {\bibfnamefont {E.~T.}\ \bibnamefont
  {Jaynes}}\ and\ \bibinfo {author} {\bibfnamefont {F.~W.}\ \bibnamefont
  {Cummings}},\ }\href {\doibase 10.1109/PROC.1963.1664} {\bibfield  {journal}
  {\bibinfo  {journal} {Proceedings of the IEEE}\ }\textbf {\bibinfo {volume}
  {51}},\ \bibinfo {pages} {89} (\bibinfo {year} {1963})}\BibitemShut {NoStop}%
\bibitem [{\citenamefont {Lehmberg}(1970)}]{Lehmberg1970_Radia}%
  \BibitemOpen
  \bibfield  {author} {\bibinfo {author} {\bibfnamefont {R.~H.}\ \bibnamefont
  {Lehmberg}},\ }\href {\doibase 10.1103/PhysRevA.2.883} {\bibfield  {journal}
  {\bibinfo  {journal} {Phys. Rev. A}\ }\textbf {\bibinfo {volume} {2}},\
  \bibinfo {pages} {883} (\bibinfo {year} {1970})}\BibitemShut {NoStop}%
\bibitem [{\citenamefont {Hagenm\"uller}\ \emph {et~al.}(2019)\citenamefont
  {Hagenm\"uller}, \citenamefont {Sch\"utz}, \citenamefont {Pupillo},\ and\
  \citenamefont {Schachenmayer}}]{AB_long_paper}%
  \BibitemOpen
  \bibfield  {author} {\bibinfo {author} {\bibfnamefont {D.}~\bibnamefont
  {Hagenm\"uller}}, \bibinfo {author} {\bibfnamefont {S.}~\bibnamefont
  {Sch\"utz}}, \bibinfo {author} {\bibfnamefont {G.}~\bibnamefont {Pupillo}}, \
  and\ \bibinfo {author} {\bibfnamefont {J.}~\bibnamefont {Schachenmayer}},\
  }\href@noop {} {\bibfield  {journal} {\bibinfo  {journal} {ArXiv e-prints}\ }
  (\bibinfo {year} {2019})},\ \Eprint {http://arxiv.org/abs/1912.12703}
  {arXiv:1912.12703 [quant-ph]} \BibitemShut {NoStop}%
\bibitem [{\citenamefont {Gross}\ and\ \citenamefont
  {Haroche}(1982)}]{GROSS1982301}%
  \BibitemOpen
  \bibfield  {author} {\bibinfo {author} {\bibfnamefont {M.}~\bibnamefont
  {Gross}}\ and\ \bibinfo {author} {\bibfnamefont {S.}~\bibnamefont
  {Haroche}},\ }\href {\doibase 10.1016/0370-1573(82)90102-8} {\bibfield
  {journal} {\bibinfo  {journal} {Physics Reports}\ }\textbf {\bibinfo {volume}
  {93}},\ \bibinfo {pages} {301 } (\bibinfo {year} {1982})}\BibitemShut
  {NoStop}%
\bibitem [{\citenamefont {Zwanzig}(1960)}]{doi:10.1063/1.1731409}%
  \BibitemOpen
  \bibfield  {author} {\bibinfo {author} {\bibfnamefont {R.}~\bibnamefont
  {Zwanzig}},\ }\href {\doibase 10.1063/1.1731409} {\bibfield  {journal}
  {\bibinfo  {journal} {The Journal of Chemical Physics}\ }\textbf {\bibinfo
  {volume} {33}},\ \bibinfo {pages} {1338} (\bibinfo {year}
  {1960})}\BibitemShut {NoStop}%
\bibitem [{\citenamefont {Reiter}\ and\ \citenamefont
  {S\o{}rensen}(2012)}]{PhysRevA.85.032111}%
  \BibitemOpen
  \bibfield  {author} {\bibinfo {author} {\bibfnamefont {F.}~\bibnamefont
  {Reiter}}\ and\ \bibinfo {author} {\bibfnamefont {A.~S.}\ \bibnamefont
  {S\o{}rensen}},\ }\href {\doibase 10.1103/PhysRevA.85.032111} {\bibfield
  {journal} {\bibinfo  {journal} {Phys. Rev. A}\ }\textbf {\bibinfo {volume}
  {85}},\ \bibinfo {pages} {032111} (\bibinfo {year} {2012})}\BibitemShut
  {NoStop}%
\bibitem [{\citenamefont {Sch\"utz}\ \emph {et~al.}(2013)\citenamefont
  {Sch\"utz}, \citenamefont {Habibian},\ and\ \citenamefont
  {Morigi}}]{Schuetz2013}%
  \BibitemOpen
  \bibfield  {author} {\bibinfo {author} {\bibfnamefont {S.}~\bibnamefont
  {Sch\"utz}}, \bibinfo {author} {\bibfnamefont {H.}~\bibnamefont {Habibian}},
  \ and\ \bibinfo {author} {\bibfnamefont {G.}~\bibnamefont {Morigi}},\ }\href
  {\doibase 10.1103/PhysRevA.88.033427} {\bibfield  {journal} {\bibinfo
  {journal} {Phys. Rev. A}\ }\textbf {\bibinfo {volume} {88}},\ \bibinfo
  {pages} {033427} (\bibinfo {year} {2013})}\BibitemShut {NoStop}%
\bibitem [{\citenamefont {Bienaim\'{e}}\ \emph {et~al.}(2011)\citenamefont
  {Bienaim\'{e}}, \citenamefont {Petruzzo}, \citenamefont {Bigerni},
  \citenamefont {Piovella},\ and\ \citenamefont
  {Kaiser}}]{doi:10.1080/09500340.2011.594911}%
  \BibitemOpen
  \bibfield  {author} {\bibinfo {author} {\bibfnamefont {T.}~\bibnamefont
  {Bienaim\'{e}}}, \bibinfo {author} {\bibfnamefont {M.}~\bibnamefont
  {Petruzzo}}, \bibinfo {author} {\bibfnamefont {D.}~\bibnamefont {Bigerni}},
  \bibinfo {author} {\bibfnamefont {N.}~\bibnamefont {Piovella}}, \ and\
  \bibinfo {author} {\bibfnamefont {R.}~\bibnamefont {Kaiser}},\ }\href
  {\doibase 10.1080/09500340.2011.594911} {\bibfield  {journal} {\bibinfo
  {journal} {Journal of Modern Optics}\ }\textbf {\bibinfo {volume} {58}},\
  \bibinfo {pages} {1942} (\bibinfo {year} {2011})}\BibitemShut {NoStop}%
\bibitem [{\citenamefont {Lesanovsky}\ \emph {et~al.}(2019)\citenamefont
  {Lesanovsky}, \citenamefont {Olmos}, \citenamefont {Guerin},\ and\
  \citenamefont {Kaiser}}]{lesano}%
  \BibitemOpen
  \bibfield  {author} {\bibinfo {author} {\bibfnamefont {I.}~\bibnamefont
  {Lesanovsky}}, \bibinfo {author} {\bibfnamefont {B.}~\bibnamefont {Olmos}},
  \bibinfo {author} {\bibfnamefont {W.}~\bibnamefont {Guerin}}, \ and\ \bibinfo
  {author} {\bibfnamefont {R.}~\bibnamefont {Kaiser}},\ }\href@noop {}
  {\bibfield  {journal} {\bibinfo  {journal} {ArXiv e-prints}\ } (\bibinfo
  {year} {2019})},\ \Eprint {http://arxiv.org/abs/1902.02989} {arXiv:1902.02989
  [physics.atom-ph]} \BibitemShut {NoStop}%
\bibitem [{\citenamefont {Liu}\ \emph {et~al.}(2014)\citenamefont {Liu},
  \citenamefont {Luan}, \citenamefont {Li}, \citenamefont {Gong}, \citenamefont
  {Wong},\ and\ \citenamefont {Xiao}}]{PhysRevLett.112.213602}%
  \BibitemOpen
  \bibfield  {author} {\bibinfo {author} {\bibfnamefont {Y.-C.}\ \bibnamefont
  {Liu}}, \bibinfo {author} {\bibfnamefont {X.}~\bibnamefont {Luan}}, \bibinfo
  {author} {\bibfnamefont {H.-K.}\ \bibnamefont {Li}}, \bibinfo {author}
  {\bibfnamefont {Q.}~\bibnamefont {Gong}}, \bibinfo {author} {\bibfnamefont
  {C.~W.}\ \bibnamefont {Wong}}, \ and\ \bibinfo {author} {\bibfnamefont
  {Y.-F.}\ \bibnamefont {Xiao}},\ }\href {\doibase
  10.1103/PhysRevLett.112.213602} {\bibfield  {journal} {\bibinfo  {journal}
  {Phys. Rev. Lett.}\ }\textbf {\bibinfo {volume} {112}},\ \bibinfo {pages}
  {213602} (\bibinfo {year} {2014})}\BibitemShut {NoStop}%
\bibitem [{\citenamefont {Najer}\ \emph {et~al.}(2019)\citenamefont {Najer},
  \citenamefont {S\"ollner}, \citenamefont {Sekatski}, \citenamefont {Dolique},
  \citenamefont {L\"obl}, \citenamefont {Riedel}, \citenamefont {Schott},
  \citenamefont {Starosielec}, \citenamefont {Valentin}, \citenamefont {Wieck},
  \citenamefont {Sangouard}, \citenamefont {Ludwig},\ and\ \citenamefont
  {Warburton}}]{Najer2019}%
  \BibitemOpen
  \bibfield  {author} {\bibinfo {author} {\bibfnamefont {D.}~\bibnamefont
  {Najer}}, \bibinfo {author} {\bibfnamefont {I.}~\bibnamefont {S\"ollner}},
  \bibinfo {author} {\bibfnamefont {P.}~\bibnamefont {Sekatski}}, \bibinfo
  {author} {\bibfnamefont {V.}~\bibnamefont {Dolique}}, \bibinfo {author}
  {\bibfnamefont {M.~C.}\ \bibnamefont {L\"obl}}, \bibinfo {author}
  {\bibfnamefont {D.}~\bibnamefont {Riedel}}, \bibinfo {author} {\bibfnamefont
  {R.}~\bibnamefont {Schott}}, \bibinfo {author} {\bibfnamefont
  {S.}~\bibnamefont {Starosielec}}, \bibinfo {author} {\bibfnamefont {S.~R.}\
  \bibnamefont {Valentin}}, \bibinfo {author} {\bibfnamefont {A.~D.}\
  \bibnamefont {Wieck}}, \bibinfo {author} {\bibfnamefont {N.}~\bibnamefont
  {Sangouard}}, \bibinfo {author} {\bibfnamefont {A.}~\bibnamefont {Ludwig}}, \
  and\ \bibinfo {author} {\bibfnamefont {R.~J.}\ \bibnamefont {Warburton}},\
  }\href {\doibase 10.1038/s41586-019-1709-y} {\bibfield  {journal} {\bibinfo
  {journal} {Nature}\ }\textbf {\bibinfo {volume} {575}},\ \bibinfo {pages}
  {622} (\bibinfo {year} {2019})}\BibitemShut {NoStop}%
\bibitem [{\citenamefont {Michl}\ \emph {et~al.}(2014)\citenamefont {Michl},
  \citenamefont {Teraji}, \citenamefont {Zaiser}, \citenamefont {Jakobi},
  \citenamefont {Waldherr}, \citenamefont {Dolde}, \citenamefont {Neumann},
  \citenamefont {Doherty}, \citenamefont {Manson}, \citenamefont {Isoya},\ and\
  \citenamefont {Wrachtrup}}]{Michl2014}%
  \BibitemOpen
  \bibfield  {author} {\bibinfo {author} {\bibfnamefont {J.}~\bibnamefont
  {Michl}}, \bibinfo {author} {\bibfnamefont {T.}~\bibnamefont {Teraji}},
  \bibinfo {author} {\bibfnamefont {S.}~\bibnamefont {Zaiser}}, \bibinfo
  {author} {\bibfnamefont {I.}~\bibnamefont {Jakobi}}, \bibinfo {author}
  {\bibfnamefont {G.}~\bibnamefont {Waldherr}}, \bibinfo {author}
  {\bibfnamefont {F.}~\bibnamefont {Dolde}}, \bibinfo {author} {\bibfnamefont
  {P.}~\bibnamefont {Neumann}}, \bibinfo {author} {\bibfnamefont {M.~W.}\
  \bibnamefont {Doherty}}, \bibinfo {author} {\bibfnamefont {N.~B.}\
  \bibnamefont {Manson}}, \bibinfo {author} {\bibfnamefont {J.}~\bibnamefont
  {Isoya}}, \ and\ \bibinfo {author} {\bibfnamefont {J.}~\bibnamefont
  {Wrachtrup}},\ }\href {\doibase 10.1063/1.4868128} {\bibfield  {journal}
  {\bibinfo  {journal} {Applied Physics Letters}\ }\textbf {\bibinfo {volume}
  {104}},\ \bibinfo {pages} {102407} (\bibinfo {year} {2014})}\BibitemShut
  {NoStop}%
\bibitem [{Note1()}]{Note1}%
  \BibitemOpen
  \bibinfo {note} {Note that we are defining this direction as $y$ while
  defining it as $z$ is more common in the literature.}\BibitemShut {Stop}%
\bibitem [{\citenamefont {Hepp}\ \emph {et~al.}(2014)\citenamefont {Hepp},
  \citenamefont {M{\ifmmode\ddot{u}\else\"{u}\fi}ller}, \citenamefont
  {Waselowski}, \citenamefont {Becker}, \citenamefont {Pingault}, \citenamefont
  {Sternschulte}, \citenamefont
  {Steinm{\ifmmode\ddot{u}\else\"{u}\fi}ller-Nethl}, \citenamefont {Gali},
  \citenamefont {Maze}, \citenamefont {Atat{\ifmmode\ddot{u}\else\"{u}\fi}re},\
  and\ \citenamefont {Becher}}]{Hepp_Electronic_2014}%
  \BibitemOpen
  \bibfield  {author} {\bibinfo {author} {\bibfnamefont {C.}~\bibnamefont
  {Hepp}}, \bibinfo {author} {\bibfnamefont {T.}~\bibnamefont
  {M{\ifmmode\ddot{u}\else\"{u}\fi}ller}}, \bibinfo {author} {\bibfnamefont
  {V.}~\bibnamefont {Waselowski}}, \bibinfo {author} {\bibfnamefont {J.~N.}\
  \bibnamefont {Becker}}, \bibinfo {author} {\bibfnamefont {B.}~\bibnamefont
  {Pingault}}, \bibinfo {author} {\bibfnamefont {H.}~\bibnamefont
  {Sternschulte}}, \bibinfo {author} {\bibfnamefont {D.}~\bibnamefont
  {Steinm{\ifmmode\ddot{u}\else\"{u}\fi}ller-Nethl}}, \bibinfo {author}
  {\bibfnamefont {A.}~\bibnamefont {Gali}}, \bibinfo {author} {\bibfnamefont
  {J.~R.}\ \bibnamefont {Maze}}, \bibinfo {author} {\bibfnamefont
  {M.}~\bibnamefont {Atat{\ifmmode\ddot{u}\else\"{u}\fi}re}}, \ and\ \bibinfo
  {author} {\bibfnamefont {C.}~\bibnamefont {Becher}},\ }\href {\doibase
  10.1103/PhysRevLett.112.036405} {\bibfield  {journal} {\bibinfo  {journal}
  {Phys. Rev. Lett.}\ }\textbf {\bibinfo {volume} {112}},\ \bibinfo {pages}
  {036405} (\bibinfo {year} {2014})}\BibitemShut {NoStop}%
\bibitem [{\citenamefont {Evans}\ \emph {et~al.}(2018)\citenamefont {Evans},
  \citenamefont {Bhaskar}, \citenamefont {Sukachev}, \citenamefont {Nguyen},
  \citenamefont {Sipahigil}, \citenamefont {Burek}, \citenamefont {Machielse},
  \citenamefont {Zhang}, \citenamefont {Zibrov}, \citenamefont {Bielejec},
  \citenamefont {Park}, \citenamefont {Lon\v{c}ar},\ and\ \citenamefont
  {Lukin}}]{Evans662}%
  \BibitemOpen
  \bibfield  {author} {\bibinfo {author} {\bibfnamefont {R.~E.}\ \bibnamefont
  {Evans}}, \bibinfo {author} {\bibfnamefont {M.~K.}\ \bibnamefont {Bhaskar}},
  \bibinfo {author} {\bibfnamefont {D.~D.}\ \bibnamefont {Sukachev}}, \bibinfo
  {author} {\bibfnamefont {C.~T.}\ \bibnamefont {Nguyen}}, \bibinfo {author}
  {\bibfnamefont {A.}~\bibnamefont {Sipahigil}}, \bibinfo {author}
  {\bibfnamefont {M.~J.}\ \bibnamefont {Burek}}, \bibinfo {author}
  {\bibfnamefont {B.}~\bibnamefont {Machielse}}, \bibinfo {author}
  {\bibfnamefont {G.~H.}\ \bibnamefont {Zhang}}, \bibinfo {author}
  {\bibfnamefont {A.~S.}\ \bibnamefont {Zibrov}}, \bibinfo {author}
  {\bibfnamefont {E.}~\bibnamefont {Bielejec}}, \bibinfo {author}
  {\bibfnamefont {H.}~\bibnamefont {Park}}, \bibinfo {author} {\bibfnamefont
  {M.}~\bibnamefont {Lon\v{c}ar}}, \ and\ \bibinfo {author} {\bibfnamefont
  {M.~D.}\ \bibnamefont {Lukin}},\ }\href {\doibase 10.1126/science.aau4691}
  {\bibfield  {journal} {\bibinfo  {journal} {Science}\ }\textbf {\bibinfo
  {volume} {362}},\ \bibinfo {pages} {662} (\bibinfo {year}
  {2018})}\BibitemShut {NoStop}%
\bibitem [{\citenamefont {Meesala}\ \emph {et~al.}(2018)\citenamefont
  {Meesala}, \citenamefont {Sohn}, \citenamefont {Pingault}, \citenamefont
  {Shao}, \citenamefont {Atikian}, \citenamefont {Holzgrafe}, \citenamefont
  {G\"undo\u{g}an}, \citenamefont {Stavrakas}, \citenamefont {Sipahigil},
  \citenamefont {Chia}, \citenamefont {Evans}, \citenamefont {Burek},
  \citenamefont {Zhang}, \citenamefont {Wu}, \citenamefont {Pacheco},
  \citenamefont {Abraham}, \citenamefont {Bielejec}, \citenamefont {Lukin},
  \citenamefont {Atat\"ure},\ and\ \citenamefont {Lon\v{c}ar}}]{Meesala2018}%
  \BibitemOpen
  \bibfield  {author} {\bibinfo {author} {\bibfnamefont {S.}~\bibnamefont
  {Meesala}}, \bibinfo {author} {\bibfnamefont {Y.-I.}\ \bibnamefont {Sohn}},
  \bibinfo {author} {\bibfnamefont {B.}~\bibnamefont {Pingault}}, \bibinfo
  {author} {\bibfnamefont {L.}~\bibnamefont {Shao}}, \bibinfo {author}
  {\bibfnamefont {H.~A.}\ \bibnamefont {Atikian}}, \bibinfo {author}
  {\bibfnamefont {J.}~\bibnamefont {Holzgrafe}}, \bibinfo {author}
  {\bibfnamefont {M.}~\bibnamefont {G\"undo\u{g}an}}, \bibinfo {author}
  {\bibfnamefont {C.}~\bibnamefont {Stavrakas}}, \bibinfo {author}
  {\bibfnamefont {A.}~\bibnamefont {Sipahigil}}, \bibinfo {author}
  {\bibfnamefont {C.}~\bibnamefont {Chia}}, \bibinfo {author} {\bibfnamefont
  {R.}~\bibnamefont {Evans}}, \bibinfo {author} {\bibfnamefont {M.~J.}\
  \bibnamefont {Burek}}, \bibinfo {author} {\bibfnamefont {M.}~\bibnamefont
  {Zhang}}, \bibinfo {author} {\bibfnamefont {L.}~\bibnamefont {Wu}}, \bibinfo
  {author} {\bibfnamefont {J.~L.}\ \bibnamefont {Pacheco}}, \bibinfo {author}
  {\bibfnamefont {J.}~\bibnamefont {Abraham}}, \bibinfo {author} {\bibfnamefont
  {E.}~\bibnamefont {Bielejec}}, \bibinfo {author} {\bibfnamefont {M.~D.}\
  \bibnamefont {Lukin}}, \bibinfo {author} {\bibfnamefont {M.}~\bibnamefont
  {Atat\"ure}}, \ and\ \bibinfo {author} {\bibfnamefont {M.}~\bibnamefont
  {Lon\v{c}ar}},\ }\href {\doibase 10.1103/PhysRevB.97.205444} {\bibfield
  {journal} {\bibinfo  {journal} {Phys. Rev. B}\ }\textbf {\bibinfo {volume}
  {97}},\ \bibinfo {pages} {205444} (\bibinfo {year} {2018})}\BibitemShut
  {NoStop}%
\bibitem [{\citenamefont {Bradac}\ \emph {et~al.}(2017)\citenamefont {Bradac},
  \citenamefont {Johnsson}, \citenamefont {van Breugel}, \citenamefont
  {Baragiola}, \citenamefont {Martin}, \citenamefont {Juan}, \citenamefont
  {Brennen},\ and\ \citenamefont {Volz}}]{Bradac2017}%
  \BibitemOpen
  \bibfield  {author} {\bibinfo {author} {\bibfnamefont {C.}~\bibnamefont
  {Bradac}}, \bibinfo {author} {\bibfnamefont {M.~T.}\ \bibnamefont
  {Johnsson}}, \bibinfo {author} {\bibfnamefont {M.}~\bibnamefont {van
  Breugel}}, \bibinfo {author} {\bibfnamefont {B.~Q.}\ \bibnamefont
  {Baragiola}}, \bibinfo {author} {\bibfnamefont {R.}~\bibnamefont {Martin}},
  \bibinfo {author} {\bibfnamefont {M.~L.}\ \bibnamefont {Juan}}, \bibinfo
  {author} {\bibfnamefont {G.~K.}\ \bibnamefont {Brennen}}, \ and\ \bibinfo
  {author} {\bibfnamefont {T.}~\bibnamefont {Volz}},\ }\href {\doibase
  10.1038/s41467-017-01397-4} {\bibfield  {journal} {\bibinfo  {journal}
  {Nature Communications}\ }\textbf {\bibinfo {volume} {8}},\ \bibinfo {pages}
  {1205} (\bibinfo {year} {2017})}\BibitemShut {NoStop}%
\bibitem [{\citenamefont {Jahnke}\ \emph {et~al.}(2015)\citenamefont {Jahnke},
  \citenamefont {Sipahigil}, \citenamefont {Binder}, \citenamefont {Doherty},
  \citenamefont {Metsch}, \citenamefont {Rogers}, \citenamefont {Manson},
  \citenamefont {Lukin},\ and\ \citenamefont {Jelezko}}]{Jahnke_2015}%
  \BibitemOpen
  \bibfield  {author} {\bibinfo {author} {\bibfnamefont {K.~D.}\ \bibnamefont
  {Jahnke}}, \bibinfo {author} {\bibfnamefont {A.}~\bibnamefont {Sipahigil}},
  \bibinfo {author} {\bibfnamefont {J.~M.}\ \bibnamefont {Binder}}, \bibinfo
  {author} {\bibfnamefont {M.~W.}\ \bibnamefont {Doherty}}, \bibinfo {author}
  {\bibfnamefont {M.}~\bibnamefont {Metsch}}, \bibinfo {author} {\bibfnamefont
  {L.~J.}\ \bibnamefont {Rogers}}, \bibinfo {author} {\bibfnamefont {N.~B.}\
  \bibnamefont {Manson}}, \bibinfo {author} {\bibfnamefont {M.~D.}\
  \bibnamefont {Lukin}}, \ and\ \bibinfo {author} {\bibfnamefont
  {F.}~\bibnamefont {Jelezko}},\ }\href {\doibase
  10.1088/1367-2630/17/4/043011} {\bibfield  {journal} {\bibinfo  {journal}
  {New Journal of Physics}\ }\textbf {\bibinfo {volume} {17}},\ \bibinfo
  {pages} {043011} (\bibinfo {year} {2015})}\BibitemShut {NoStop}%
\bibitem [{\citenamefont {Zhang}\ \emph {et~al.}(2018)\citenamefont {Zhang},
  \citenamefont {Sun}, \citenamefont {Burek}, \citenamefont {Dory},
  \citenamefont {Tzeng}, \citenamefont {Fischer}, \citenamefont {Kelaita},
  \citenamefont {Lagoudakis}, \citenamefont {Radulaski}, \citenamefont {Shen},
  \citenamefont {Melosh}, \citenamefont {Chu}, \citenamefont {Lon\v{c}ar},\
  and\ \citenamefont {Vu\v{c}kovi\'{c}}}]{doi:10.1021/acs.nanolett.7b05075}%
  \BibitemOpen
  \bibfield  {author} {\bibinfo {author} {\bibfnamefont {J.~L.}\ \bibnamefont
  {Zhang}}, \bibinfo {author} {\bibfnamefont {S.}~\bibnamefont {Sun}}, \bibinfo
  {author} {\bibfnamefont {M.~J.}\ \bibnamefont {Burek}}, \bibinfo {author}
  {\bibfnamefont {C.}~\bibnamefont {Dory}}, \bibinfo {author} {\bibfnamefont
  {Y.-K.}\ \bibnamefont {Tzeng}}, \bibinfo {author} {\bibfnamefont {K.~A.}\
  \bibnamefont {Fischer}}, \bibinfo {author} {\bibfnamefont {Y.}~\bibnamefont
  {Kelaita}}, \bibinfo {author} {\bibfnamefont {K.~G.}\ \bibnamefont
  {Lagoudakis}}, \bibinfo {author} {\bibfnamefont {M.}~\bibnamefont
  {Radulaski}}, \bibinfo {author} {\bibfnamefont {Z.-X.}\ \bibnamefont {Shen}},
  \bibinfo {author} {\bibfnamefont {N.~A.}\ \bibnamefont {Melosh}}, \bibinfo
  {author} {\bibfnamefont {S.}~\bibnamefont {Chu}}, \bibinfo {author}
  {\bibfnamefont {M.}~\bibnamefont {Lon\v{c}ar}}, \ and\ \bibinfo {author}
  {\bibfnamefont {J.}~\bibnamefont {Vu\v{c}kovi\'{c}}},\ }\href {\doibase
  10.1021/acs.nanolett.7b05075} {\bibfield  {journal} {\bibinfo  {journal}
  {Nano Letters}\ }\textbf {\bibinfo {volume} {18}},\ \bibinfo {pages} {1360}
  (\bibinfo {year} {2018})}\BibitemShut {NoStop}%
\bibitem [{\citenamefont {Riedel}\ \emph {et~al.}(2017)\citenamefont {Riedel},
  \citenamefont {S{\ifmmode\ddot{o}\else\"{o}\fi}llner}, \citenamefont
  {Shields}, \citenamefont {Starosielec}, \citenamefont {Appel}, \citenamefont
  {Neu}, \citenamefont {Maletinsky},\ and\ \citenamefont
  {Warburton}}]{Riedel_Deterministic_2017}%
  \BibitemOpen
  \bibfield  {author} {\bibinfo {author} {\bibfnamefont {D.}~\bibnamefont
  {Riedel}}, \bibinfo {author} {\bibfnamefont {I.}~\bibnamefont
  {S{\ifmmode\ddot{o}\else\"{o}\fi}llner}}, \bibinfo {author} {\bibfnamefont
  {B.~J.}\ \bibnamefont {Shields}}, \bibinfo {author} {\bibfnamefont
  {S.}~\bibnamefont {Starosielec}}, \bibinfo {author} {\bibfnamefont
  {P.}~\bibnamefont {Appel}}, \bibinfo {author} {\bibfnamefont
  {E.}~\bibnamefont {Neu}}, \bibinfo {author} {\bibfnamefont {P.}~\bibnamefont
  {Maletinsky}}, \ and\ \bibinfo {author} {\bibfnamefont {R.~J.}\ \bibnamefont
  {Warburton}},\ }\href {\doibase 10.1103/PhysRevX.7.031040} {\bibfield
  {journal} {\bibinfo  {journal} {Phys. Rev. X}\ }\textbf {\bibinfo {volume}
  {7}},\ \bibinfo {pages} {031040} (\bibinfo {year} {2017})}\BibitemShut
  {NoStop}%
\bibitem [{\citenamefont {Lather}\ \emph {et~al.}(2019)\citenamefont {Lather},
  \citenamefont {Bhatt}, \citenamefont {Thomas}, \citenamefont {Ebbesen},\ and\
  \citenamefont {George}}]{Lather2018}%
  \BibitemOpen
  \bibfield  {author} {\bibinfo {author} {\bibfnamefont {J.}~\bibnamefont
  {Lather}}, \bibinfo {author} {\bibfnamefont {P.}~\bibnamefont {Bhatt}},
  \bibinfo {author} {\bibfnamefont {A.}~\bibnamefont {Thomas}}, \bibinfo
  {author} {\bibfnamefont {T.~W.}\ \bibnamefont {Ebbesen}}, \ and\ \bibinfo
  {author} {\bibfnamefont {J.}~\bibnamefont {George}},\ }\href {\doibase
  10.1002/anie.201905407} {\bibfield  {journal} {\bibinfo  {journal}
  {Angewandte Chemie International Edition}\ }\textbf {\bibinfo {volume}
  {58}},\ \bibinfo {pages} {10635} (\bibinfo {year} {2019})}\BibitemShut
  {NoStop}%
\bibitem [{\citenamefont {Viennot}\ \emph {et~al.}(2015)\citenamefont
  {Viennot}, \citenamefont {Dartiailh}, \citenamefont {Cottet},\ and\
  \citenamefont {Kontos}}]{Viennot408}%
  \BibitemOpen
  \bibfield  {author} {\bibinfo {author} {\bibfnamefont {J.~J.}\ \bibnamefont
  {Viennot}}, \bibinfo {author} {\bibfnamefont {M.~C.}\ \bibnamefont
  {Dartiailh}}, \bibinfo {author} {\bibfnamefont {A.}~\bibnamefont {Cottet}}, \
  and\ \bibinfo {author} {\bibfnamefont {T.}~\bibnamefont {Kontos}},\ }\href
  {\doibase 10.1126/science.aaa3786} {\bibfield  {journal} {\bibinfo  {journal}
  {Science}\ }\textbf {\bibinfo {volume} {349}},\ \bibinfo {pages} {408}
  (\bibinfo {year} {2015})}\BibitemShut {NoStop}%
\end{thebibliography}%


\begin{thebibliography}{9}%
\makeatletter
\providecommand \@ifxundefined [1]{%
 \@ifx{#1\undefined}
}%
\providecommand \@ifnum [1]{%
 \ifnum #1\expandafter \@firstoftwo
 \else \expandafter \@secondoftwo
 \fi
}%
\providecommand \@ifx [1]{%
 \ifx #1\expandafter \@firstoftwo
 \else \expandafter \@secondoftwo
 \fi
}%
\providecommand \natexlab [1]{#1}%
\providecommand \enquote  [1]{``#1''}%
\providecommand \bibnamefont  [1]{#1}%
\providecommand \bibfnamefont [1]{#1}%
\providecommand \citenamefont [1]{#1}%
\providecommand \href@noop [0]{\@secondoftwo}%
\providecommand \href [0]{\begingroup \@sanitize@url \@href}%
\providecommand \@href[1]{\@@startlink{#1}\@@href}%
\providecommand \@@href[1]{\endgroup#1\@@endlink}%
\providecommand \@sanitize@url [0]{\catcode `\\12\catcode `\$12\catcode
  `\&12\catcode `\#12\catcode `\^12\catcode `\_12\catcode `\%12\relax}%
\providecommand \@@startlink[1]{}%
\providecommand \@@endlink[0]{}%
\providecommand \url  [0]{\begingroup\@sanitize@url \@url }%
\providecommand \@url [1]{\endgroup\@href {#1}{\urlprefix }}%
\providecommand \urlprefix  [0]{URL }%
\providecommand \Eprint [0]{\href }%
\providecommand \doibase [0]{http://dx.doi.org/}%
\providecommand \selectlanguage [0]{\@gobble}%
\providecommand \bibinfo  [0]{\@secondoftwo}%
\providecommand \bibfield  [0]{\@secondoftwo}%
\providecommand \translation [1]{[#1]}%
\providecommand \BibitemOpen [0]{}%
\providecommand \bibitemStop [0]{}%
\providecommand \bibitemNoStop [0]{.\EOS\space}%
\providecommand \EOS [0]{\spacefactor3000\relax}%
\providecommand \BibitemShut  [1]{\csname bibitem#1\endcsname}%
\let\auto@bib@innerbib\@empty
\bibitem [{\citenamefont {Lehmberg}(1970)}]{Lehmberg1970_Radia}%
  \BibitemOpen
  \bibfield  {author} {\bibinfo {author} {\bibfnamefont {R.~H.}\ \bibnamefont
  {Lehmberg}},\ }\href {\doibase 10.1103/PhysRevA.2.883} {\bibfield  {journal}
  {\bibinfo  {journal} {Phys. Rev. A}\ }\textbf {\bibinfo {volume} {2}},\
  \bibinfo {pages} {883} (\bibinfo {year} {1970})}\BibitemShut {NoStop}%
\bibitem [{\citenamefont {Zwanzig}(1960)}]{doi:10.1063/1.1731409}%
  \BibitemOpen
  \bibfield  {author} {\bibinfo {author} {\bibfnamefont {R.}~\bibnamefont
  {Zwanzig}},\ }\href {\doibase 10.1063/1.1731409} {\bibfield  {journal}
  {\bibinfo  {journal} {The Journal of Chemical Physics}\ }\textbf {\bibinfo
  {volume} {33}},\ \bibinfo {pages} {1338} (\bibinfo {year}
  {1960})}\BibitemShut {NoStop}%
\bibitem [{\citenamefont {Bonifacio}\ \emph {et~al.}(1971)\citenamefont
  {Bonifacio}, \citenamefont {Schwendimann},\ and\ \citenamefont
  {Haake}}]{Bonifacio1971}%
  \BibitemOpen
  \bibfield  {author} {\bibinfo {author} {\bibfnamefont {R.}~\bibnamefont
  {Bonifacio}}, \bibinfo {author} {\bibfnamefont {P.}~\bibnamefont
  {Schwendimann}}, \ and\ \bibinfo {author} {\bibfnamefont {F.}~\bibnamefont
  {Haake}},\ }\href {\doibase 10.1103/PhysRevA.4.302} {\bibfield  {journal}
  {\bibinfo  {journal} {Phys. Rev. A}\ }\textbf {\bibinfo {volume} {4}},\
  \bibinfo {pages} {302} (\bibinfo {year} {1971})}\BibitemShut {NoStop}%
\bibitem [{\citenamefont {Reiter}\ and\ \citenamefont
  {S\o{}rensen}(2012)}]{PhysRevA.85.032111}%
  \BibitemOpen
  \bibfield  {author} {\bibinfo {author} {\bibfnamefont {F.}~\bibnamefont
  {Reiter}}\ and\ \bibinfo {author} {\bibfnamefont {A.~S.}\ \bibnamefont
  {S\o{}rensen}},\ }\href {\doibase 10.1103/PhysRevA.85.032111} {\bibfield
  {journal} {\bibinfo  {journal} {Phys. Rev. A}\ }\textbf {\bibinfo {volume}
  {85}},\ \bibinfo {pages} {032111} (\bibinfo {year} {2012})}\BibitemShut
  {NoStop}%
\bibitem [{\citenamefont {Sch\"utz}\ \emph {et~al.}(2013)\citenamefont
  {Sch\"utz}, \citenamefont {Habibian},\ and\ \citenamefont
  {Morigi}}]{Schuetz2013}%
  \BibitemOpen
  \bibfield  {author} {\bibinfo {author} {\bibfnamefont {S.}~\bibnamefont
  {Sch\"utz}}, \bibinfo {author} {\bibfnamefont {H.}~\bibnamefont {Habibian}},
  \ and\ \bibinfo {author} {\bibfnamefont {G.}~\bibnamefont {Morigi}},\ }\href
  {\doibase 10.1103/PhysRevA.88.033427} {\bibfield  {journal} {\bibinfo
  {journal} {Phys. Rev. A}\ }\textbf {\bibinfo {volume} {88}},\ \bibinfo
  {pages} {033427} (\bibinfo {year} {2013})}\BibitemShut {NoStop}%
\bibitem [{\citenamefont {Hagenm\"uller}\ \emph {et~al.}(2018)\citenamefont
  {Hagenm\"uller}, \citenamefont {Sch\"utz}, \citenamefont {Schachenmayer},
  \citenamefont {Genes},\ and\ \citenamefont {Pupillo}}]{PhysRevB.97.205303}%
  \BibitemOpen
  \bibfield  {author} {\bibinfo {author} {\bibfnamefont {D.}~\bibnamefont
  {Hagenm\"uller}}, \bibinfo {author} {\bibfnamefont {S.}~\bibnamefont
  {Sch\"utz}}, \bibinfo {author} {\bibfnamefont {J.}~\bibnamefont
  {Schachenmayer}}, \bibinfo {author} {\bibfnamefont {C.}~\bibnamefont
  {Genes}}, \ and\ \bibinfo {author} {\bibfnamefont {G.}~\bibnamefont
  {Pupillo}},\ }\href {\doibase 10.1103/PhysRevB.97.205303} {\bibfield
  {journal} {\bibinfo  {journal} {Phys. Rev. B}\ }\textbf {\bibinfo {volume}
  {97}},\ \bibinfo {pages} {205303} (\bibinfo {year} {2018})}\BibitemShut
  {NoStop}%
\bibitem [{\citenamefont {Hagenm\"uller}\ \emph {et~al.}(2019)\citenamefont
  {Hagenm\"uller}, \citenamefont {Sch\"utz}, \citenamefont {Pupillo},\ and\
  \citenamefont {Schachenmayer}}]{AB_long_paper}%
  \BibitemOpen
  \bibfield  {author} {\bibinfo {author} {\bibfnamefont {D.}~\bibnamefont
  {Hagenm\"uller}}, \bibinfo {author} {\bibfnamefont {S.}~\bibnamefont
  {Sch\"utz}}, \bibinfo {author} {\bibfnamefont {G.}~\bibnamefont {Pupillo}}, \
  and\ \bibinfo {author} {\bibfnamefont {J.}~\bibnamefont {Schachenmayer}},\
  }\href@noop {} {\bibfield  {journal} {\bibinfo  {journal} {ArXiv e-prints}\ }
  (\bibinfo {year} {2019})},\ \Eprint {http://arxiv.org/abs/1912.12703}
  {arXiv:1912.12703 [quant-ph]} \BibitemShut {NoStop}%
\bibitem [{\citenamefont {Navarrete-Benlloch}(2015)}]{Navarrete-Benlloch}%
  \BibitemOpen
  \bibfield  {author} {\bibinfo {author} {\bibfnamefont {C.}~\bibnamefont
  {Navarrete-Benlloch}},\ }\href@noop {} {\bibfield  {journal} {\bibinfo
  {journal} {ArXiv e-prints}\ } (\bibinfo {year} {2015})},\ \Eprint
  {http://arxiv.org/abs/1504.05266} {arXiv:1504.05266 [quant-ph]} \BibitemShut
  {NoStop}%
\bibitem [{\citenamefont {James}(1993)}]{PhysRevA.47.1336}%
  \BibitemOpen
  \bibfield  {author} {\bibinfo {author} {\bibfnamefont {D.~F.~V.}\
  \bibnamefont {James}},\ }\href {\doibase 10.1103/PhysRevA.47.1336} {\bibfield
   {journal} {\bibinfo  {journal} {Phys. Rev. A}\ }\textbf {\bibinfo {volume}
  {47}},\ \bibinfo {pages} {1336} (\bibinfo {year} {1993})}\BibitemShut
  {NoStop}%
\end{thebibliography}%

\end{document}


\title{Supplemental material for ``Ensemble-induced strong light-matter coupling of a single quantum emitter''}

\author{S.~Sch\"utz}
\affiliation{ISIS (UMR 7006) and icFRC, University of Strasbourg and CNRS, 67000 Strasbourg, France}
\author{J.~Schachenmayer}
\affiliation{ISIS (UMR 7006) and icFRC, University of Strasbourg and CNRS, 67000 Strasbourg, France}
\author{D.~Hagenm\"uller}
\affiliation{ISIS (UMR 7006) and icFRC, University of Strasbourg and CNRS, 67000 Strasbourg, France}
\author{G.~K.~Brennen}
\affiliation{Department of Physics \& Astronomy and ARC Centre of Excellence for Engineered Quantum Systems, Macquarie University, NSW 2109, Australia}
\author{T.~Volz}
\affiliation{Department of Physics \& Astronomy and ARC Centre of Excellence for Engineered Quantum Systems, Macquarie University, NSW 2109, Australia}
\author{V.~Sandoghdar}
\affiliation{Max Planck Institute for the Science of Light, Staudtstra{\ss}e 2,
	D-91058 Erlangen, Germany}
\affiliation{Department of Physics, University of Erlangen-Nuremberg, Staudtstra{\ss}e 7,
	D-91058 Erlangen, Germany}
\author{T.~W.~Ebbesen}
\affiliation{ISIS (UMR 7006) and icFRC, University of Strasbourg and CNRS, 67000 Strasbourg, France}
\author{C.~Genes}
\affiliation{Max Planck Institute for the Science of Light, Staudtstra{\ss}e 2,
	D-91058 Erlangen, Germany}
\affiliation{Department of Physics, University of Erlangen-Nuremberg, Staudtstra{\ss}e 7,
	D-91058 Erlangen, Germany}
\author{G. Pupillo}
\affiliation{ISIS (UMR 7006) and icFRC, University of Strasbourg and CNRS, 67000 Strasbourg, France}

\maketitle


In this supplemental material, we provide details of the calculations serving as a basis for the numerical simulations and backing up the statements of the main text. In Sec.~\ref{fQME}, we start from the full quantum master equation including the coupling between the two-level systems (TLSs) and the cavity mode, as well as the coherent and incoherent parts of the dipole-dipole interactions. In Sec.~\ref{eff_QME}, we present an effective quantum master equation for the reduced density operator of the subsystem $A$-cavity when the degrees of freedom associated to the $B$ ensemble are traced out, showing that this effective master equation involves new effective parameters. A coupled-oscillator model based on the equations of motion of the quantum emitters and the cavity mode under the assumption of low spin excitations is used in Sec.~\ref{sec:cav_spec} to compute the cavity transmission spectrum. In Sec.~\ref{sec:disorder}, we present a mode analysis of the $B$ ensemble in the presence of either positional disorder or inhomogeneous broadening.  
 
\section{Full quantum master equation}
\label{fQME}

We consider one TLS $A$ (frequency $\omega_{A}$, decay rate $\gamma_{A}$) at position $\vec{r}_A=(x_{A},y_{A},z_{A})$ coupled via dipole-dipole interactions to a collection of $N$ TLSs $B$ (frequency $\omega_{B}$, decay rate $\gamma_{B}$) at positions $\vec{r}_1, \dots, \vec{r}_N$, and interacting with a cavity mode (frequency $\omega_{c}$, decay rate $\kappa$). We denote the $A-B$ separation as $\vec{r}_{j A} = \vec{r}_{j} - \vec{r}_A $ and the $B-B$ separation as $\vec{r}_{j \ell} = \vec{r}_{j} - \vec{r}_{\ell}$ with $j,\ell=1, \dots, N$ and $\vec{r}_{j}=(x_{j},y_{j},z_{j})$. In the frame rotating with $\omega_A$, the full quantum master equation of the system reads
\begin{align}
\partial_t \rho = - \mi [ H_0 + H_{\text{JC}} + H_{\text{DD}} , \rho ] + \mathcal{L}\rho \,,
\label{FQME}
\end{align}
where $\rho$ is the density operator and $H_0 = \Delta_c a^{\dagger} a + \Delta_{B} \sum_{j=1}^{N} \sigma_j^+ \sigma_j^-$ is the bare Hamiltonian with $\Delta_c = \omega_{c}-\omega_{A}$ and $\Delta_B = \omega_{B}-\omega_{A}$. The bosonic operators $a$ and $a^{\dagger}$ annihilate and create a cavity photon, and $\sigma^{\pm}_{A}$ and $\sigma^{\pm}_{j}$ are the spin ladder operators of the TLS $A$ and the $j^{\textrm{th}}$ TLS $B$. The light-matter interaction Hamiltonian reads $H_{\text{JC}} = a^{\dagger} \Big( g_A \sigma_A^- + \sum_{j=1}^{N} g_{j} \sigma_j^- \Big) + \textrm{h.c.}$ with the (real-valued) coupling strengths $g_{A}=g^{(0)}_{A}\cos\left(k y_{A}\right)$ and $g_{j}=g^{(0)}_{B}\cos\left(k y_{j}\right)$ (we choose $g^{(0)}_{A},g^{(0)}_{B}>0$). The photon wave vector $k$ (along the $y$-direction) is related to the cavity mode wavelength $\lambda$ by $k=2\pi/\lambda=\omega_{c}/c$ with $c$ the speed of light in vacuum. The dipole-dipole interaction Hamiltonian reads $H_{\text{DD}} = \sum_{j=1}^{N} \Omega_{j A} \big( \sigma_{j}^+ \sigma_A^- + \sigma_A^+ \sigma_{j}^- \big) + \sum_{j \neq \ell}^{N} \Omega_{j \ell} \sigma_{j}^+ \sigma_{\ell}^{-}$, where the $A$-$B$ and $B$-$B$ coupling strengths $\Omega_{j A} = \Re \big[ V_{j A}\big]$ and $\Omega_{j\ell} = \Re \big[ V_{j\ell} \big]$ correspond to the real parts of the functions~\cite{Lehmberg1970_Radia}
\begin{align}
V_{j A} &= - \frac{3 \sqrt{\gamma_A \gamma_B}}{2} \Big( \sin^2(\Theta_{j A}) \frac{\exp(\mi \xi_{j A})}{\xi_{j A}} + \Big[ 3 \cos^2(\Theta_{j A}) - 1 \Big] \Big[ \frac{\exp(\mi \xi_{j A})}{\xi^{3}_{j A}} - \mi \frac{\exp(\mi \xi_{j A})}{\xi^{2}_{j A}} \Big] \Big), \label{G_funAB}\\
V_{j \ell} &= - \frac{3 \gamma_B}{2} \Big( \sin^2(\Theta_{j \ell}) \frac{\exp(\mi \xi_{j \ell})}{\xi_{j \ell}} + \Big[ 3 \cos^2(\Theta_{j \ell}) - 1 \Big] \Big[ \frac{\exp(\mi \xi_{j \ell})}{\xi^{3}_{j \ell}} - \mi \frac{\exp(\mi \xi_{j \ell})}{\xi^{2}_{j \ell}} \Big] \Big). 
\label{G_funBB}
\end{align}
Here, $\xi_{j A} = k |\vec{r}_{j A}|$ and $\xi_{j \ell} = k |\vec{r}_{j \ell}|$, the angles $\Theta_{j A} = \arccos\left( z_{j A}/|\vec{r}_{j A}| \right)$ and $\Theta_{j \ell} = \arccos\left(z_{j \ell}/|\vec{r}_{j \ell}| \right)$ are defined with respect to the $z$-axis, and we assume that $k_{A}\approx k_{B}\approx k$, where $k_{A}=\omega_{A}/c$ and $k_{B}=\omega_{B}/c$. The dissipative term entering Eq.~(\ref{FQME}) takes the form
\begin{align*}
\mathcal{L} \rho 
= &- \kappa \mathcal{D} (a^{\dagger},a) \rho - \gamma_A \mathcal{D} (\sigma_A^+, \sigma_A^-) \rho - \sum_{j=1}^{N} \gamma_{j A} \Big( \mathcal{D} (\sigma_{j}^+ ,\sigma_A^-) \rho + \mathcal{D} (\sigma_{A}^+,\sigma_j^-) \rho \Big)  
- \sum_{j,\ell= 1}^{N} \gamma_{j \ell} \mathcal{D} (\sigma_{j}^+,\sigma_{\ell}^-) \rho,
\end{align*}
with $\mathcal{D} (x,y) \rho = [x, y \rho] + [\rho x, y]$, and where $\gamma_{j A} = - \Im \big[ V_{j A}\big]$ and $\gamma_{j \ell} = - \Im \big[ V_{j \ell}\big]$ are given by the imaginary parts of Eqs.~(\ref{G_funAB}) and (\ref{G_funBB}), respectively.
The anisotropic dimensionless functions $g(\vec{r})$ and $f(\vec{r})$ in the main text are given by $g(\vec{r}_{j A}) = \Omega_{j A} / \sqrt{\gamma_A \gamma_B}$ and $f(\vec{r}_{jA}) = \gamma_{jA} / \sqrt{\gamma_A \gamma_B}$ (or alternatively $g(\vec{r}_{j \ell}) = \Omega_{j \ell} / \gamma_B$ and $f(\vec{r}_{j \ell}) = \gamma_{j \ell} / \gamma_B$).

\vspace*{0.5cm}
\noindent {\bf Simulation of Rabi oscillations with the full master equation} (Fig.~2({\bf b}) \& ({\bf d}) of the main text). Any expectation value $\langle O \rangle_t = \text{Tr} [ O \rho(t) ]$ for an arbitrary operator $O$ can be determined by computing the density operator $\rho (t)$ for a given initial condition $\rho(0)$ according to the dynamics encoded in Eq.~\eqref{FQME}. Here, $\text{Tr} [\bullet]$ denotes the trace over the full Hilbert space. For instance, the values
$$
\bar{n} (t) = \langle a^{\dagger} a \rangle_t = \text{Tr} [a^{\dagger} a \rho(t)],
\quad
P_A (t) = \langle \sigma_A^+ \sigma_A^- \rangle_t = \text{Tr} [\sigma_A^+ \sigma_A^- \rho(t)],
\quad \text{and} \quad
P_j(t) = \langle \sigma_j^+ \sigma_j^- \rangle_t = \text{Tr} [\sigma_j^+ \sigma_j^- \rho(t)] 
$$ 
correspond to the mean photon number $\bar{n}(t)$, the mean population/excitation $P_A(t)$ of emitter $A$, and the mean population/excitation $P_j(t)$ of emitter $j$ ($j=1, \dots, N$) of the $B$ ensemble as a function of time, respectively. The time evolution corresponding to the lines in Fig.~2({\bf b}) \& ({\bf d}) of the main text is computed with the initial condition $\rho(0) = \ket{\psi_0} \bra{\psi_0}$, where the initial state $\ket{\psi_0}$ corresponds to a single photon in the cavity with the emitter $A$ and all emitters $B$ in their ground states. 

\section{Effective quantum master equation}
\label{eff_QME}

In this section, we present the effective quantum master equation for the subsystem $A$-cavity when the TLSs $B$ are traced out. We assume that the TLSs $B$ remain close to their ground state and that the dynamics of the reduced system is slow compared to that of the fast evolving $B$ ensemble, which can be adiabatically eliminated~\cite{doi:10.1063/1.1731409,Bonifacio1971,PhysRevA.85.032111,Schuetz2013,PhysRevB.97.205303}. The parameters entering the effective master equation are given below. A detailed derivation including a discussion on the conditions for the validity of the adiabatic elimination can be found in Ref.~[\onlinecite{AB_long_paper}].

\vspace*{0.5cm}
\noindent {\bf Effective master equation.} We can write the effective master equation in the form 
\begin{align}
{\partial_t} v = \mathfrak{L}^{\text{eff}}v = - \mi[H_0^{\text{eff}} + H_{\rm JC}^{\rm eff}, v] + \mathcal{L}^{\text{eff}}v \,,
\label{final_mast_eq}
\end{align}
where $v = \rho_{gg} \ket{g} \bra{g}$ with $\rho_{gg} = \langle g | \rho |g \rangle$ denotes the density operator $\rho$ projected on the state $| g \rangle $ with {\it all} emitters of the $B$ ensemble in their ground states. Its dynamics is governed by the effective Hamiltonian $H^{\text{eff}} = H_0^{\text{eff}} + H_{\rm JC}^{\rm eff}$ with $H_0^{\text{eff}}=\Delta^{\text{eff}}_{A} \sigma^{+}_{A}\sigma^{-}_{A} + \Delta^{\text{eff}}_{c} a^{\dagger} a$ and $H_{\rm JC}^{\rm eff} = g_A^{\rm eff} \Big( a^{\dagger} \sigma_A^- + \sigma_A^+ a \Big)$, and by the effective dissipator
\begin{align*}
\mathcal{L}^{\text{eff}} v =-\kappa^{\text{eff}} \mathcal{D}(a^{\dagger}, a) v -\gamma_A^{\text{eff}} \mathcal{D}(\sigma_A^+, \sigma_A^-) v - \mu \Big( \mathcal{D} (a^{\dagger}, \sigma_A^-) v + \mathcal{D} (\sigma_A^+, a) v \Big)\,.
\end{align*}
The effective parameters are 
\begin{alignat*}{3}
\Delta_c^{\text{eff}} &= \Delta_c - \Re [\vec{g}^{\rm T} {\bf M}^{-1} \vec{g}]
&\qquad 
\Delta_A^{\text{eff}} &= - \Re [\vec{v}^{\rm T} {\bf M}^{-1} \vec{v}]  
&\qquad g_A^{\text{eff}} &= g_A - \Re [\vec{g}^{\rm T} {\bf M}^{-1} \vec{v}] \nonumber  \\
\kappa^{\text{eff}} &= \kappa + \Im [\vec{g}^{\rm T} {\bf M}^{-1} \vec{g}] & \gamma_A^{\text{eff}} &= \gamma_A + \Im [\vec{v}^{\rm T} {\bf M}^{-1} \vec{v}] & \mu &= \Im [\vec{g}^{\rm T} {\bf M}^{-1} \vec{v}] \,,
\end{alignat*}
where $(\vec{g})_j = g_j$, $(\vec{v})_j = \Omega_{j A} - \mi \gamma_{j A}$ and $({\bf M})_{j \ell} = (\Delta_B - \mi \gamma_B) \delta_{j\ell} + (1-\delta_{j\ell}) (\Omega_{j \ell} - \mi \gamma_{j \ell})$. Using $v=\rho_{gg}| g \rangle \langle g |$ and the reduced density operator $\rho^{\rm eff} \approx \rho_{gg}$, the effective master equation \eqref{final_mast_eq} is identical to Eq.~(2) of the main text.

\vspace*{0.5cm}
\noindent {\bf Simulation of Rabi oscillations with the effective master equation} (Fig.~2({\bf b}) and ({\bf d}) of the main text). The expectation values of the observables of the subsystem $A$-cavity $$\bar{n}^{(\text{eff})} (t) = \langle a^{\dagger} a \rangle_t^{(\text{eff})} = \text{Tr}^{(\text{eff})} [a^{\dagger} a \rho^{\text{eff}}(t)],
\quad \text{and} \quad
P_A (t)^{(\text{eff})} = \langle \sigma_A^+ \sigma_A^- \rangle^{(\text{eff})}_t = \text{Tr}^{(\text{eff})} [\sigma_A^+ \sigma_A^- \rho^{\text{eff}}(t)]
$$ (points in Fig.~2({\bf b}) and ({\bf d}) of the main text) are determined by computing $\rho^{\text{eff}} (t)$ according to the dynamics encoded in Eq.~\eqref{final_mast_eq}. Here, $\text{Tr}^{(\text{eff})} [\bullet]$ denotes the trace over the reduced Hilbert space of the subsystem $A$-cavity, and the initial condition is $\rho^{\text{eff}}(0) = \ket{\psi_0^{\text{eff}}} \bra{\psi_0^{\text{eff}}}$ where the initial state $\ket{\psi_0^{\text{eff}}}$ corresponds to a single photon in the cavity and the emitter $A$ in its ground state. Since Eq.~\eqref{final_mast_eq} is derived under the assumption that the emitters $B$ are close to their ground states, the total excitation $P_B = \sum_{\ell=1}^{N} \langle \sigma_\ell^+ \sigma_\ell^- \rangle$ of the $B$ ensemble must remain small, i.e. $P_B \ll 1$. In the adiabatic regime, we estimate $P_B$ as
\begin{align*}
P_B 
= \sum_{\ell=1}^{N} \langle \sigma_\ell^+ \sigma_\ell^- \rangle 
&\approx 
\vec{g}^T ({\bf M} {\bf M}^{\dagger})^{-1} \vec{g} \langle a^{\dagger} a \rangle
+
(\vec{v}^{*})^{T} ({\bf M} {\bf M}^{\dagger})^{-1} \vec{v} \langle \sigma_A^+ \sigma_A^- \rangle \\
&+
\vec{g}^{T} ({\bf M} {\bf M}^{\dagger})^{-1} \vec{v} \langle a^{\dagger} \sigma_A^- \rangle
+
(\vec{v}^{*})^{T} ({\bf M} {\bf M}^{\dagger})^{-1} \vec{g} \langle \sigma_A^+ a \rangle \,.
\nonumber
\end{align*}
Note that the expectation value $\langle \bullet \rangle$ is evaluated with respect to the operator $\rho_{gg} \approx \rho^{\text{eff}}$.

\vspace*{0.5cm}
\noindent {\bf Effective master equation with cavity drive.} In the presence of a weak laser driving the cavity with strength $\phi$ (see also Sec.~\ref{sec:cav_spec}), the system's dynamics can be expressed in a frame rotating with the laser frequency $\omega_L$. This frequency is assumed to be close to the typical frequencies of the probed subsystem $A$-cavity (e.g. cavity frequency and transition frequency of $A$). In this case, the effective master equation for the operator $\tilde{v}$ in the laser frame reads
\begin{align}
{\partial_t} \tilde{v} = \widetilde{\mathfrak{L}}^{\text{eff}} \tilde{v} = - \mi[ \widetilde{H}_0^{\text{eff}} + \widetilde{H}_{\rm JC}^{\rm eff}+ \widetilde{H}_{L}, \tilde{v}] + \widetilde{\mathcal{L}}^{\text{eff}} \tilde{v}
\label{eff_mast_eq_laser}
\end{align}
with $\widetilde{H}_0^{\text{eff}}=\widetilde{\Delta}^{\text{eff}}_{A} \sigma^{+}_{A}\sigma^{-}_{A} + \widetilde{\Delta}^{\text{eff}}_{c} a^{\dagger} a$, $\widetilde{H}_{\rm JC}^{\rm eff} = \widetilde{g}_A^{\rm eff} \Big( a^{\dagger} \sigma_A^- + \sigma_A^+ a \Big)$, $\widetilde{H}_{L}=\phi (a+a^{\dagger})$, and 
\begin{align*}
\widetilde{\mathcal{L}}^{\text{eff}} \tilde{v} =
-\widetilde{\kappa}^{\text{eff}} \mathcal{D}(a^{\dagger}, a) \tilde{v} - \widetilde{\gamma}_A^{\text{eff}} \mathcal{D}(\sigma_A^+, \sigma_A^-) \tilde{v} - \widetilde{\mu} \Big( \mathcal{D} (a^{\dagger}, \sigma_A^-) \tilde{v} + \mathcal{D} (\sigma_A^+, a) \tilde{v} \Big).
\end{align*}
The effective parameters are
\begin{alignat}{3}
\widetilde{\Delta}_c^{\text{eff}} &= \widetilde{\Delta}_c - \Re [\vec{g}^{\rm T} \widetilde{{\bf M}}^{-1} \vec{g}]
&\qquad 
\widetilde{\Delta}_A^{\text{eff}} &= \widetilde{\Delta}_A - \Re [\vec{v}^{\rm T} \widetilde{{\bf M}}^{-1} \vec{v}]  
&\qquad \widetilde{g}_A^{\text{eff}} &= g_A - \Re [\vec{g}^{\rm T} \widetilde{{\bf M}}^{-1} \vec{v}] \nonumber  \\
\widetilde{\kappa}^{\text{eff}} &= \kappa + \Im [\vec{g}^{\rm T} \widetilde{{\bf M}}^{-1} \vec{g}] & \widetilde{\gamma}_A^{\text{eff}} &= \gamma_A + \Im [\vec{v}^{\rm T} \widetilde{{\bf M}}^{-1} \vec{v}] & \widetilde{\mu} &= \Im [\vec{g}^{\rm T} \widetilde{{\bf M}}^{-1} \vec{v}].
\label{eff_param_laser}
\end{alignat}
Here, $\widetilde{\Delta}_c = \omega_c - \omega_L$, $\widetilde{\Delta}_A = \omega_A - \omega_L$, and $(\widetilde{{\bf M}})_{j \ell}= \big(\widetilde{\Delta}_{B} - \mi \gamma_B\big) \delta_{j\ell}+ (1-\delta_{j\ell})(\Omega_{j\ell} - \mi \gamma_{j \ell})$ with $\widetilde{\Delta}_B = \omega_B - \omega_L$. 

\vspace*{0.5cm}
\noindent {\bf Transmission spectrum and intensity-intensity correlation function} (Figs.~1({\bf b}) \& ({\bf c}) and 3({\bf c}) of the main text). The steady state $\tilde{v}_{\infty} = \lim_{t \rightarrow \infty} \tilde{v}(t)$ can be obtained by simulating Eq.~\eqref{eff_mast_eq_laser} in the long-time limit, or alternatively by constructing a matrix associated to the superoperator $\widetilde{\mathfrak{L}}^{\text{eff}}$ and determining its null space (corresponding to the zero eigenvalue \cite{Navarrete-Benlloch}). The normalized steady-state cavity transmission spectra shown in Fig.~1({\bf b}) of the main text are computed as
\begin{align}
\mathcal{T}_c (\omega_L) \equiv (\kappa/\phi)^2 \langle a^{\dagger} a \rangle_{\infty} = (\kappa/\phi)^2 \text{Tr}^{(\text{eff})} [a^{\dagger} a \tilde{v}_{\infty}] \,,
\label{eq:steady_transmission}
\end{align}
and the steady-state equal-time intensity-intensity correlation functions [Figs.~1({\bf c}) and 3({\bf c}) of the main text] as
\begin{align*}
g^{(2)}(0) \equiv \frac{\langle a^{\dagger} a^{\dagger} a a \rangle_{\infty}}{\langle a^{\dagger} a \rangle_{\infty}^2} = \frac{\text{Tr}^{(\text{eff})} [a^{\dagger} a^{\dagger} a a \tilde{v}_{\infty}]}{\left(\text{Tr}^{(\text{eff})} [a^{\dagger} a \tilde{v}_{\infty}]\right)^2} \,.
\end{align*}
%
In Figs.~1({\bf b}) \& ({\bf c}) and 3({\bf c}) of the main text we choose a cavity drive with strength $\phi = 0.1 \kappa$ and a cutoff $N_{\rm ph}^{(\rm max)} =3$ in the photon number, for which sufficient convergence (in the considered range) has been found.

\section{Coupled-oscillator model for the cavity transmission spectrum}
\label{sec:cav_spec}

In this section, we present the equations of motion for the expectation values of the cavity mode and the spin operators under the assumption of low spin excitations (low-saturation limit). In this limit, the resulting linear equations of motion are equivalent to that of coupled harmonic oscillators. We then use this model to compute the cavity transmission spectrum in the presence of a weak laser probe. 

\vspace*{0.5cm}
\noindent {\bf Equations of motion in the low-saturation limit}. 
We consider a weak laser probe with frequency $\omega_L$ driving the cavity with strength $\phi$ (we choose $\phi \in \mathbb{R}$). This laser drive is described by the (additional) time-dependent Hamiltonian $H_{L} = \phi (a e^{\mi \omega_L t} + a^{\dagger} e^{-\mi \omega_L t})$
and can be included in the full dynamics of Eq. \eqref{FQME} by adding the term $- i [H_{L}, \rho]$ on its right-hand side.
The explicit time-dependence can be removed by choosing a convenient frame that rotates at the laser frequency $\omega_L$. 
%
%
In this frame, the equations of motion for
$\alpha (t) = \langle a \rangle_t$, $\beta_A (t) = \langle \sigma_A^- \rangle_t$, and $(\vec{\beta})_{\ell} (t) = \langle \sigma_\ell^- \rangle_t$ ($\ell = 1, \dots, N$) can be derived as
\begin{align}
\label{equ_exact_ss}
\partial_t \alpha &= - \mi \big[ \widetilde{\Delta}_c - \mi \kappa \big] \alpha - \mi g_A \beta_A - \mi \vec{g}^{\rm T} \vec{\beta} - \mi \phi \,, \\
\partial_t \beta_A &= - \mi \big[ \widetilde{\Delta}_A - \mi \gamma_A \big] \beta_A - \mi g_A \alpha - \mi \vec{v}^{\rm T} \vec{\beta} \,, \nonumber \\
\partial_t \vec{\beta} &= -\mi \widetilde{{\bf M}} \vec{\beta} - \mi \vec{g} \alpha -\mi \vec{v} \beta_A \,. \nonumber
\end{align}
These equations are obtained under the assumption of weak excitation, i.e.
$[\sigma_A^-, \sigma_A^+] \approx 1$ and $[\sigma_{\ell}^-, \sigma_{\ell}^+] \approx 1$ (harmonic-oscillator approximation~\cite{PhysRevA.47.1336}). The parameters are given in the laser frame with $\widetilde{\Delta}_c = \omega_c - \omega_L$, $\widetilde{\Delta}_A = \omega_A - \omega_L$, and $(\widetilde{{\bf M}})_{j \ell}= \big(\widetilde{\Delta}_{B} - \mi \gamma_B\big) \delta_{j\ell}+ \big( 1-\delta_{j\ell} \big) \big( \Omega_{j\ell} - \mi \gamma_{j \ell} \big)$ with $\widetilde{\Delta}_B = \omega_B - \omega_L$. The steady-state solution ($t \rightarrow \infty$) of Eq.~\eqref{equ_exact_ss} reads
\begin{align}
\begin{pmatrix*}
\alpha^{\rm st} \\ \\
\beta^{\rm st}_A
\end{pmatrix*}
=
-
\begin{pmatrix*}
\widetilde{\Delta}_{c}^{\text{eff}} - \mi \widetilde{\kappa}^{\text{eff}} 
& \widetilde{g}_A^{\text{eff}} - \mi \widetilde{\mu} \\ \\
\widetilde{g}_A^{\text{eff}} - \mi \widetilde{\mu}
& \widetilde{\Delta}_{A}^{\text{eff}} - \mi \widetilde{\gamma}_A^{\text{eff}}
\end{pmatrix*}^{-1}
\begin{pmatrix*}
\phi \\ \\
0 
\end{pmatrix*}
\,,
\label{mat_laser}
\end{align}
where $\alpha^{\rm st} = \lim\limits_{t \rightarrow \infty} \alpha (t)$, $\beta_A^{\rm st} = \lim\limits_{t \rightarrow \infty} \beta_A (t)$, and where the parameters are given by Eq.~\eqref{eff_param_laser}. In the limit of a weak laser drive, the mean photon number in the steady-state $\bar{n}^{\rm st} \simeq |\alpha^{\rm st}|^2$ can be obtained from Eq.~\eqref{mat_laser}. This allows to compute the normalized steady-state cavity transmission spectrum as
\begin{align}
\mathcal{T}_c (\omega_L) = (\kappa/\phi)^2 |\alpha^{\rm st} |^2\,.
\label{eq:steady_trans_osc}
\end{align}
In the weak-drive limit ($\phi \rightarrow 0$), we expect that $\mathcal{T}_c$ computed from Eq.~\eqref{eq:steady_trans_osc} provides a good approximation of the transmission that would be obtained from an operator expectation value as in Eq.~\eqref{eq:steady_transmission}. Note that while the adiabatic elimination leading to the effective master equation Eq.~\eqref{eff_mast_eq_laser} does require a time-scale separation, the derivation of the steady-state solution Eq.~\eqref{eq:steady_trans_osc} does not. This allows to use Eq.~\eqref{eq:steady_trans_osc} to compute the steady-state cavity transmission spectra for disordered configurations, as shown in Sec.~\ref{sec:disorder}.

\section{Impact of disorder}
\label{sec:disorder}

In this section, we study the robustness of the coupling-enhancement scheme against both inhomogeneous broadening and positional disorder of the $B$ ensemble. 

\begin{figure}[ht]
	\centering
	\includegraphics[width=1\columnwidth]{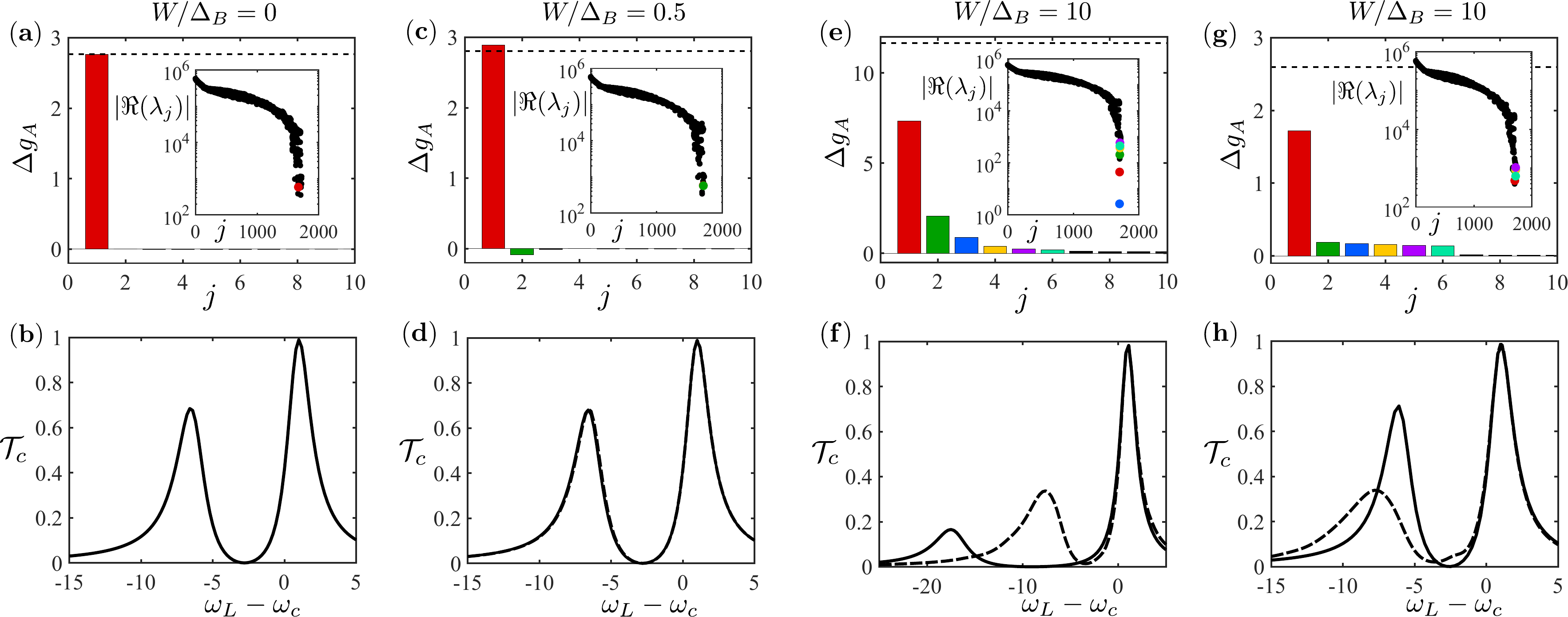}
	\caption{(\textbf{a}), (\textbf{c}), (\textbf{e}), (\textbf{g}) Modal decomposition of the coupling strength enhancement $\Delta g_{A}\equiv g^{\rm eff}_{A}-g_{A}=-\sum_j \Re ( \vec{g}^T \vec{x}_j \vec{x}^T_j \vec{v} /\lambda_j)$ related to the eigenvectors $\vec{x}_{j}$ and eigenvalues $\lambda_{j}$ ($\vert \Re (\lambda_{j}) \vert$ is shown in the inset) of the matrix ${\bf M}$. The total coupling enhancement $\Delta g_{A}$ is displayed as a dashed line. (\textbf{b}), (\textbf{d}), (\textbf{f}), (\textbf{h}) Steady-state cavity transmission spectrum $\mathcal{T}_{c}$ (single disorder realization). The transmission averaged over $100$ realizations is displayed as a dashed line. Vertically, (\textbf{a}) \& (\textbf{b}) correspond to $W=0$ (no disorder), (\textbf{c}) \& (\textbf{d}) to $W/\Delta_{B}=0.5$, and (\textbf{e}) \& (\textbf{f}) and (\textbf{g}) \& (\textbf{h}) correspond to $W/\Delta_{B}=10$ and two different disorder realizations. Parameters are the same as in Fig.~1 of the main text ($N=12^3$).}
\label{fig:weak}
\end{figure}

\vspace*{0.5cm}
\noindent {\bf Inhomogeneous broadening}. We first consider inhomogeneous broadening of the $B$ ensemble and replace $\Delta_B$ by $\Delta_{j}\in [\Delta_{B}-W/2,\Delta_{B}+W/2]$ (uniform distribution) with $W$ the disorder strength. Figure \ref{fig:weak} shows a modal decomposition of the coupling strength enhancement $$\Delta g_{A}\equiv g^{\rm eff}_{A}-g_{A}=-\sum_{j=1}^{N} \Re ( \vec{g}^T \vec{x}_j \vec{x}^T_j \vec{v} /\lambda_j)$$ together with the transmission spectrum $\mathcal{T}_{c}$ computed from Eq.~\eqref{eq:steady_trans_osc} using the steady state solution of Eq.~(\ref{equ_exact_ss}). Here, $\vec{x}_j$ are the eigenvectors and $\lambda_{j}$ the eigenvalues of the symmetric matrix ${\bf M} = \sum_{j}\lambda_{j} \vec{x}_{j}\vec{x}^{\rm T}_{j}$, which is defined in Sec.~\ref{eff_QME}, together with $\vec{g}$ and $\vec{v}$. The eigenvectors here fulfill the completeness relation $\sum_{j}\vec{x}_{j} \vec{x}^{\rm T}_{j}= {\bm 1}$ and $\vec{x}^{\rm T}_{j} \vec{x}_{\ell}=\delta_{j,\ell}$.

For $W=0$ [Fig.~\ref{fig:weak}(\textbf{a}) \& (\textbf{b})], the single-mode approximation is very well suited as $\Delta g_{A}$ is completely dominated by the contribution of a collective mode which features an overlap of $95\%$ with the symmetric mode $\vec{x}_S = (1,1,\dots,1)/\sqrt{N}$. The corresponding eigenvalue $\Re (\lambda_{S})\approx 575$ is relatively close to $\Delta_{B}=1000$ as compared to the other ones which are shifted to very large values due to the dipole-dipole interactions within the emitter ensemble. Figure \ref{fig:weak}(\textbf{c}) \& (\textbf{d}) shows $\Delta g_{A}$ and $\mathcal{T}_{c}$ for a single realization in the case of a small disorder $W/\Delta_{B}=0.5$. We find that the single-mode approximation is still well justified, as the overlap between the dominant mode (red) and the symmetric one $\vec{x}_S$ is $\approx 97\%$, while this overlap for the next mode (green) is only $\approx 10\%$. This picture is confirmed by the small fluctuations of the transmission spectrum. Furthermore, the conditions for adiabatic elimination of the $B$ ensemble are well satisfied owing to the small population of the ensemble $P_{B}=\sum_{j}\langle \sigma^{+}_{j}\sigma^{-}_{j}\rangle \equiv \vert \vec{\beta}\vert^{2}\simeq 0.005$, where we used $P_{B}\approx \vec{g}^T ({\bf M}{\bf M}^{\dagger})^{-1} \vec{g}$ assuming one cavity photon.

Two different realizations obtained for a large disorder strength $W/\Delta_{B}=10$ are shown in Fig.~\ref{fig:weak}(\textbf{e})-(\textbf{h}). In panels (\textbf{e}) \& (\textbf{f}), the first three modes with the largest contributions to $\Delta g_{A}$ feature overlaps with the symmetric mode of $43\%$ (red), $48\%$ (green), and $0.03\%$ (blue), with eigenvalues (real parts) $\approx 43$, $\approx 200$, and $\approx 2.5$, respectively. Such small eigenvalues indicate a breakdown of the conditions for adiabatic elimination, and the effective coupling strength therefore loses its physical meaning. In this case, the latter is $g^{\rm eff}_{A}\approx 12.6$ and does not correspond to the half-splitting $\approx 9.2$ between the polariton peaks observed in the transmission spectrum. The breakdown of adiabatic elimination can be also inferred from the increased population of the ensemble $P_{B} \approx 0.2$. In panels (\textbf{g}) \& (\textbf{h}), we show an example of configuration where the single-mode approximation is still rather well justified. In this case, the red mode with eigenvalue $\approx 468$ and with the largest contribution to $\Delta g_{A}$ features an overlap with the symmetric mode of $70\%$, while the next one (green) with eigenvalue $\approx 860$ only exhibits an overlap of $\approx 30\%$. The computed effective coupling strength is $g^{\rm eff}_{A}\approx 3.6$, and corresponds to the half-splitting of (\textbf{h}). The adiabatic elimination of the emitter ensemble is here also well justified due to the large eigenvalues and small population $P_{B}\approx 0.005$. Note that for the single-mode model, in the limit $W/\Delta_{B}\ll 1$, a second-order expansion of $\Delta g_{A}$ gives a disorder-averaged enhancement $$\overline{\Delta g_{A}}\approx -\frac{\vec{g}^T \vec{x}_S \vec{x}^T_S \vec{v}}{\lambda_{S}}\left[1+\mathcal{O}\left(W/\lambda_{S}\right)^{2}\right],$$ indicating a slight increase with $W$.

\begin{figure}[ht]
	\centering
	\includegraphics[width=0.7\textwidth]{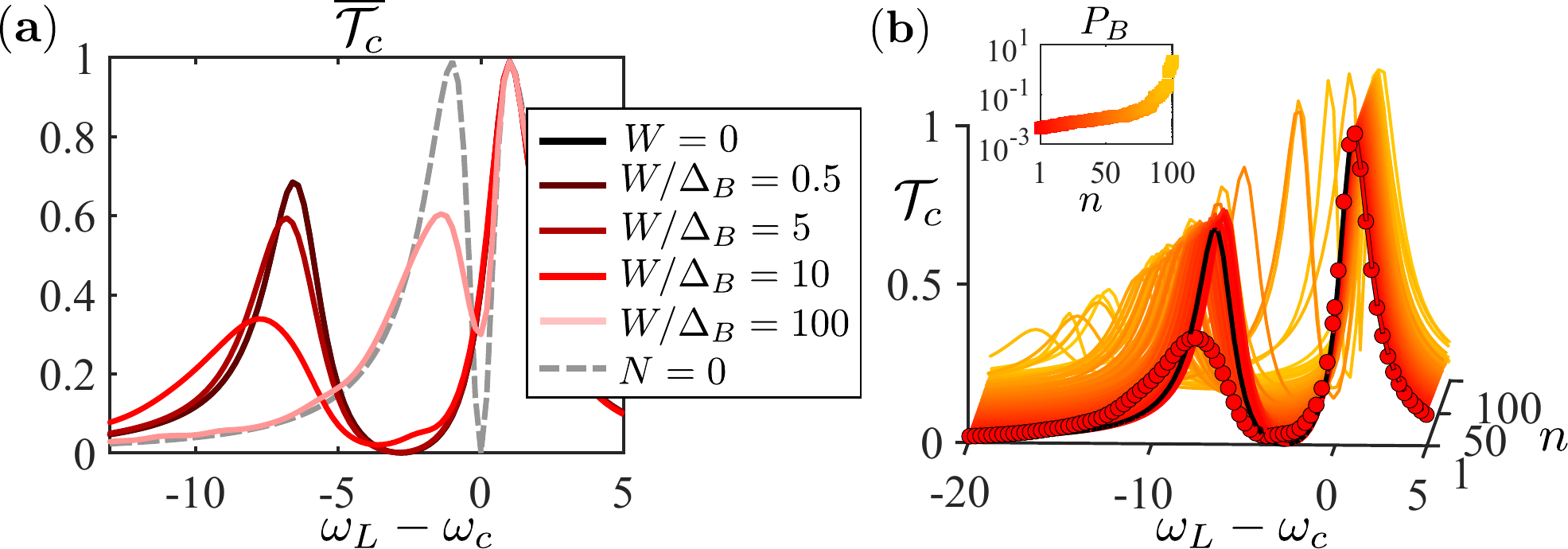}
	\caption{(\textbf{a}) Transmission spectrum $\overline{\mathcal{T}_c}$ averaged over $100$ disorder realizations of various strengths $W$, for the same parameters as in Fig.~1 of the main text ($N=12^3$). (\textbf{b}) $100$ realizations of the transmission spectrum $\mathcal{T}_c$ for $W/\Delta_{B}=10$. Realizations ($n~=~1,\dots,100$) are sorted from light red to yellow according to increasing ensemble population $P_{B}$ (inset), with $P_{B}\approx \vec{g}^T ({\bf M}{\bf M}^{\dagger})^{-1} \vec{g}$ assuming one cavity photon. The averaged spectrum $\overline{\mathcal{T}_c}$ is displayed as red circles, and the spectrum for $W=0$ as a black line.} 
	\label{fig:coll}
\end{figure}

The transmission spectra $\mathcal{T}_c$ obtained in the presence of inhomogeneous broadening and for different disorder strengths are shown in Fig.~\ref{fig:coll}({\bf a}). For $W/\Delta_{B} \gtrsim 1$, $\mathcal{T}_c$ features large realization-dependent fluctuations, and the breakdown of the adiabatic elimination (indicated by an increasing population of the ensemble) [Fig.~\ref{fig:coll}({\bf b})]. However, we note that some realizations can still lead to large $g_A^{\rm eff}$ (indicated by a large splitting with low ensemble populations). Also the disorder averaged $\overline{\mathcal{T}_c}$ still exhibits a well defined splitting even for $W/\Delta_{B} \gtrsim 1$, and recovers the spectrum of the bare target emitter only for $W/\Delta_{B} \gg 1$ [Fig.~\ref{fig:coll}({\bf a})]. 

\begin{figure}[ht]
\centering
\includegraphics[width=0.7\columnwidth]{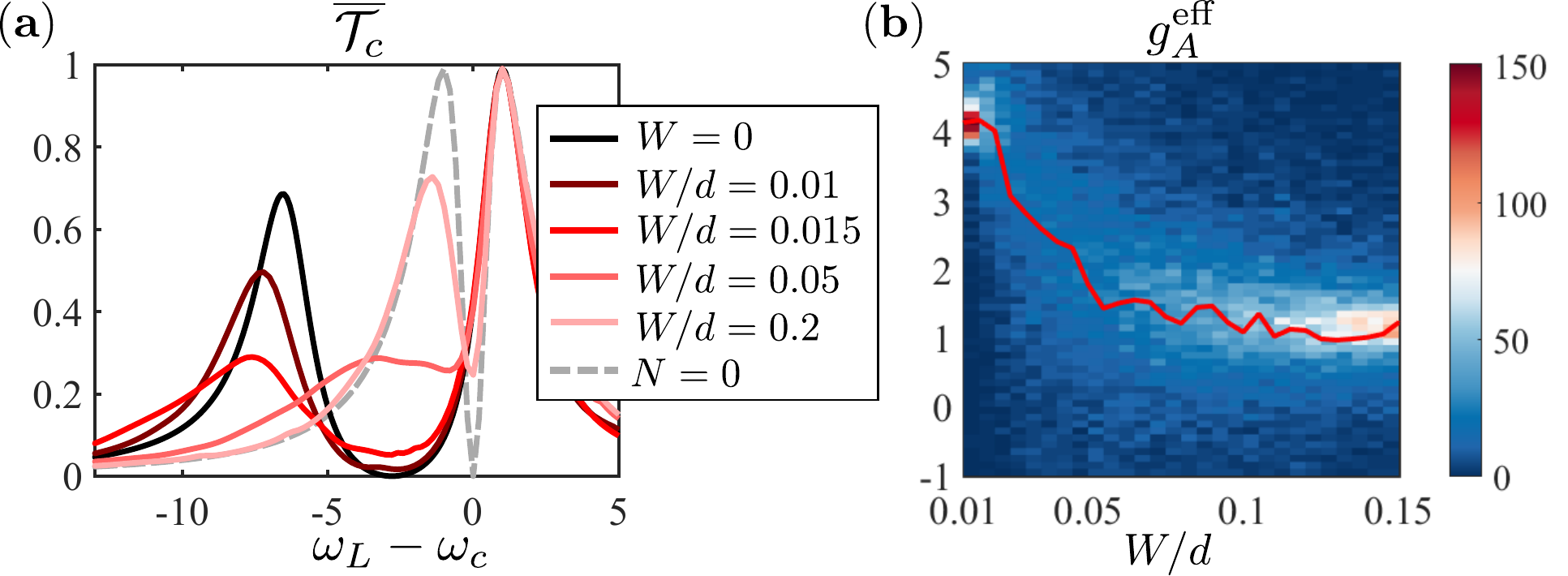}
\caption{(\textbf{a}) $\mathcal{T}_c$ averaged over $100$ disorder realizations of various strengths $W$, for the same parameters as in Fig.~1 of the main text ($N=12^3$). (\textbf{b}) Histograms of $g_A^{\rm eff}$ for various $W$, and $1000$ disorder realizations each (color coded) and corresponding mean-values (red line).}
\label{fig:pos}
\end{figure}

\vspace*{0.5cm}
\noindent {\bf Positional disorder.} Remarkably, we find similar robustness for $\overline{\mathcal{T}_c}$ averaged over positional disorder. In this case, fluctuations in the lattice positions of the cubic lattice are taken into account by picking random displacements around the regular lattice positions $X^{(0)}_{j}$, i.e. $X_j \in [X^{(0)}_{j}-W/2,X^{(0)}_{j} + W/2]$ along all dimensions ($X = x,y,z$). The averaged transmission spectrum $\overline{\mathcal{T}_{c}}$ is shown in Fig.~\ref{fig:pos}(\textbf{a}) for various disorder strengths in units of the lattice parameter $d$. These results are very similar to the case of inhomogeneous broadening [see Fig.~\ref{fig:coll}(\textbf{a})]. In Fig.~\ref{fig:pos}(\textbf{a}), for $W/d < 0.02$, we find a robust coupling enhancement, even with a tendency of slightly increased $g^{\rm eff}_{A}$. However, larger $W/d > 0.02$ destroy the enhanced Rabi splitting and one recovers the situation of the bare system $A$-cavity. The reason why the crossover between strong and weak coupling takes place at rather small disorder strengths compared to the case of inhomogeneous broadening stems from the large variations of the dipole-dipole interactions between neighboring sites. Indeed, the latter scales as $\sim 1/d^{3}$ since $d\ll \lambda$. An histogram of $g_A^{\rm eff}$ for various $W$ and $1000$ disorder realizations each is represented on Fig.~\ref{fig:pos}(\textbf{b}), showing large fluctuations for $0.02 \lesssim W \lesssim 0.1$, in the regime where the adiabatic elimination is likely to break down. However, even for large $W \gg 0.02$, realizations with large $g_A^{\rm eff}$ remain possible.

\bibliographystyle{apsrev4-1}
\bibliography{paperAB_arXiv}